\documentclass[11pt,preprint]{aastex}
\usepackage{natbib}
                                                                                                                            
\newcommand{\be}{\begin{equation}}
\newcommand{\ee}{\end{equation}}

\renewcommand{\vec}[1]{ {\bf #1} }

\newcommand{\Lbox}{L_{\rm box}}

\newcommand{\hinv}{{h^{-1}}}

\newcommand{\Msun}{\,\rm M_{\odot}}

\newcommand{\lya}{\mbox{Ly$\alpha$}~}
\newcommand{\laf}{\mbox{Ly$\alpha$ forest~}}
\newcommand{\hi}{\mbox{H\,{\scriptsize I}\ }}
\newcommand{\hii}{\mbox{H\,{\scriptsize II}\ }}
\newcommand{\hei}{\mbox{He\,{\scriptsize I}\ }}
\newcommand{\heii}{\mbox{He\,{\scriptsize II}\ }}
\newcommand{\heiii}{\mbox{He\,{\scriptsize III}\ }}
                                                                                                                            
\begin{document}

                                                                                                                            
\title{Late Reheating of the IGM by Quasars: A Radiation Hydrodynamical Simulation
of Helium II Reionization}
\author{Pascal Paschos \altaffilmark{1}, Michael L. Norman\altaffilmark{1,2} James O. Bordner\altaffilmark{1} 
\& Robert Harkness\altaffilmark{3} }
                                                                                                                            
\altaffiltext{1}{Laboratory for Computational Astrophysics, \\
Center for Astrophysics and Space Sciences,\\
University of California at San Diego, La Jolla, CA 92093, U.S.A.\\
Email:  ppaschos, mlnorman, jbordner@ucsd.edu}
\altaffiltext{2}{Physics Department, UCSD}
\altaffiltext{3}{San Diego SuperComputer Center}


\begin{abstract} 

We study the ionization and thermal evolution of the intergalactic medium
during the epoch of \heii reionization by means of radiation
hydrodynamical cosmological simulations. We post-process baryonic density
fields from a standard optically-thin IGM simulation with a homogeneous
galaxy-dominated UV background (UVB) which reionizes \hi and \hei at
z=6.5 but does not have any contribution to the ionization of \heii.
Therefore, we suppress the \heii photoheating contribution to the gas
temperature due to the homogeneous UVB. Quasars are introduced as point
sources throughout the 100 Mpc simulation volume located at cold dark
matter (CDM) density peaks consistent with the Pei luminosity function.
We assume an intrinsic quasar spectrum $J(\nu) \propto \nu^{-1.8}$ and a
luminosity proportional to the halo mass. We evolve the spatial
distribution of the \heii ionizing radiation field at h$\nu$ = 4, 8, and
16 Ryd using a time-implicit variable tensor Eddington factor radiative
transfer scheme. Simultaneously, we also solve for the local ionization
of \heii to \heii and the associated photoheating of the gas including
opacity effects.  We find that the percolation of the \heiii regions is
essentially complete by z=2.5. When comparing to a self-consistent
optically thin simulation at the same redshift, in which \heii is also
ionized by the uniform UVB, we find that inclusion of opacity effects
results in higher IGM temperature by a factor of approximately 1.7 at the
mean gas density level. We construct synthetic absorption line spectra
from which we derive statistical parameters of the \heii \laf. We use 300 
long ($\Delta z = 0.2$) random lines of sight to compute at $\bar{z} = 2.5 \pm 0.1$ a 
mean \heii \lya line transmission of $\bar{F} = 0.304 \pm 0.002$. 
The error corresponds to a significant one standard deviation 
in the transmitted flux due to the sightline to sightline variance equal to $\simeq 11\%$ 
the mean value. The opacity effect on the gas temperature is shown by
comparing the broadening width of the \hi and \heii \lya lines to the
results from the self-consistent optically thin simulation. We find a shift by
approximately 1.25 km/s to higher b-parameter values for both \hi and
\heii. Finally, we estimate the relative broadening width between the two
forests and find that the \heii median b-parameter is about 0.8 times the
median \hi broadening width. This implies that the \heii absorbers are
physically extended consistent with conclusions from observed lines of
sight.

\end{abstract}

\keywords{cosmology: hydrodynamical simulations: radiation transfer}

\section{Introduction}




The thermal evolution of the IGM after reionization is governed chiefly
by photoheating (Efstathiou 1992; Miralda-Escud\'{e} \& Rees 1994).  
Models for the thermal evolution of the IGM (Miralda-Escud\'{e} \&
Ostriker 1990; Giroux \& Shapiro 1996; Abel \& Haehnelt 1999)  generally
assume that \hi and \hei are nearly fully ionized by $z \sim 6$ by star
forming galaxies (Fan, Carilli \& Keating 2006), while \heii ionizes
somewhat later at z $\sim$ 3 by quasars due to \heii's higher ionization
potential and recombination rate (Sokasian, Abel \& Hernquist 2002;
hereafter SAH). Observational support for late \heii reionization is
summarized in SAH. Abel \& Haehnelt (1999) emphasized the importance of
opacity effects during reionization in establishing the post-reionization
temperature of the gas. They showed that models using the optically thin
expression for the photoheating rate during \heii reionization
underestimate the IGM temperature at mean density by a factor of $\sim$
2.

Hydrodynamical cosmological simulations of the \laf (Cen et al. 1994;
Zhang, Anninos \& Norman 1995; Hernquist et al. 1996; Miralda-Escud\'{e}
et al. 1996; Zhang et al. 1997, 1998; Theuns et al. 1998) generally adopt
the optically thin expression for photoheating for simplicity and
computational economy. This is a reasonable assumption for the \hi and
\hei photoheating at $z\sim3$ due to the low opacity of the IGM, but not
for the \heii photoheating as \heii is in the process of being ionized by
the percolation of \heiii spheres (SAH). These simulations therefore
underestimate the temperature of the IGM during the epoch of helium
reionization. The standard approach taken in these simulations is to
assume a homogeneous photoionizing background which evolves with redshift
consistent with observed quasar and galaxy counts, such as that by Haardt
\& Madau (2001). The ionization and thermal state of the baryonic gas is
then computed self-consistently with its dynamics by solving the
equations of hydrodynamic cosmology including radiative heating and
cooling (Cen 1992; Katz, Weinberg \& Hernquist 1996; Anninos et al.
1997). It is found that the temperature of the low density IGM is
determined almost entirely by photoheating balancing adiabatic cooling
due to cosmic expansion. This results in a tight relationship between gas
temperature and density, the so-called equation of state of the IGM (Hui
\& Gnedin 1997):  \begin{equation}
	T=T_0\Delta^\beta 
\end{equation}
where $\Delta=\rho/\bar{\rho}$ is the gas overdensity and $\beta$ is
redshift dependent but in the range $0 \leq \beta(z) \leq 0.6$. Abel \&
Haehnelt (1999) show that opacity effects during \heii reionization
raises $T_0$ and reduces $\beta$ relative to an optically thin
calculation. A technique often employed to include opacity effects within
hydrodynamic simulations is to simply multiply the \heii photoheating
rate by a constant factor $X_{\heii}\geq 1$, with a value of 2-4 being
sufficient to match observed temperatures (e.g., Bryan \& Machacek 2000;
Jena et al. 2005).
                                                                                                                                                                               
The combination of high spectral resolution quasar absorption line
observations and hydrodynamical cosmological simulations provide a means
for measuring the thermal evolution of the IGM.  The thermal state of the
gas is generally deduced from \hi \lya linewidths (b-values) in high
resolution spectra
(Rauch et al. 1997; Schaye et al. 1999, 2000; Bryan \& Machacek 2000;
Theuns et al. 2000; Bolton et al. 2005; Jena et al. 2005) although the
flux power spectrum has also been employed (Zaldarriaga, Hui \& Tegmark
2001). If the temperature of the IGM alone determined the b-values, then
it would be straightforward to measure the temperature from high
resolution spectra. However, Hubble broadening is always of the same
order as the thermal broadening in \laf absorption lines (Schaye 1999),
hence the need for comparison with simulations. Theuns, Schaye \&
Haehnelt (2000) used the b-parameter distribution to measure the
temperature of the IGM at z=3.25, finding $T_0 \geq 15,000$ K. Schaye et
al. (2000) used the lower cutoff in the linewidth--column density scatter
diagram to measure the temperature evolution of the IGM over the redshift
range 2-4.5. They found evidence of late reheating at z$\sim$3, which
they ascribed to late \heii reionization by quasars. Bryan \& Machacek
(2000) independently explored the same diagnostics and found that
temperature estimates were sensitive to the assumed cosmology, in
particular the amplitude of mass fluctuations on a few Mpc scales.
Zaldarriaga, Hui \& Tegmark (2001) used the falloff of the \lya forest
flux power spectrum at small scales to measure the IGM temperature, and
found $T_0 \sim 2 \times 10^4$ K, dependent on their assumed $\beta$.

A proper calculation of \heii reionization would treat quasars as point
sources within the simulation volume and solve the equation of radiative
transfer for the spatial distribution of the ionizing background coupled
self-consistently to the dynamical, thermal, and ionization evolution of
the gas. This was done in an approximate way by SAH, who post-processed a
series of density fields taken from a SPH hydrodynamical simulation of
the IGM using the GADGET code. The hydrodynamical simulation used the
optically thin prescription for ionizing and heating the gas, assuming
all ionization states of hydrogen and helium were in ionization
equilibrium with the UVB of Haardt \& Madau (1996). In the
post-processing step, the helium reionization calculation was recomputed
for an evolving quasar source population treated as point sources within
the volume. Quasars were assumed to have a constant lifetime of $10^7$
yr. Every $10^7$ yr, peaks within the density field of a suitably chosen
data dump from the hydrodynamic simulation were populated with quasars
consistent with an empirical luminosity function. Around each point
source the static equation of radiative transfer for \heii ionizing
photons was solved using a photon-conserving ray-casting scheme. The
ionization state of \heii was then updated ignoring thermal feedback to
the gas. SAH found that for reasonable parameter choices, \heii
reionization occurred in the range $3 \leq z \leq 4$ consistent with
observations.

In this paper we present a simulation which is similar in spirit to SAH,
but with several important differences. We also postprocess a series of
snapshots from a hydrodynamic cosmological simulation with a radiative
transfer code, however we keep track of the photoheating of the IGM as
expanding \heiii spheres percolate and eventually merge. Our hydrodynamic
simulation was performed on an Eulerian grid of $512^3$ cells versus
SAH's $224^3$ in the same volume, giving us $\sim 12$ times the mass
resolution and $\sim 2.3$ times the spatial resolution in the low density
IGM. We have also separated the effects of stellar and quasar populations
of the ionization evolution of the IGM differently from SAH. Our
hydrodynamic simulation computes the ionization of \hi and \hei due to
stellar sources only using the homogeneous UVB of Haardt \& Madau (2001)
arising from galaxies "GAL". In the post-processing step, quasars are
treated as point sources, which ionize \heii to \heiii. Although our
radiative transfer scheme is based on completely different spatial and
angular discretizations as SAH, the important difference for the purpose
of this paper is that we solve the RT equation at three frequencies
$h\nu$=4, 8 and 16 Ryd in order to evaluate, albeit crudely, the local
\heii photoheating rate taking the processed QSO spectrum into account.

The outline of this paper is as follows. In \S2 we describe the
cosmological realization for the hydrodynamic simulation of homogeneous
\hi and \hei reionization described in \S3. In \S2 we also introduce some
physical concepts relating to late \heii reionization and also describe
our treatment of quasars in the simulation volume. In \S4 we describe our
method for simulating inhomogeneous \heii reionization. In \S4.1 our
radiative transfer scheme is detailed, in \S4.2 we present our single
species ionization model for \heii, and in \S4.3 we present results in
terms of globally integrated quantities.  Since we have split the
calculation into two phases, the homogeneous reionization of \hi and
\hei, followed by the inhomogeneous reionization of \heii, our
calculation is not fully self-consistent. Although we keep track of the
late reheating, we do not modify the underlying density fields nor do we
alter temperature-dependent recombination rates, which will affect the
detailed ionization state of the gas. In \S5 we present an analysis of
our main result, which is the late reheating of the IGM due to
inhomogeneous \heii reionization, as well as the importance of neglecting
these coupling effects. We show that these effects, while present, are
small, and do not seriously undermine our estimate of the reheating. In
\S6 we present observational signatures of \heii reionization based on
synthetic \hi and \heii \lya absorption line spectra derived from our
simulation. In \S7 we summarize our main results and conclude. In a
series of appendices we derive the rate equation for species fraction in
an expanding universe (Appendix A), provide more detail on the affect of
neglecting \heii photoheating on the ionization state of the gas
(Appendix B), and document the reaction rate coefficients we use
(Appendix C).
                                                                                                           
\section{Simulations}

\subsection{Cosmic Realization}

\begin{figure*}
\includegraphics[width=3in,height=3in]{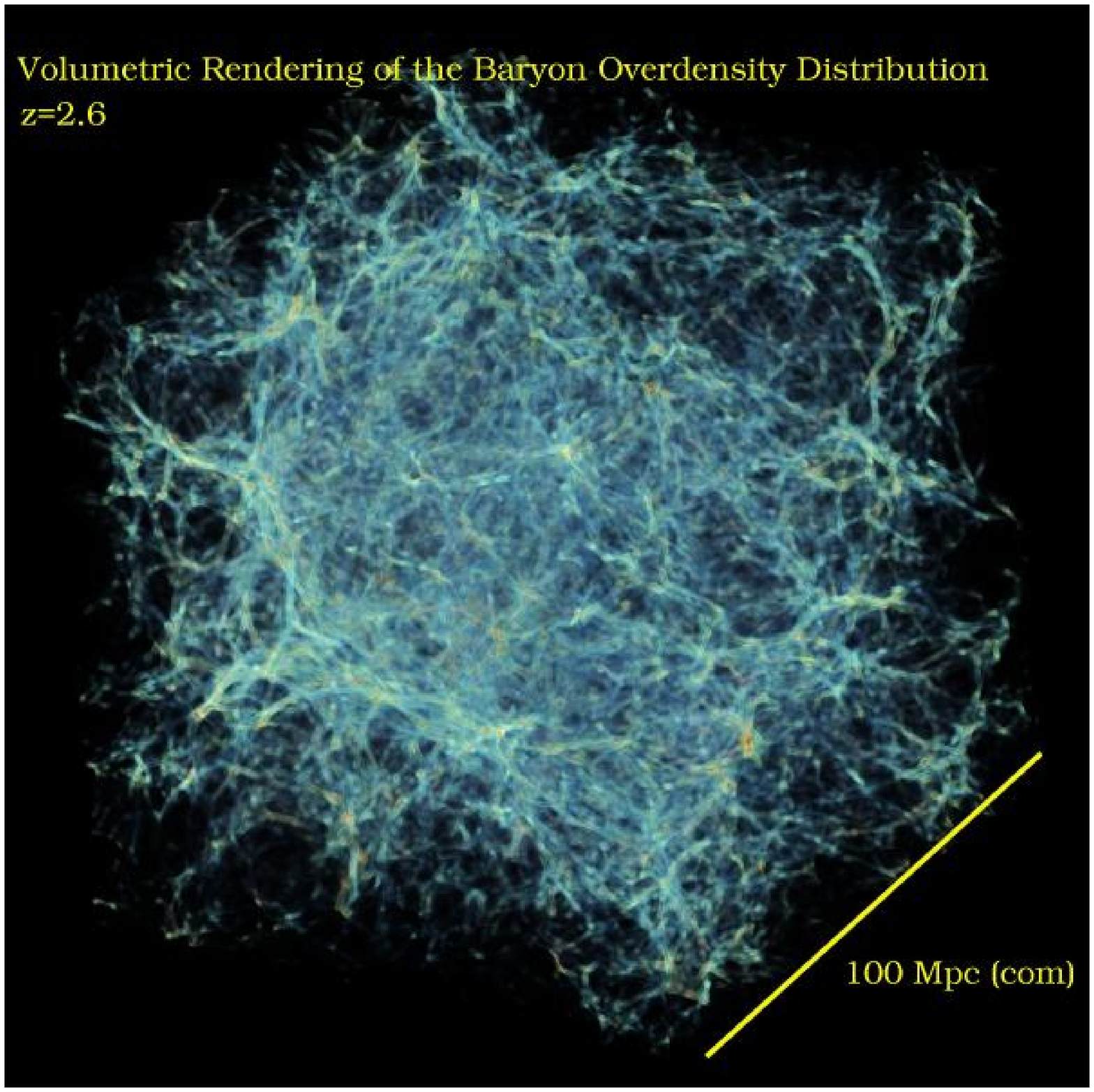}
\includegraphics[width=3in,height=3in]{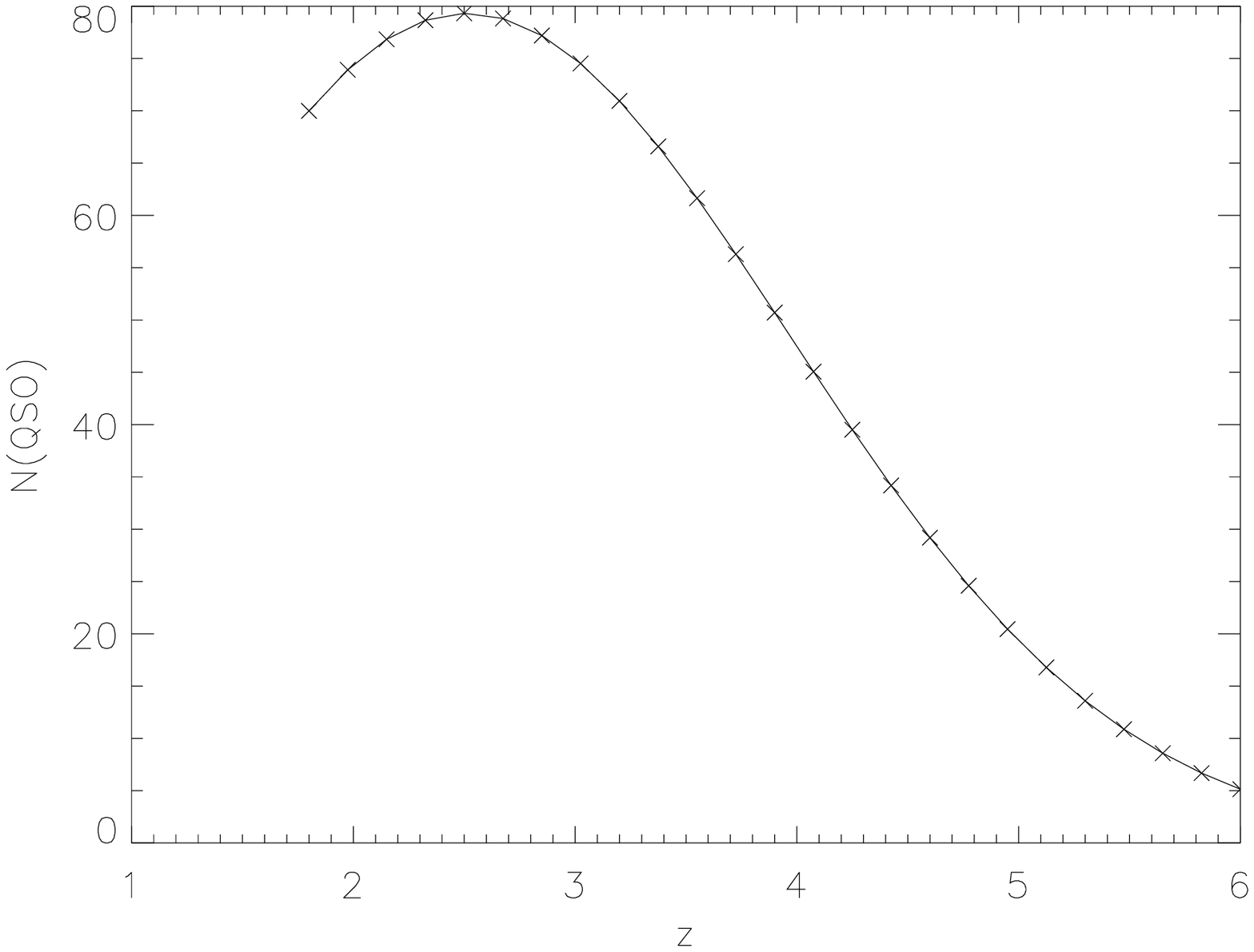}
\caption{Left Panel: Volumetric rendering of the logarithm of baryon overdensity
at z=2.6. Right Panel: Evolution in the number of QSO sources versus redshift in this simulation} \label{ch5_volfig_nqso}
\end{figure*}

In this work, we present the results from post-processing the redshift
evolution of a cosmic realization computed with the Eulerian cosmological
hydrodynamic code Enzo (Bryan \& Norman 1997; Norman \& Bryan 1999;
O'Shea et al. 2004). The box of size 67$h^{-1}$ Mpc comoving was evolved
in a flat ($\Omega =1$) $\Lambda$CDM cosmology on a unigrid mesh of
$512^{3}$ grid cells and $512^{3}$ dark matter particles from z=99 to $z
\simeq 2$. We used the initial power spectrum of matter fluctuations by
Eisenstein \& Hu (1999) to initialize the calculation which was then
computed forward under the Zel'dovich approximation (Bertschinger \& Gelb
1991). Our choice of cosmological parameters is $\sigma_{8}=0.8$,
$\Omega_{\Lambda} =0.7$, $\Omega_{b} = 0.04, n_s=1$ with a present day
Hubble constant of $h=0.67$ in units of 100 km/s/Mpc. With these
parameters our cosmic realization has a mesh resolution of $130h^{-1}
\simeq 200$ kpc within the 100 Mpc cube, and $\simeq 1.6h^{-1} \times
10^{8}$ $M_{\odot}$ dark matter particle mass. In
Figure~(\ref{ch5_volfig_nqso}) we show a volumetric rendering of the
cosmic gas distribution as mapped by the baryon overdensity in our
simulation at z=2.6.

One of the fundamental simplifications used here, which places limits on
the validity of our results is that the ionization of \hi and \hei are
treated entirely differently than \heii. The motivation is that of
numerical simplicity and calculation speed.  From the IGM cosmological
evolution standpoint, hydrogen and neutral helium are believed to be
globally ionized at an earlier cosmic epoch than \heii. In the simulation
discussed here, the photo--ionization/heating rates of \hi and \hei are
computed self-consistently during the hydrodynamical calculation in the
optically thin regime. The premise of our argument is that, if one adopts
a picture where neutral hydrogen and neutral helium reionization is
completed by $z \simeq 6$, such species will have large ionization
fractions by the time \heii reionization occurs, which theoretical models
and observations place at $z \simeq 3-2$.

The reionization of the \hi/\hei species is achieved by the evolving
uniform metagalactic flux due to stellar sources as computed in Haardt \&
Madau (2001).  The uniform ultra-violet background photo-ionizes and
photo-heats the IGM, however it is prevented by hand to radiatively alter
the \heii abundance. We do so by suppressing the ionization/heating rates
in Enzo that control the \heii $\leftrightarrow$ \heiii chemistry.  The
latter is computed separately by our inhomogeneous point-source
distribution of QSOs which is discussed in \S\ref{subqso}.

The initial calculation represents an undisturbed cosmological ensemble
of gas fields ionized by a distributed galaxy population. The local QSO
component then acts as a perturbation in the amplitude of the radiative
energy that further ionizes and photoheats the diffuse IGM. The end
result is that the \heii reionization proceeds, under the limitations
discussed above, via the mergers of individual \heiii I-fronts during the
cosmic epoch that spans the redshift interval $z=6-2$.

\subsection{Why Late \heii reionization}\label{whylate}

In this section we introduce some definitions and basic results related
to \heii reionization that we will refer to later on. The softness
parameter of the cosmic radiation is defined as $S =
\frac{\Gamma_{HI}}{\Gamma_{HeII}}$. In the optical thin limit under a
single power law profile for the radiation spectrum, $J_{\nu} = J_{912}
(\frac{\nu}{\nu_{912}})^{-\alpha_{q}}$, where $J_{912}$ denotes the
volume averaged mean intensity at the hydrogen ionization threshold, we
can estimate the photoionization rates to be $\Gamma_{HI} \equiv
\Gamma_{1} = \frac{4 \pi J_{912} \sigma^{o}_{HI}}{h}
\frac{1}{1+\alpha_{q}}$ and $\Gamma_{HeII} \equiv \Gamma_{2} = \frac{4
\pi J_{912} \sigma^{o}_{HeII}}{h} \frac{1}{1+\alpha_{q}}
4^{-\alpha_{q}}$. Dividing the last two relations yields an estimate for
the softness parameter in the optical thin limit equal to $S =
4^{1+\alpha_{q}}$. Therefore, the ability of a background radiation field
to ionize the \hi and \heii species depends on the spectral slope. A soft
radiation field would have large spectral slope and would primarily favor
hydrogen ionization. \heii ionization therefore requires a hard ionizing
spectrum.

The available sources of radiation at redshifts $ z \leq 10$ are of two
types.  Stellar sources, associated with radiation from galaxies forming
in the cosmic medium at such redshifts, have large softness parameter
values of $S_{stellar} \sim 4 \times 10^{3}$. QSOs on the other hand,
have much smaller softness parameter values of $S_{QSO} \sim 50$ which
makes them ideal for \heii ionization. However, the number density of
QSOs, as measured by observations sharply rises in the interval $6 > z >
3$ and reaches a peak at about $z \approx 3$ (Pei 1995). The highest
redshift QSO observed to date is at $z_{em} = 6.56$ (Hu, Cowie \& McMahon
2002). Direct observations of QSO in Ly$\alpha$ spectra are very
difficult because of the hydrogen Gunn-Peterson effect, the optical depth
manifestation of hydrogen neutrality at high redshifts (Fan, Carilli \&
Keating 2006). This suggests that either the quasars have a well
established population by $z \approx 6$ but nevertheless obscured by the
large optical depth or that their formation began at that epoch. However,
when their emissivity is computed at redshifts just past the
Gunn-Peterson trough, their numbers and spectrum shape is insufficient to
produce the hydrogen reionization thought to be primarily achieved by a
stellar component in the UVB which ionizes hydrogen and neutral helium to
the singly ionized state. Because the stellar component of the radiation
background is not an efficient \heii ionizer, \heii reionization is
expected to occur later; at times when a sufficient number hard photons
becomes available.

The Gunn-Peterson optical depth of \heii for a uniform IGM can be
expressed in the same way as the corresponding \hi and that leads to the
ratio $\frac{\tau_{HeII}(z)}{\tau_{HI}(z)} = \frac{n_{HeII}}{n_{HI}}
\frac{\sigma_{HeII}} {\sigma_{HI}} = \frac{1}{4}
\frac{n_{HeII}}{n_{HI}}$. The ratio of optical depths is known as the
R-factor and is used in observations to measure the relative properties
between the \heii and \hi Ly$\alpha$ forest spectra. Observations find
that in transmission spectra longward of quasars seen at emission
redshifts $z \sim 3$ typically measure values of the R-factor between $R
\approx 10 - 100$ within $\delta z \sim 0.5$ (Reimers et al. 1997; Heap
et al. 2000; Kriss et al. 2001). This suggests the presence of a HeII
Gunn-Peterson trough at redshifts $ z \gtrsim 2.5$. In the optical thin
limit the R-factor and the softness parameter relation can be derived.

In ionization equilibrium the ratio of $n_{HeII}/n_{HI}$ can be computed
if we assume that the higher ionization states of both species (\heiii,
\hii) dominate.  In that case, for hydrogen we get $\chi_{1} \Gamma_{1} =
n_{e} \alpha_{1} (1-\chi_{1}) $ where $\chi_{1} = n_{HI}/n_{H}$ and
$\alpha_{1}$ is \hii the recombination coefficient. Similarly for \heii
we can also write $\psi_{2} \Gamma_{2} = n_{e} \alpha_{2} (1-\psi_{2})$,
where $\psi_{2} = n_{HeII}/n_{He}$ and $\alpha_{2}$ is the \heii
recombination coefficient. The above two relations yield
$\chi_{1}/\psi_{2} \simeq \frac{\Gamma_{2}}{\Gamma_{1}}
\frac{\alpha_{1}}{\alpha_{2}}$, for $\chi_{1} \ll 1$ and $\psi_{2} \ll
1$. The last equation allows for the determination of the ratio
$n_{HI}/n_{HeII} = \frac{\chi_{1} n_{H}}{\psi_{2} n_{He}} \simeq 12
\times \frac{ \Gamma_{2} \alpha_{1} } {\Gamma_{1} \alpha_{2}}$.
Therefore, for $\frac{\alpha_{2}}{\alpha_{1}} \sim 5$ (at $T \approx
10^{4}$ K) the ratio of number densities becomes $\frac{n_{HeII}}{n_{HI}}
\simeq \frac{5}{12} S$. From the last relation it follows that $R \simeq
0.1~S$. For $S \simeq 4 \times 10^{3}$ (stellar radiation) the relation
predicts $ R \simeq 400$. Similarly for quasar radiation, with $S \simeq
50$, $R \simeq 5$. In addition, we can compute that since the quantity
$n_{e} \alpha(T)/\Gamma = \tau^{ion}/\tau^{rec}$, the ratio of the
ionization to recombination time scales,
$\frac{\tau^{ion}_{HeII}}{\tau^{rec}_{HeII}} =
5S~\frac{\tau^{ion}_{HI}}{\tau^{rec}_{HI}}$

Typical softness parameter values $S \geq 100$ then yield
$\frac{\tau^{ion}_{He_{II}}}{\tau^{rec}_{He_{II}}} \geq 5\times10^{2}
\frac{\tau^{ion}_{H_{I}}}{\tau^{rec}_{H_{I}}}$. The last relation shows
that, due to the larger recombination coefficient and the fewer number of
photons available for ionization, it is more difficult to ionize \heii
compared to \hi. The shorter (longer) recombination (ionization)  time
scale effectively restricts \heii reionization to take place at a slower
rate.

Large values in the R-factor is suggestive of large optical depths in
\heii and/or small optical depths in neutral hydrogen. Comparable optical
depths at late redshifts, which correlate to small cosmic neutral
hydrogen fraction and therefore small cosmic fraction in \heii, yield
small values for the R-factor. The observations can therefore infer a
range in the observed softness parameter from measuring the ratio of
optical depths in the \heii and \hi line forest.  The observed range of
R=10-100 consequently suggests that S=100-1000 between z=2-3 although the
upper limit is an overestimate because we assumed $\psi_{2} \ll 1$.
Because such observed values are sampled in transmission spectra that
probe the ionization phase transition of \heii to \heiii, we can infer
that at that epoch the galaxy dominated ultraviolet background is
gradually being replaced by a quasar dominated type.

In \S\ref{obs}, we show that the R-factor evolves with redshift from
large to small values as it is computed in transmission through the
computational volume. The quantity that is actually computed is the
$\eta$ parameter defined as the column density ratio between \heii and
\hi, $\eta = \frac{N_{HeII}}{N_{HI}}$. The R-factor evolution and value
follows directly as $R = \frac{\eta}{4}$.  The gradual evolution of the
R-factor and $\eta$ parameter from large values is indicative of the
cosmic evolution of the softness parameter from a stellar to a quasar
dominated type.

\subsection{Quasar Placement and Evolution Pre-processing}\label{subqso}

For the quasar placement and evolution in the simulation we use the
quasar luminosity function by Pei et al. (1995), shown in
Equation~(\ref{peieq}).  This luminosity function was also used by Haardt
\& Madau (1996,2001) to derive the quasar emissivity and the volume
average photoionization and heating rates used in the simulation for
every species other than \heii.

\begin{equation}
\phi(L,z) = \frac{\phi_{*}/L_{*}(z)}
{[L/L_{*}(z)]^{\beta_{1}}+[L/L_{*}(z)]^{\beta_2}}
\label{peieq}
\end{equation}

\[
L_{*}(z) = L_{*}(0)(1+z)^{\alpha_q-1} 
\frac{e^{\zeta z}
(1+e^{\xi z_{*}})}{e^{\xi z}+e^{\xi z_{*}}}
\]

Our first requirement is the placement of a single QSO in the
computational volume at z=6.5. The redshift is for all practical purposes
a matter of choice, since there is no accurate prediction of when the
first QSO appears in the universe. We adopt a scenario where quasars
become visible in observations after the epoch of hydrogen reionization
is completed by $z \simeq 6.5$ due a soft component in the ultra-violet
background most likely associated with dwarf galaxy formation. In
addition, we are constrained by the mass resolution of our simulation,
which cannot resolve halos smaller than $ \simeq 5 \times 10^{9}$
$M_{\odot}$. Therefore, quasar point sources have to be placed in the
centers of halos above this cutoff. We do so by identifying the evolution
of a list of halo centers through the computational volume from $z=6.5$
to $z=2$. The number of quasars per redshift interval is determined by
the luminosity function in Equation~(\ref{peieq}) and the functional form
of the mass-halo to quasar luminosity relation.

We assume an intrinsic quasar spectrum
$J_{\nu}=J_{912}(\frac{\nu}{\nu_{912}})^{-\alpha_{q}}$ for the LyC part
of the spectrum with a spectral index $\alpha_{q} = 1.8$. Therefore, the
number flux of emitted LyC photons, in $photons/s/cm^{2}$, is then
$\dot{n}_{ph} = \frac{4 \pi J_{912}}{h} \int_{1}^{\infty}
\frac{\epsilon^{-\alpha_{q}}} {\epsilon} d\epsilon$. In this relation,
$\epsilon = \frac{h\nu}{h\nu_{912}}$. The integration yields
$\dot{n}_{ph} = \frac{4 \pi J_{912}}{h \alpha_{q}}$. Similarly the LyC
energy flux is $l_{LyC} = \frac{ 4 \pi \nu_{912} J_{912}}{\alpha_{q}-1}$,
which yields $l_{LyC} = (h \nu_{912}) \dot{n}_{ph}
\frac{\alpha_{q}-1}{\alpha_{q}} \Rightarrow L_{1} = (h \nu_{912})
\dot{N}_{ph} \frac{\alpha_{q}-1}{\alpha_{q}}$. In the last equation,
$L_{1}$ and $\dot{N}_{ph}$ represent the total LyC luminosity (ergs/s)
and emitted photon rate (photons/s) respectively per point source (QSO)
above the \hi ionization threshold of 1 Ryd (13.6 eV).  This allows for a
parametrization of the emitted ionizing flux based on the number of LyC
photons rather than energy. The motivation is entirely for consistency
with the photon-conserving schemes in simulating reionization (Abel \&
Haehnelt 1999; Sokasian, Abel \& Haehnelt 2001, 2002;  Ciardi et al.
2003; Whalen, Abel \& Norman 2004). The total luminosity for the \heii
ionizing radiation above 4 Ryd (54.4 eV) is then obtained through $L_{4}
= 4^{-(\alpha_{q}+1)} \times L_{1}$. The luminosity of each source is
determined by the dark matter mass of the halo that initially creates it.
For a dark matter halo of mass $M_{halo}$ we put in a UV source emitting
$\dot{N}_{ph} = 10^{51} \times \frac{M_{halo}}{10^{8}}$ \hi ionizing
photons/s. This is equivalent to placing a $\dot{N}_{ph} = 10^{51}$
$ph/s$ mini-quasar inside a $10^{8}$ $M_{\odot}$ dark matter halo which
is the prescription used in Abel \& Haenhelt (1999). Sokasian et al.
(2001) investigated an array of QSO placement methods in the
computational volume and found that the results are largely insensitive
to the choice. We adopt the linear relation for convenience. In this
work, the most massive halo computed at at z=2.5 has mass $M_{halo}
\approx 6 \times 10^{13}$ $M_{\odot}$. If a QSO source is placed there
then it will emit LyC photons at a rate of $\dot{N}_{ph} = 6 \times
10^{56}$ $s^{-1}$. For an input spectrum with slope $\alpha_{q} = 1.8$
the photon rate corresponds to a LyC luminosity of $L_{1} = 2.9 \times
10^{46}$ ergs/s and $L_{4} = 6.1 \times 10^{44}$ ergs/s for \hi and \heii
respectively.
                                                                                
The placement of the point sources in the volume is dynamical in nature.
The list of dark matter halos is assigned a quasar source with a
luminosity value determined by our phenomenological prescription. The
location of the quasar from that point on is locked to the position of
the dark matter particle closest to the center of the halo. The
distribution of luminosities is then integrated from the higher value to
the smallest up to the point where the average luminosity per unit volume
reproduces the distribution fit given by Equation~(\ref{peieq}) at each
redshift. For simplicity, we do not evolve the luminosity in each dark
matter halo, which remains the same at initialization.  As the luminosity
function increases with decreasing redshift additional sources are
spawned in the simulation, leading to an overall increase of their
numbers. At $z \lesssim 3$ the flattening and subsequent decrease in the
luminosity function is modeled by randomly removing point sources from
the quasar list. In the right panel of Figure~(\ref{ch5_volfig_nqso})  
we show the redshift evolution in the number of QSO sources in the
volume.

The location each point source is used to compute the 3D distribution of
the tensor Eddington factor (\S4.1) which is stored at each cosmic time
step.  The local value $f_{ij}(\bold{x})$ at $t^{n}$ is an interpolated
value between the local values of the two data dumps whose redshifts
bound the instantaneous time ($z_{1} \leq t^{n} \leq z_{2}$).  As we will
discuss in the next sections, the cosmic evolution of the helium
reionization is decoupled from the corresponding hydrogen one.
Consequently, the placement of a \heii ionizing source on a density peak
with a precomputed hydrogen and therefore electron density can yield
significant discrepancy between our calculation and a self-consistent
one, particularly in close proximity to the source.

\section{Homogeneous Hydrogen Ionization}\label{h1r}

As mentioned above, in the numerical and physical setup of this present
work, we treat the ionization of \hi and \hei separately from \heii. The
metagalactic flux we use from Haardt \& Madau (2001) tabulates the
contributions due to galaxies (GAL) and quasars (QSO) separately. In Enzo
we have the option of running a simulation with GAL only, QSO only, or
GAL+QSO. In a standard optically thin simulation the GAL+QSO UVB would be
used. We have carried out such a simulation, hereafter called Simulation
A, for comparison with the inhomogeneous reionization simulation. For the
latter, we first run a hydro simulation using the GAL UVB to ionize \hi
and \hei. Then \heii reionization is accomplished by treating quasars as
point sources as detailed in \S4. Since our quasar population follows the
same luminosity function and intrinsic spectrum as assumed by Haardt \&
Madau, we are able to compare the homogeneous and inhomogeneous
simulations directly.  We consider the effect of the $\epsilon \geq 54.4$
eV photon field as a perturbation on the previously ionized gas which
affects only the \heii $\Leftrightarrow$ \heiii chemistry while the \hi
and \hei ionization states remain unaffected.  The treatment is by all
measures an approximation that allows the problem to be solved as a
single species ionization problem only and therefore it remains
conceptually simple. In reality, the ionization of \heii has two
inter-dependent effects; it releases additionally one electron per \heii
ionization which in turn, when thermalized, raises the mean temperature
of the IGM. These two effects combined would in principle shift the
ionization balance of the \hi/\hii and \hei/\heii species. However, we
show in this section that because by the time this takes place hydrogen
and helium have already large ionization fractions the aforementioned
effects are small.

Starting with the rate equation for hydrogen and the cosmic mean density,
we can safely ignore the collisional contributions and write the chemical
balance in the proper frame of reference as follows:

\begin{equation}\label{hieqn}
\dot{n}_{HI} = -3H(z)n_{HI} -n_{HI} \Gamma_{1} + n_{e} n_{HII} \alpha_1(T).
\end{equation}

In Equation~(\ref{hieqn}), $n_{e} = n_{HII} + n_{HeI} + 2n_{HeII}$,
$\Gamma_{1}$ is the integrated \hi photoionization rate, and
$\alpha_1(T)$ is the radiative recombination coefficient in the $\hii + e
\rightarrow \hi + \gamma$ reaction and is a function of temperature. In a
cosmic medium, $n_{He} = f n_{H}$, where $f \simeq \frac{1}{12}$ for a
mass of fraction of $\rho_{H} = 0.75 \rho$ and $\rho_{He} = 0.25 \rho$
respectively. The \heii number density can then be rewritten as $n_{HeII}
= n_{He} - n_{HeI} - n_{HeIII}$ which in turn allows the electron density
to be expressed as follows:

\begin{equation}\label{neeqn}
n_{e} = n_{H} [ \chi_{HII} + \frac{ ( 1-\chi_{HeI}+\chi_{HeIII})}{12}]
\end{equation}

Changing slightly the notation, we can rewrite Equation~(\ref{hieqn}) according
to Appendix~(A) in terms of the 
ionization fractions $\chi_{2} = n_{HII}/n_{H}$, $\psi_{1} = n_{HeI}/n_{He}$,
$\psi_{2} = n_{HeII}/n_{He}$ and $\psi_{3} = n_{HeIII}/n_{He}$ as:

\[
\dot{\chi}_{2} = \Gamma_{1}(1-\chi_{2}) - \chi_{2} n_{H} \alpha_1(T) [ \chi_{2} + \frac{1}{12}
(1-\psi_{1}+\psi_{3})] \Rightarrow 
\]
\begin{equation}\label{hieqn2}
\dot{\chi}_{2} \approx \Gamma_{1}(1-\chi_{2}) -\chi_{2} n_{H} \alpha_1(T)
[\chi_{2} + \frac{1+\psi_{3}}{12}]
\end{equation}

In the last equation, we assumed that almost all of the helium is highly
ionized to the \heii state $\psi_{1} \rightarrow 0$. In ionization
equilibrium ($\dot{\chi}_{2} = 0$), after setting $A \equiv
(1+\psi_{3})/12$ we can rewrite Equation~(\ref{hieqn2}) as
$(1+\frac{\chi_{2}}{A}) \frac{\chi_{2}}{1-\chi_{2}} = \Gamma_{1}/(A
n_{H}\alpha_1)$. On one hand, the ionization of the \heii would increase
quantity A due to the increase in the \heiii abundance. On the other
hand, the increased temperature would decrease the recombination
coefficient which can be approximated as $\alpha_1 \propto T^{-\beta}$,
for temperatures in the range of $T \simeq 10^{3}-10^{5}$ K and $\beta =
0.51$.  The approximation is based on expression fits by Bugress (1964).

Therefore, we will investigate two extreme cases.  Case~(I) represents
zero \heiii abundance, $\psi_{3} \rightarrow 0$ and would correspond to
the unperturbed temperature $T_{I}$. Case~(II) represents a limit of
almost full \heii ionization, $\psi_{3} \rightarrow 1$, that would
correspond to a new temperature $T_{II}$. In both cases, the hydrogen
ionization rate is the same because our treatment of the individual QSO
sources placement has a distribution that is statistically identical to
the global average used for \hi and \hei ionization.  Therefore, we can
form the ratio:

\begin{equation}\label{hieqn3}
\frac{(1+\chi^{II}_{2}/A_{II}) \chi^{II}_{2} (1-\chi^{I}_{2})}
{(1+\chi^{I}_{2}/A_{I}) \chi^{I}_{2} (1-\chi^{II}_{2})}  = 
(\frac{T_{II}}{T_{I}})^{\beta} \frac{A_{I}}{A_{II}}
\end{equation}

Equation~(\ref{hieqn3}) is solved in Appendix~(B), for $\chi^{II}_{2}$ in
the range of $1-\chi^{I}_{2}=10^{-6}-10^{-1}$ and for $A_{II}=1/6$,
$A_{I} = 1/12$, $T_{II}/T_{I} = 1.5-2$. The ratio of temperatures is
based on estimates of the temperature increase due to the photoelectrons
injected in the IGM from the ionized \heii atoms (Haehnelt \& Steinmetz
1998; Abel \& Haehnelt 1999) and will be discussed in more detail in
\S\ref{he2tm}. The results in the Appendix figures show that when
hydrogen is highly ionized then the shift in the ionization balance due
to the \heii ionization results in a decrease of the neutral hydrogen
fraction primarily due to the decrease of the ionized hydrogen's
recombination coefficient.

The $\approx 15-25$\% reduction depends on the temperature increase,
where the largest increase (a factor of 2) produces the biggest shift.
The fractional decrease in the neutral hydrogen fraction is smallest at
low fractions of ionized hydrogen.  For a typical neutral hydrogen
fraction of $\chi_{1} \simeq 10^{-5}$ at post-reionization redshifts, we
can then estimate that our treatment (of not updating the hydrogen
abundances)  would overestimate the neutral hydrogen fraction by about
25\% if the gas temperature increase is between 1.5-2 times the
non-perturbed value. Consequently, we anticipate an overestimate by the
same amount in the optical depth of hydrogen Ly$\alpha$ radiation and an
underestimate in the mean transmitted flux. The problem is similar to the
one described in Zhang et al. (1997), Bryan et al. (1999) and Jena et al.
(2005). An unmodified UVB by Haardt \& Madau (1996), applied in those
simulations, would not yield an agreement with the observed b-parameter
of the Ly$\alpha$ forest. An adjustment by a factor of 1.5-2.0 ~in the
\heii photoheating rate was then necessary to match the observed results.
Therefore, we caution against a strict interpretation of the resulting
hydrogen Ly$\alpha$ forest in the original calculation in which we
suppressed the \heii photoionization and photoheating processes
altogether.

\section{Inhomogeneous Helium II Reionization}\label{he2r}

We compute the inhomogeneous \heii reionization due to an evolving
distribution of local QSO-type sources by solving for the time and
spatial evolution of the ionizing radiation energy density
$E_{\nu}(\bold{r},t)$ as shown in following section. In \S\ref{chem} we
describe our simplified chemistry model, and in \S\ref{chemresults} we
present results. We defer a discussion of late photoheating to
\S\ref{he2tm}.

\subsection{Radiation Transfer Equation}\label{crte}

We consider simulation volumes of box length $L$ much smaller than the horizon scale $L\ll L_H= c/H(z)$. Also, prior to bubble overlap, the time between emission and absorption of a random \heii ionizing photon will be much shorter than a Hubble time. In this limit, the cosmological radiative transfer equation reduces to the familiar one (Norman, Paschos \& Abel 1998):
\begin{equation}\label{rte}
\frac{1}{c}\frac{\partial I_{\nu}}{\partial t}+\frac{\hat{n}\cdot\bold{\nabla}I_{\nu}}{a}=\eta_{\nu}-\chi_{\nu}I_{\nu}
\end{equation}
where $I_{\nu}$ is the monochromatic specific intensity, $\eta_{\nu}, \chi_{\nu}$ are emission and extinction coefficients, and $\nu$ is the instantaneous, comoving frequency. In equation~(\ref{rte}) the gradient is comoving, and hence we divide by the cosmological scale factor $a$ to convert to proper distances. The zeroth and first angular moments of equation~(\ref{rte}) yield
\begin{equation}\label{ch3_eq4}
\frac{\partial E_{\nu} }{\partial t}+\frac{1}{a}\bold{\nabla}\cdot\bold{F}_{\nu}
=\epsilon_{\nu}-ck_{\nu}E_{\nu}
\end{equation}
and 
\begin{equation}\label{ch3_eq5}
\frac{1}{c^2} \frac{\partial \bold{F}_{\nu}}{\partial t}+\frac{1}{a}{\nabla}\cdot\bold{P}_{\nu}=
-\frac{1}{c}k_{\nu}\bold{F}_{\nu},
\end{equation}
where $E_{\nu}, \bold{F}_{\nu}$ and $\bold{P}_{\nu}$ are the radiation
energy, flux vector, and pressure tensor, respectively.  All quantities
are measured in the fluid (proper) frame, where $\epsilon_\nu =
4\pi\eta_\nu$ is the emissivity, and $k_\nu$ is the absorption
coefficient. The UV photons are scattered by the free electrons and for
the range of electron number densities in the IGM (in the redshift
interval we are interested in) we can estimate that $\lambda_{s}$ (the
scattering mean free path) $ = \frac{1}{n_{e}\sigma_{Th}}>L_H$.
Therefore, we ignore the scattering coefficient.

The radiation pressure tensor is coupled to the energy density
$\bold{P}_{\nu} \rightarrow E_{\nu}$ in the moment equations through the
tensor Eddington factor, $
\bold{f}_{\nu}=\frac{\bold{P}_{\nu}}{E_{\nu}}$. The latter guarantees the
correct direction of the flux vector (Mihalas \& Mihalas, 1984). It can
be shown that the time derivative term in equation~(\ref{ch3_eq5}) is
small compared to the rest if we integrate using a timestep long compared
to the light crossing time of a computational cell, as we do. Therefore,
dropping the time derivative in equation~(\ref{ch3_eq5}) and combining it
with equation~(\ref{ch3_eq4}) we get
\begin{equation}\label{eneqn0}
\frac{\partial E_{\nu}}{\partial t}=
\frac{1}{a^{2}}
\nabla\cdot[\frac{c}{k_{\nu}}\nabla\cdot(\bold{f}_{\nu} E_{\nu})]
+\epsilon_{\nu}-ck_{\nu}E_{\nu}
\end{equation}
where
\[
\nabla\cdot(\bold{f}_{\nu}E_{\nu})\equiv\frac{\partial}{\partial x_i}(f^{ij}_{\nu} E_{\nu}).
\]
In Equation~(\ref{eneqn0}), $\epsilon_{\nu}$ is the spatially discrete
monochromatic emissivity at the locations of the emitting sources and
$k_{\nu} = n_{HeII} \sigma^{HeII}_{\nu}$ is the local opacity, where the
functional form of $\sigma^{HeII}_{\nu}$ is given by Osterbrock (1989)
and in Appendix C. Equation \ref{eneqn0} can be solved for
$E_{\nu}(\bold{r},t)$ for a given source distribution provided the
spatially-dependent Eddington tensor is known. Formally, $f_{ij}$ is
obtained from angular quadratures of the specific intensity (e.g., Hayes
\& Norman 2003). However, we wish to avoid solving the full angle-- and
frequency--dependent equation of radiative transfer. Instead we employ a
geometric closure introduced by Gnedin \& Abel (2001) in which we
calculate the radiation pressure tensor assuming the medium is optically
thin. In this limit

\begin{equation}\label{tensor}
P_{{ij}_{\nu}}(\bold{r})=\frac{1}{4 \pi c} \sum_{k}^N
\frac{L_{k_{\nu}}}{|\bold{r}-\bold{s_k}|^2}
\frac{(\hat{n}_i(\bold{r}-\bold{s}_k))(\hat{n}_j(\bold{r}-
\bold{s}_k))}
{|\bold{r}-\bold{s_k}|^2}
\end{equation}

\noindent
where $\bold{s}_k$ are the positions of the ionizing sources, and
$\hat{n}_{i,j}$ are the direction vectors (basis) at the point
$\bold{r}$. Equation \ref{tensor} describes pure radial streaming
radiation from a collection of point sources in a transparent medium.
Until \heiii bubbles begin to overlap, this is an excellent approximation
inside the \heiii regions but a poor one outside. However, since there is
very little ionizing radiation in the \heii regions, it makes little
difference what one chooses for $\bold{f}$. As discussed by Gnedin \&
Abel, the greatest error is when two bubbles begin to overlap and the two
ionizing sources begin to ``see one another." If one source is much more
luminous that the other, this can lead to a $\sim 10\%$ error in the
expansion rate of the smaller I-front.

To solve Equation~(\ref{eneqn0}) we employ a finite volume method by
rewriting the zeroth moment equation in a conservative form and
integrating over a grid cell.  The energy density, emissivity and opacity
are zone centered quantities, therefore $\int_{V_{g}} E_{\nu} dV =V_{g}
E_{\nu}$, $\int_{V_{g}} \epsilon_{\nu} dV =V_{g} \epsilon_{\nu}$,
$\int_{V_{g}} k_{\nu} dV =V_{g} k_{\nu}$, where $V_{g}$ is the volume of
a grid cell and $E_{\nu}, \epsilon_{\nu}, k_{\nu}$ are now understood to
be cell averages. Equation~(\ref{eneqn0}) then becomes:
\begin{equation}\label{eneqn}
\frac{\partial E_{\nu}
}{\partial t}+\frac{1}{a V_{g}} \oint_{cell} \bold{F_{\nu}}\cdot d
\bold S_{cell\, surf} =\epsilon_{\nu}-ck_{\nu} E_{\nu} 
\end{equation} 
\noindent
\[ F_{\nu}^i=-\frac{c}{a k_{\nu}}\frac{\partial}{\partial
x_j}(f^{ij}_{\nu} E_{\nu}) \] 
Our time-implicit discretization scheme will be discussed in a follow up
paper. Briefly, equation~(\ref{eneqn}) is discretized on a uniform
cartesian mesh and integrated using backward Euler time differencing.
Spatial discretization of the RHS of equation~(\ref{eneqn}) yields a
19-point stencil. The resulting sparse-banded system of linear equations
is solved using the stabilized biconjugate gradient (BiCGstab) algorithm
implemented in the $\tt{MGMPI}$
package\footnote{lca.ucsd.edu/portal/software/mgmpi} developed by the
Laboratory for Computational Astrophysics (Bordner 2002).

For this problem, we compute the radiative energy density at three
frequency values above the ionization threshold value, $E_{1}$, $E_{2}$,
$E_{3}$, corresponding to photon energies $\epsilon_{1} = 4$,
$\epsilon_{2} = 2 \times \epsilon_{1}$ and $\epsilon_{3} = 4 \times
\epsilon_{1}$ in Rydberg units ($\epsilon = \frac{h\nu}{h \nu_{912}}$)
respectively.  The three points plus a fourth one at $\epsilon_{4} = 32$
Ryd with $E_{4} = E_{3}
(\frac{\epsilon_{4}}{\epsilon_{3}})^{-\alpha_{q}}$, are used to infer the
interpolated profile of a 4th degree polynomial $E(\epsilon)/E_{1} =
\sum_{k=0}^{k=4} c_{k} (\frac{\epsilon}{4})^{-k}$ between $\epsilon = 4 -
16$ Ryd. At $\epsilon \geq 16$ Ryd, we assume that $E(\epsilon) = E_{3}
(\frac{\epsilon}{\epsilon_{3}})^{-\alpha_{q}}$, which follows from the
reasonable assumption that at four times the ionization threshold energy,
$\approx 0.2~keV$ already in the soft X-ray energy band, there is little
effect in the attenuation of the radiative energy due to opacity. The
interpolation is necessary in order to be able to compute the ionization
and heating rates which involve an integration in frequency space (photon
energy).  The use of three frequencies and locally computing $c_{k}$
bypasses the requirement more many frequency bins or the rewrite of the
moment equations in frequency groups. Our motivation is that our
calculation does not aim in computing the reprocessing of the radiation
field spectrum, but rather in a reasonable and spatially inhomogeneous
estimate of the photoionization/heating rates that will give rise to the
3D \heii reionization process.

Upon obtaining the solution $E_{1..3}$, $E$ hereafter, we proceed to
compute the local ionization and heating rates as described in
\S\ref{chem}. After that point, the coefficients $c_{k}$ are no longer
necessary and are discarded. The obvious advantage of this method is
computational speed in the derivation of the photo-ionization/heating
rates through an analytical formula. The disadvantage is that the
accuracy is as only good as the cubic interpolation scheme. However, we
note that our interpolation scheme does a reasonably good job of
describing the processed quasar spectrum obtained with the full
multifrequency calculation of Abel \& Haehnelt (1999). In fact, our
choice of frequency points was strongly guided by the inset spectra in
their Figure 1.

The distribution of local sources determine the point source emissivity
(source function) and the 3D distribution of the Eddington factor.
Because in our numerical setup all sources emit radiation with identical
spectral slope $\alpha_{q}$, the calculation of the Eddington tensor via
equation~(\ref{tensor}) yields a quantity that does not depend on
frequency ($f^{ij}_{\nu} \equiv f^{ij}$).  The pre-processing involves
keeping track of the sources position and luminosity during the
calculation and updating the Eddington factor functional form and
emissivity source function at the beginning of each data dump
calculation. The Eddington factor is initialized at $f^{ij}_{source} =
1/3\delta^{ij}$ within each grid cell containing a source. Between time
steps, assumed to be the redshift interval between the data dumps
($\delta z =0.1$), we solve for the $E(\bold{r},t_{n})$ which we use to
update the \heii photo-ionization rate $\Gamma^{rad}_{2}$ described in
\S\ref{chem}. The photo-ionization rates are then used to compute the
next time-step for solving the transfer equation. Because the latter is
solved implicitly, the solution is not sensitive to any particular choice
of the time-step, beyond obvious concerns of numerical convergence.

\subsection{Chemistry Implementation}\label{chem}

Our simple single species chemistry determines the time step of the
evolution. We update the ionization fraction
$\psi_{3}(\bold{x},t)=\frac{n_{He_{III}}}{n_{He}}$ through the rate
equation:  \begin{equation} \dot{n}_{HeIII}=-3H(z)n_{HeIII} + n_{HeII}
\Gamma_{2} - n_{e} n_{HeIII} \alpha_2(T) \label{abund} \end{equation} In
Equation~(\ref{abund}), $\Gamma_{2} = \Gamma^{rad}_{2}+\Gamma^{col}_{2}$
is the sum of the radiative and collisional ionization rates.  The
collisional ionization is due to collisions of the \heii ion with
electrons, $He_{II} + e \rightarrow He_{III} + 2e$, and therefore is
proportional to the electron density $n_{e}$. The collisional ionization
rate per electron $\Gamma^{col}_{2}/n_{e}$ is a function of the gas
temperature. An analytic fit to the temperature dependence is provided in
Appendix C where we also provide the functional form for the
recombination coefficient (both quantities are measured in $cm^{3}
s^{-1}$. The photoionization rate per baryon is then given by the
equation (Osterbrock 1989)  $\Gamma^{rad}_{2} \equiv
\int\limits_{\nu_{224}}^{\infty} c \frac{E_{\nu}}
{h{\nu}}\sigma^{HeII}_{\nu} d\nu$.  Because we find a parametric fit of
the energy density in frequency space the photoionization can be directly
computed as follows. $\Gamma^{rad}_{2} = \frac{c}{\sigma^{HeII}_{224}}{h}
[E_{3} \frac{4^{3}}{\alpha_{q}+3} + E_{1} \sum_{k=0}^{4}
\frac{c_{k}}{k+3} (1-4^{-(k+3)})]$, where we approximate the ionization
cross section with the power law $\sigma_{\epsilon} = \sigma^{HeII}_{224}
(\frac{\epsilon}{4})^{-3}$.

To determine the time-step $\delta t^{n}$ for the advancement of the
radiation solution between $t^{n}$ and $t^{n+1}$, we follow the procedure
below. We collect the photoionization time-scales, $\tau_{chem}=
(\Gamma^{rad}_{2})^{-1}$, from grid cells that lie in the vicinity of the
ionization front and then calculate $\delta t$ through $\delta
t=max(\tau_{chem},\tau_{lc})$, where $\tau_{lc}$ is the cell light
crossing time-scale, $\tau_{lc}=\delta x c^{-1}$.  Cells close the
I-front interface are ``captured" by the criterion $\frac{\Delta
n_{HeIII}}{n_{HeIII}}(t^{n}) \gtrsim 0.1$. Comparison between the
light-crossing and chemical time scales is necessary in the initial
moments of the evolution, because in proximity to the source the I-front
propagates close to the speed of the light. Evolving the energy density
with the light-crossing time-scale constrains the I-front expansion to
subluminal speeds. In addition, the use of a chemical time-scale close to
the source would yield very small time-steps, due to the large number of
photons, that would in turn slow down the overall calculation.

Before we proceed, we need to make the following very important
clarification. In this work, we do not consider contributions to the
radiative energy density by \heii recombinations despite including them
in the update of the \heii abundance.  The significance of the
contributing \heii recombinations to the energy density are important
when a consistent ionization calculation of all species is performed. In
that case, radiative recombinations to energies $\epsilon < 54.4$ eV
would contribute to the ionization balance of hydrogen and helium.
However, these species are already at high ionization fractions and
therefore such contributions are not expected to have a significant
effect in our scheme. In a consistent calculation, the photon flux from
the central source creates a concentric set of Str\"{o}mgen regions where
the ionized hydrogen and helium II extend ahead of the corresponding
helium III volume. In such a scheme, recombinations at the \heiii
ionization front yield photons with energies $\epsilon < 54.4$ eV which
can freely propagate forward through the already ionized \hei and \hi and
assist in the advancement of the \heii and \hii fronts respectively. In
our setup, there are no \heii and \hii Str\"{o}mgen regions, only \heiii
ones. Therefore, the photon flux from recombinations on the \heiii
I-fronts are ignored and we only consider the chemical effects resulting
from the attenuation and percolation of the individual \heii ionizing
flux.

However, recombinations are included in the abundance update primarily
because it can be an important effect for the diffuse helium in proximity
to a local source that shut off.  The lack of direct photons could lead
to a rapid recombination of the \heiii bubble, if no additional radiative
flux reaches that region from another QSO source. Equation~ (\ref{abund})
can then be rewritten as follows.
\begin{equation}
\dot{\psi}_{3} = \frac{(1-\psi_{3})}{\tau^{ion}_{2}} - \frac{\psi_{3}}{\tau^{rec}_{2}}
\label{eqnchm3}
\end{equation}
In Equation~(\ref{eqnchm3}), $\tau^{ion}_{2} = (\Gamma_{2})^{-1}$ is
total ionization time scale and $\tau^{rec}_{2} = (\alpha_2(T)
n_{e})^{-1}$ is the local \heiii recombination time scale and
$\Gamma_{2}$ is the sum of all ionization processes that lead to the
forward reaction $\heii \rightarrow \heiii$ and we have assumed $\psi_1
\approx 0$. These processes in our scheme involve the combination of the
direct photoionizations from the QSO sources $\Gamma^{rad}_{2}$ and
collisional ionization $\Gamma^{col} = n_{e} k^{col}(T)$. We integrate
equation \ref{eqnchm3} using backward Euler time differencing where all
source terms are computed at the advance time $t^{n+1}$.
\begin{equation}
\psi_3^{n+1} = \frac{\psi_3^n+\delta t^{n} / (\tau^{ion}_{2})^{n+1} }
{1 + \delta t^{n} [ \frac{1}{(\tau^{ion}_{2})^{n+1}} + \frac{1}{(\tau^{rec}_{2})^{n+1}}]}
\label{discr.y3}
\end{equation}

In Equation~(\ref{discr.y3}), only the photoionization rate
$\Gamma^{rad}_{2}$ is available at the advanced time. Therefore, we are
forced to initialize the abundance update by computing
$(\frac{1}{\tau^{ion}_{2}})^{n+1} \simeq \frac{1}{
(\Gamma^{rad}_{2})^{n+1} + n^{n}_{e} (k^{col}(T))^{n} }$ and substitute
$(\frac{1}{\tau^{rec}_{2}})^{n+1} \rightarrow
(\frac{1}{\tau^{rec}_{2}})^{n}$. However, upon obtaining the \heiii
number density at $t^{n+1}$, $n_{HeIII} = \psi^{n+1}_{3} n_{He}$ we can
update the electron density at $t^{n+1}$ and insert it back in
Equation~(\ref{discr.y3}). We therefore can improve upon the original
estimate by iterating Equation~(\ref{discr.y3}) until $\frac{\Delta
n_{e}}{n_{e}} \leq 0.1$. Unfortunately, our methodology of updating the
temperature, as described in \S\ref{he2tm}, is crude and is not used in
the iterative scheme. The local helium number density at any time is
computed from the gas density as $n_{He} = \frac{\rho_{He}}{4m_{H}}
\simeq \frac{\rho}{16m_{H}}$. The electron density is given by the charge
conservation equation $n_{e} = n_{HII} + n_{HeII} + 2 n_{HeIII}$. Our
calculation assumes no change in the ionized hydrogen density between the
value in the original simulation ($^{o}$), and the post-processed value
($^{1}$). Therefore, $n^{1}_{e} - n^{o}_{e} = (n^{1}_{HeII} -
n^{o}_{HeII}) + 2 (n^{1}_{HeIII} - n^{o}_{HeIII}) = (n^{1}_{HeIII} -
n^{o}_{HeIII})$. which allows the calculation of the local electron
density at any time from $n_{e} = n^{o}_{e} + (n_{HeIII} -
n^{o}_{HeIII})$.

\subsection{Results}\label{chemresults}

\begin{figure*}
\includegraphics[width=3in,height=3.5in]{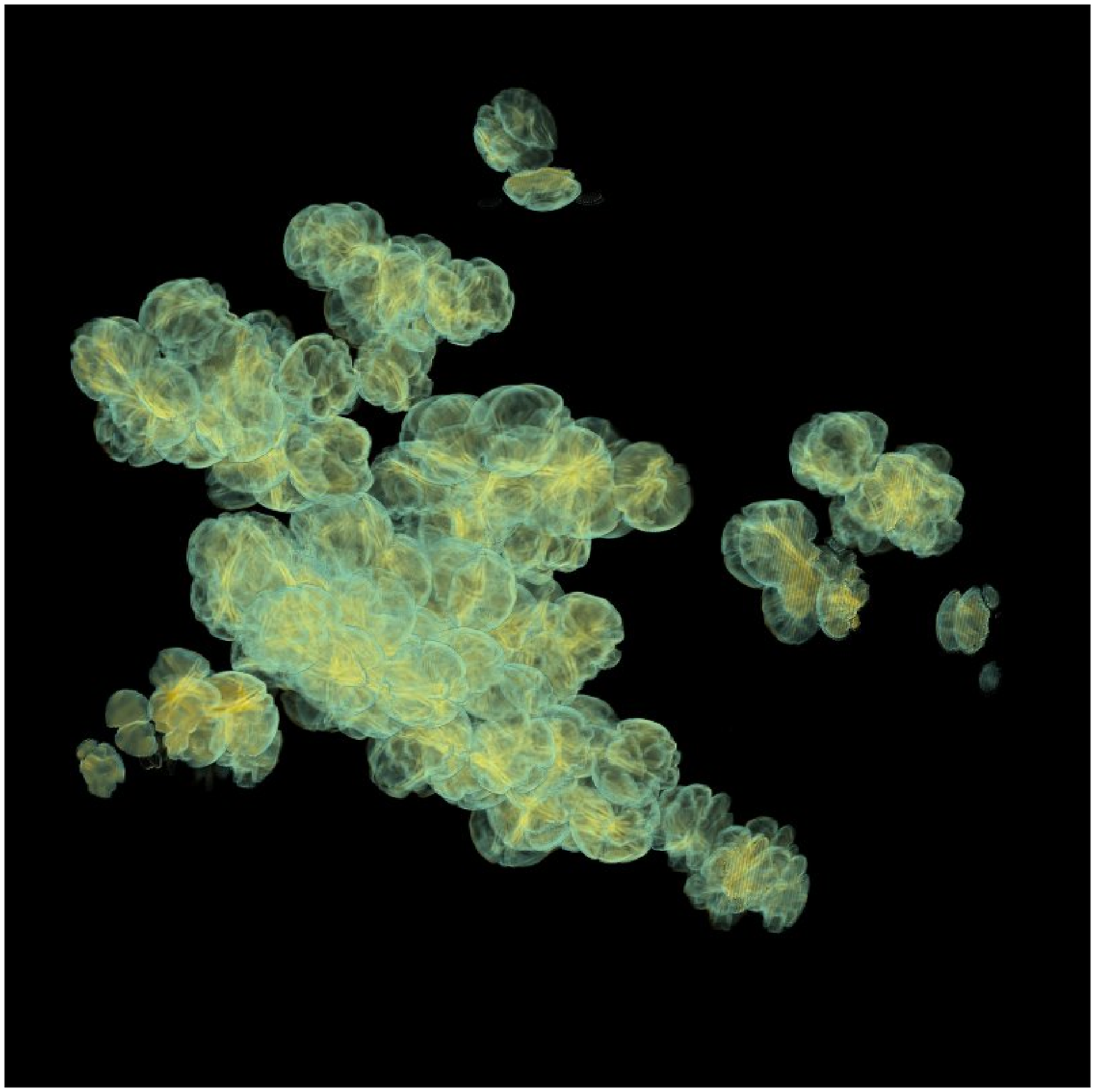}
\includegraphics[width=3in,height=3.5in]{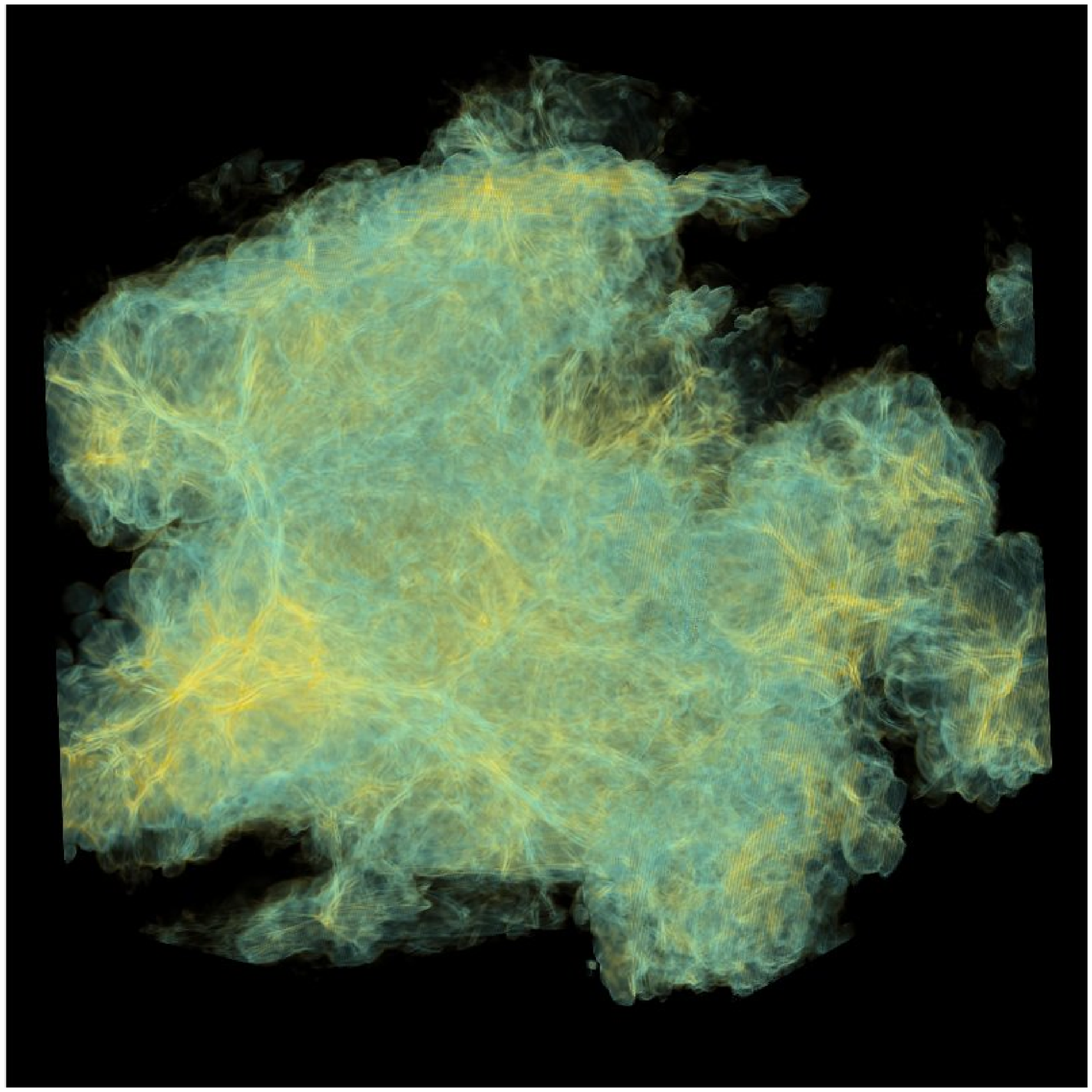}
\caption{Volumetric renderings of the \heiii distribution mapped
by the Log$n_{HeIII}$ at z=3 and z=2.6. The ionization
cutoff in this visualization is set to $\psi_{3} \equiv \frac{n_{HeIII}}{n_{He}} =
10^{-5}$ (dark regions).
\label{volnheiii}}
\end{figure*}

\begin{figure*}
\includegraphics[width=3in,height=3in]{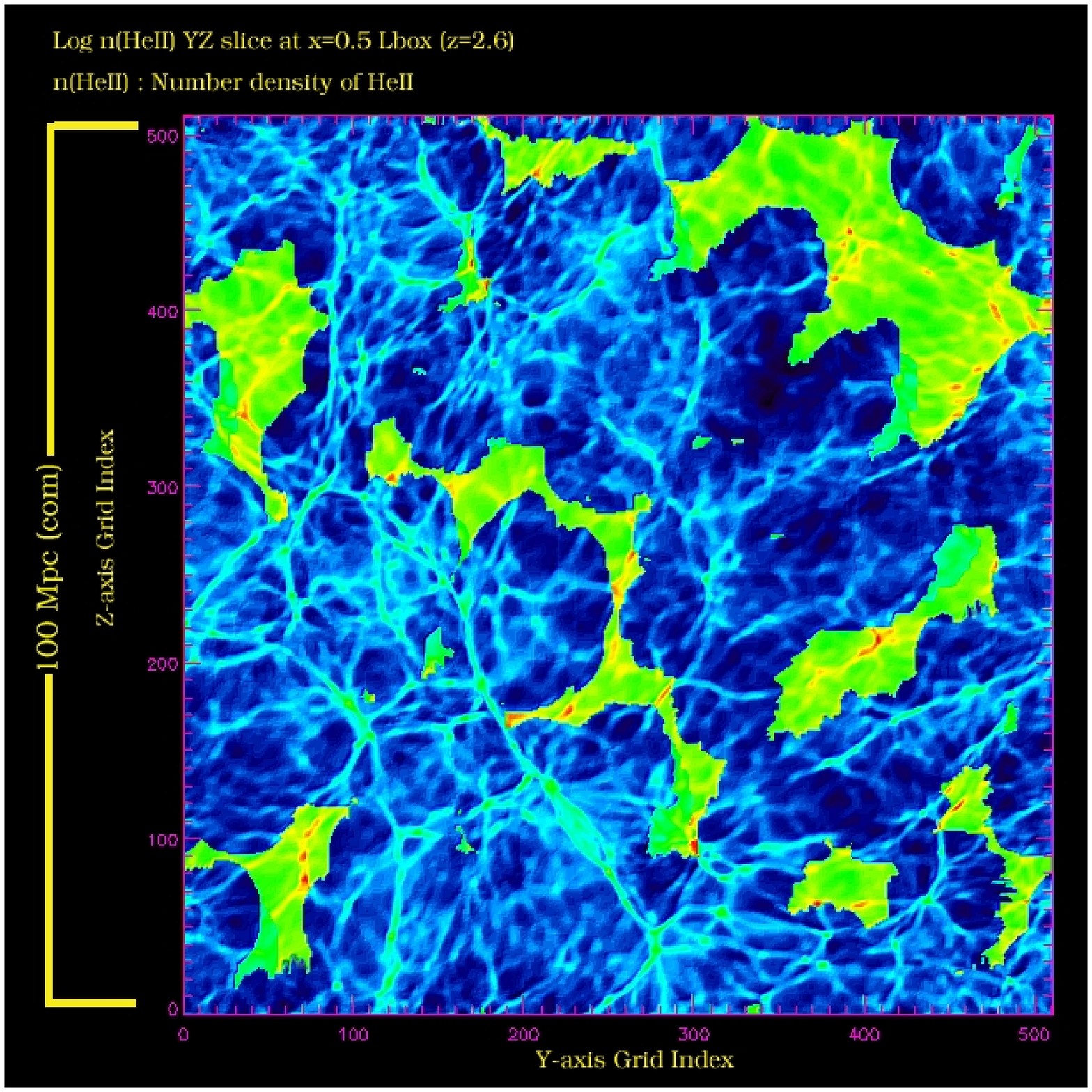}
\includegraphics[width=3in,height=3in]{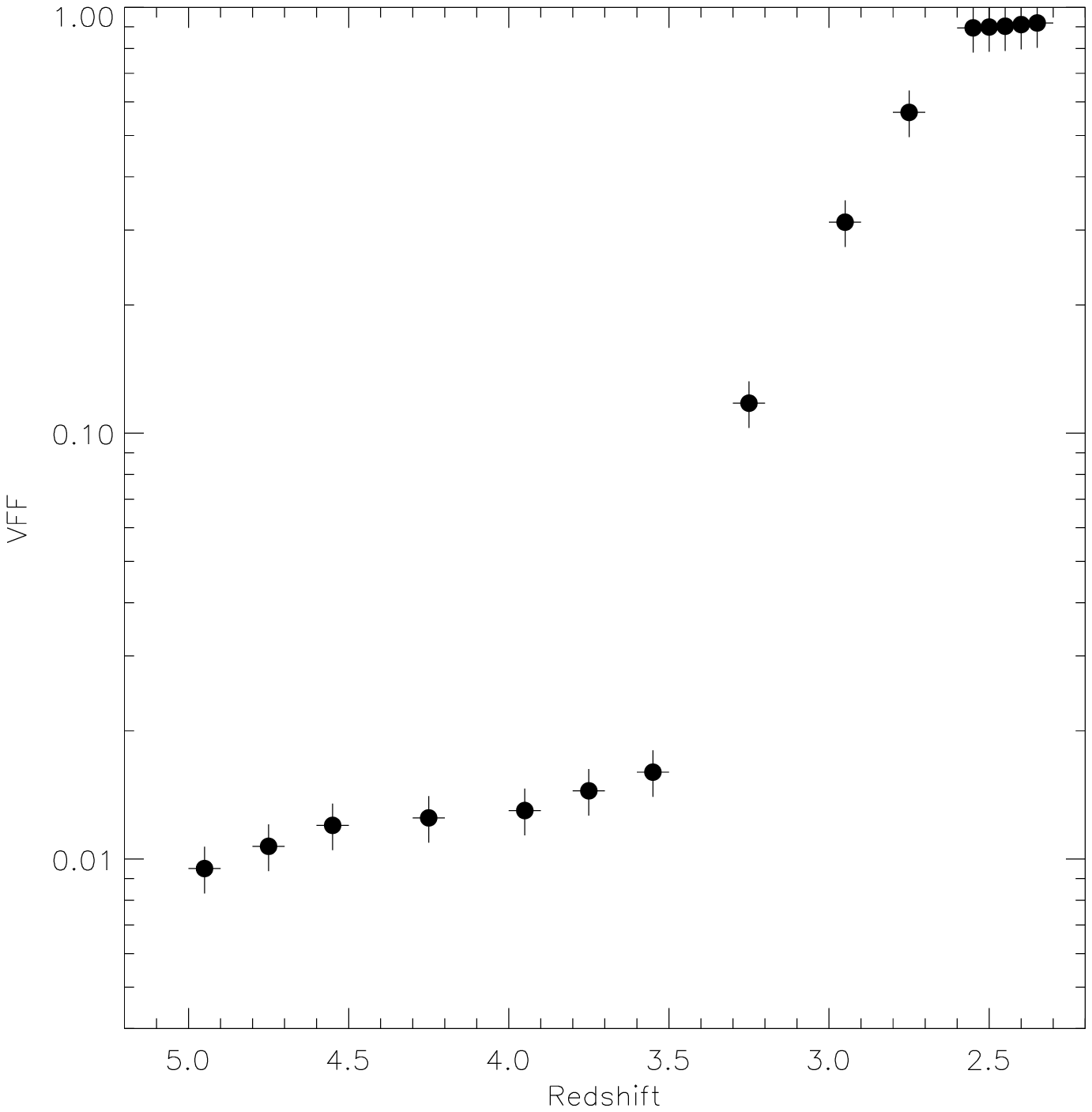}
\caption{Left Panel: Slice of the \heiii (darkened regions) and \heii
3D mass distribution at z=2.6. At that redshift, the volume filling fraction of the \heiii
is $VFF \simeq 0.90$. Right Panel: Redshift Evolution of the VFF for
$\chi_{HeIII} \geq 10^{-5}$.
The error bars result from the uncertainty in the location of the
cumulative I-front due to finite grid cell size.
\label{projvff}}
\end{figure*}

In Figure~(\ref{volnheiii}), we show volumetric renderings of
$n_{HeIII}(\bold{r},z)$ at two redshift instances. The 3D visualization
shows the expanding ionized bubbles filling up the cosmic volume due to
the combined effect of the radiative energy transport and sources being
turned on at different parts of the volume at later redshifts. Individual
bubbles of \heiii may stagnate as they reach their Str\"{o}mgren radii
due to recombinations. However, overall the volume filling factor (VFF)
of \heiii increases as more quasars are placed in the computational
volume and the percolation between the I-fronts increases the mean free
path of the ionizing photon flux. In left panel of
Figure~(\ref{projvff}), we show a slice through the cosmic volume at
$z=2.6$ of the \heii, \heiii density distributions. Ionized regions have
percolated through the cosmic medium to ``open up" the IGM to $ \gtrsim
54.4$ eV radiation, effectively completing \heii reionization by such
redshifts. In the right panel of Figure~(\ref{projvff}), we show the
redshift evolution of the VFF, as measured by the fraction of the grid
cells with ionized helium at \heiii abundance of $\psi_{3} \geq 10^{-5}$.  
As the ionized regions begin to merge, assisted by the increase in the
QSO number density, the VFF(z) rapidly increases, leading to a value of $
> 68$\% at $z \leq 2.8$. The redshift of significant merging, which we
define as $VFF \approx 0.90$ is achieved by $ z \approx 2.5$ where we
point to a statistical global \heii reionization. This redshift value
compares well with the observed determination by Kriss et al. (2001). The
solid line in the VFF evolution figure is a spline fit through the
computed data ($\delta z = 0.1$). For reference, we include the derived
values every $\delta z = 0.5$, along with the error estimates based on
the location uncertainty of the I-front. The uncertainty is simply due to
the fact that computing the radiative energy density in the zone centers
yields no information on the profile of the field across the grid cell.
Therefore, we assigned an error estimate in the ionized volume fraction
equal $(\frac{\delta x}{2})^{3}/{V}$, where $\delta x$ is the grid
resolution and V is the volume of the computational box. The evolution of
the VFF shows a rapid increase in \heiii at redshifts $z \lesssim 4$,
following an earlier epoch of apparent stagnation.

\begin{figure*}
\includegraphics[width=3in,height=3in]{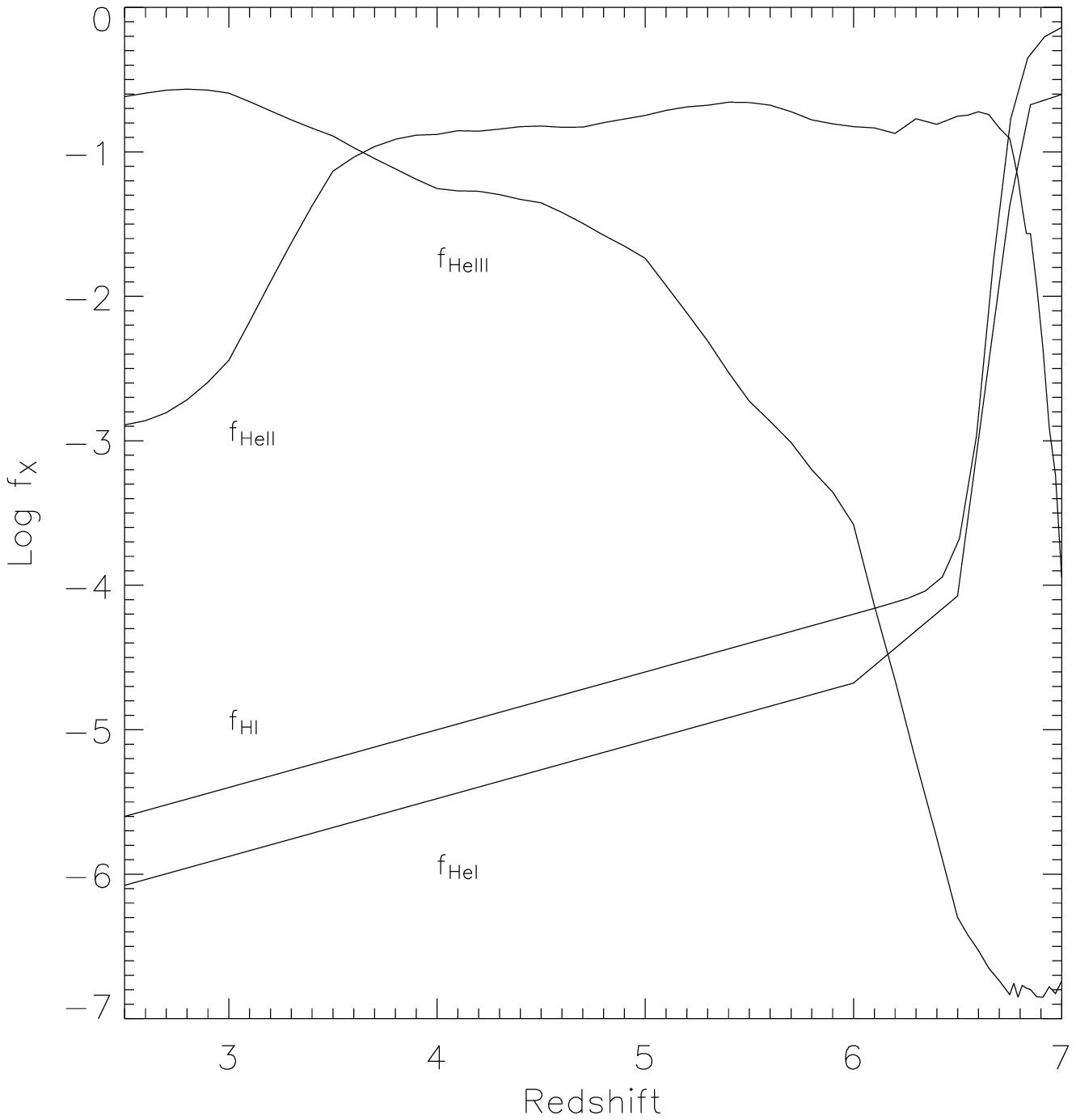}
\includegraphics[width=3in,height=3in]{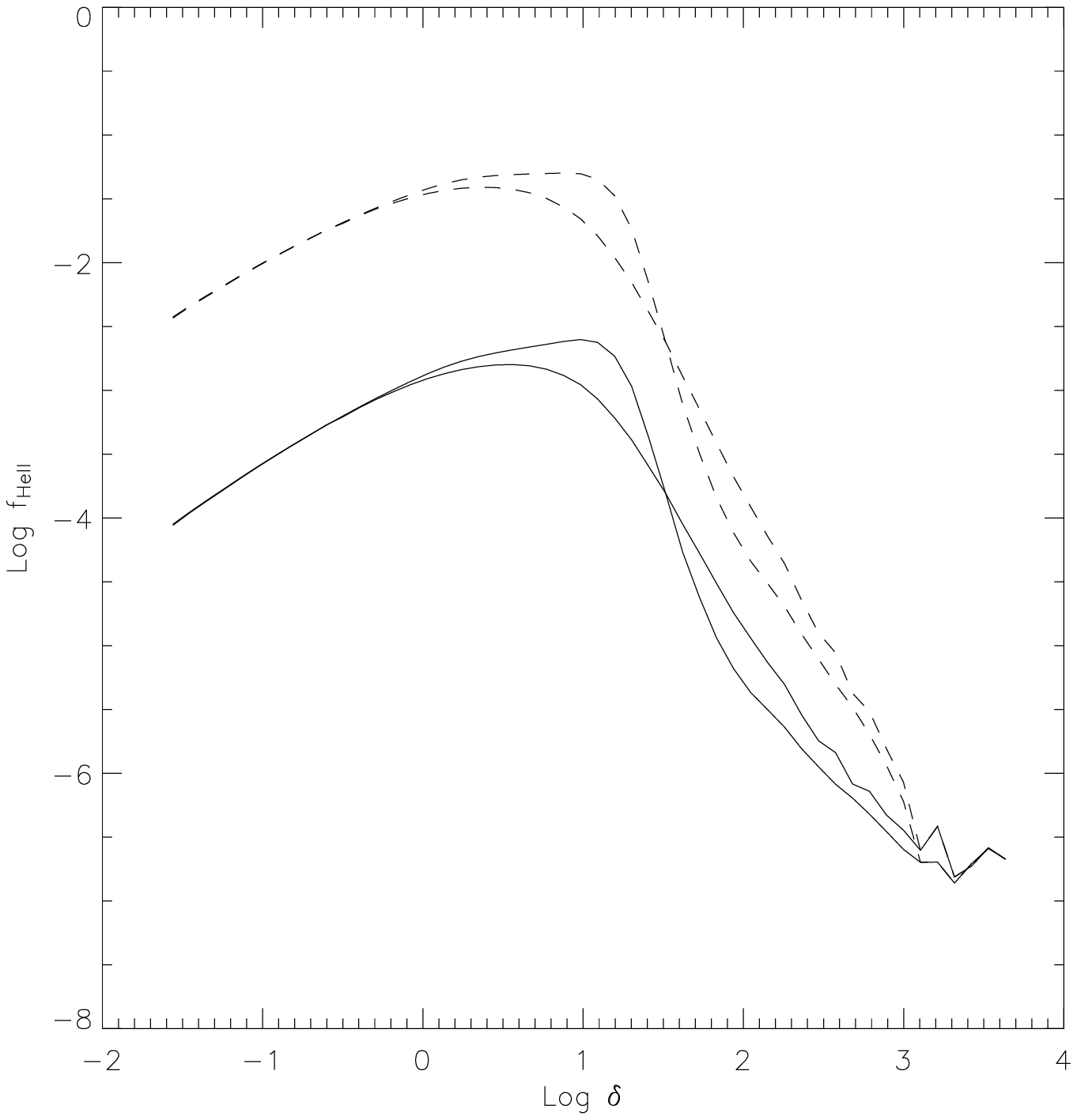}
\caption{Left Panel: Redshift evolution of the volume averaged mass fraction of singly helium.
Overplotted is the corresponding neutral hydrogen mass fraction.
Right Panel: The scatter plot between \heii fraction and local gas overdensity at z=2.5
yields the mean and median value of $Log f_{HeII}$ per logarithmic bin of gas overdensity.
Solid (dashed) curves refer to the inhomogeneous (homogeneous) case.
The median profiles correspond to the curves that peak at $Log \delta \simeq 1$
in each case (the mean profiles peak at a lower overdensity).
\label{zevlab}}
\end{figure*}

An alternative way to illustrate \heii reionization is to plot the
redshift evolution of the volume averaged abundance fraction. In the left
panel of Figure~(\ref{zevlab}), we show that the mean mass fraction in
\heii $\rho_{He_{II}}/\rho$, drops significantly at $z \lesssim 4$. For
reference, we also show the mean fraction in \hi $\rho_{H_{I}}/\rho$,
undergoes a steep drop at $z \simeq 6.5$ and continues to decrease under
a smooth redshift profile, the properties of which are discussed in
Paschos \& Norman (2005). One notable difference in examining the two
evolution profiles emerges between the slopes of the two curves. Beyond
the fundamental differences in the ionization calculation of the two
species, the result shown in Figure~(\ref{zevlab}) shows that the \heii
reionization epoch is much more extended than that of \hi. A rapid drop
in the \hi fraction occurs between $z = 7 - 6.5$ (Razoumov et al. 2002;
Paschos \& Norman 2005).  In that redshift range the mean \hi mass
fraction drops by 4 dex.  By visual inspection of the mean fraction in
\heii from Figure~(\ref{zevlab}) we can determine that it drops by $\sim
2$ dex between $z = 3.5 - 2.5$.  When the difference in the redshift
interval is converted to cosmic time it yields that a similar drop in the
mass fraction takes approximately eight times longer to occur in the case
of \heii when compared to \hi. This is consistent with the conclusion
reached in \S\ref{whylate}, where the latency in the \heii reionization
is attributed to higher recombination time scales and less available
ionizing photons per baryon for a stellar dominating UV background.

In the right panel of Figure~(\ref{zevlab}), we plot the \heii mass
fraction versus the local overdensity by logarithmically binning the
latter quantity and computing the mean and median \heii from the cells
with gas density within the bin. Such a graph is intended to show the
trend between the two quantities. For reference, we plot the trend
between the two quantities from a standard cosmological simulation, where
all ionizations are computed self-consistently due a uniform UV
background in the optically thin approximation.

The simulation is terminated at $z \simeq 2.5$ at $VFF \simeq 0.9$. The
reason for suspending the calculation is that when the cosmic volume is
effectively transparent to the ionizing radiation, the assumptions
underlying equation~(\ref{rte}) no longer apply.  Specifically, free
streaming photons may cross the volume unimpeded by absorption and their
mean free path can become comparable to the size of the horizon. The
latter effect requires keeping the cosmological dilution term in
Equation~(\ref{eneqn}), which was ignored. From the numerical
perspective, solving a parabolic equation when the conditions call for a
hyperbolic one, can create local superluminal speeds of the ionizing
front if the chemical time scale becomes smaller than the cell light
crossing time.  Concluding the calculation at the end of the opaque phase
of \heii still addresses the main question investigated in this work;
what is the primary mechanism that leads to the \heii reionization. We
conclude, that a rising population of QSO sources, assisted by the gas
dilution due to cosmic expansion, clumpiness due to structure formation
and the pre-ionization of neutral hydrogen which allows for the almost
exclusive usage of the \heii ionizing radiation are a set of physical
conditions that reproduce the epoch of \heii reionization by redshifts $z
\leq 2.5$.

\section{Late Heating due to \heii Reionization}\label{he2tm}

\subsection{Physical Considerations} Photoionization of \heii at an epoch
later than that of \hi releases an additional one photo-electron per
ionization. We can estimate the mean energy of such electrons due to
absorption of photons with energies $\geq 54.4$ eV$\equiv 4$ Ryd by \heii
and compare it to the value obtained in the case of $\geq 13.6$ eV$\equiv
1$ Ryd absorption by \hi atoms. For simplicity, we will ignore the
geometric attenuation and optical depth effects in the radiative flux due
to local sources and gas opacity and assume that the local mean intensity
of the radiation is the same as the emitted, with a spectrum
$J_{\epsilon} = J_{912} \epsilon^{-\alpha_{q}}$. As before, in this
notation $\epsilon = \frac{h\nu}{h\nu_{912}}$. The photo-heating rate, in
ergs/s, due to electrons ejected from \heii atoms, can then be computed
from $\bar{G}_{HeII} = \nu_{912} \int_{4}^{\infty} (4\pi
J_{\epsilon}/\epsilon) (\epsilon-4)\sigma_{\epsilon} d\epsilon$.
Substituting for the power law of the mean intensity, for
$\sigma_{\epsilon} = \sigma^{o}_{HeII} 4^{3} \epsilon^{-3}$ we get
$\bar{G}_{HeII} = \frac{ 4 \pi \nu_{912} J_{912} \sigma^{o}_{HeII}
4^{1-\alpha_{q}}} {(\alpha_{q}+2)(\alpha_{q}+3)}$. In a similar manner,
we can derive $\bar{G}_{HI} = \frac{ 4\pi \nu_{912} J_{912}
\sigma^{o}_{HI}} {(\alpha_{q}+2)(\alpha_{q}+3)}$. If the sources
responsible for the \heii and \hi ionization are the same then the
$J_{912}$ amplitude and the $\alpha_{q}$ spectral slope are identical in
both equations which allows for the derivation of the following ratio:

\begin{equation}
\frac{\bar{G}_{HeII}}{\bar{G}_{HI}} = \frac{\sigma^{o}_{HeII}}{\sigma^{o}_{HI}} 4^{1-\alpha_{q}}
\label{ratio1}
\end{equation}

Substituting for $\sigma^{o}_{HeII} = \frac{\sigma^{o}_{HI}}{4}$ we get
$\frac{\bar{G}_{HeII}}{\bar{G}_{HI}} = 4^{-\alpha_{q}}$. The total
photoheating rate per unit volume is proportional to the number density
of the absorbers which yields the ratio between \heii and \hi
photoheating rates to be $G_{HeII}/G_{HI} = \frac{n_{HeII}}{n_{HI}}
4^{-\alpha_{q}}$. The ratio of number densities is estimated in
\S\ref{whylate} to be $n_{HeII}/n_{HI} = \frac{5}{12} S$, where S is the
softness parameter. The ratio of the photoheating rates then becomes
$G_{HeII}/G_{HI} = \frac{5}{12} S 4^{-\alpha_{q}} \simeq \frac{5}{12}
4^{\alpha_{q}+1} 4^{-\alpha_{q}} = 5/3$. The latter factor is indicative
of the degree of temperature increase due to the thermalization of the
photoelectrons ejected by \heii ionizations when compared to the
temperature inferred by HI ionization alone.

In deriving the above value, we made the assumption that hydrogen and
singly ionized helium are photoionized simultaneously by the same
ultraviolet field. The effects of distinct ionization epochs can however
be modeled in the above relation if we adopt the scenario of hydrogen
ionization by a soft radiation background ($\alpha^{(2)} \sim 5$) and of
\heii ionization by a hard one ($\alpha^{(1)} \sim 1.5$). At a point in
time when the populations of galaxies and QSOs have the same energy
output at the LyC limit ($z \simeq 4$), we can derive an estimate of the
photoheating rates ratio to be $\frac{G_{HeII}}{G_{HI}} = \frac{5}{3}
\frac{(\alpha^{(2)}_{q}+2)}{(\alpha^{(1)}_{q}+2)} \simeq \frac{10}{3}$
for $\alpha^{(1)}_{q} = 1.5$ and for $\alpha^{(2)}_{q} = 5$. In
conclusion, the thermalization of the photoelectron in the $He_{II} +
\gamma \rightarrow He_{III} + e^{-}$ reaction can be an important
determinant of the intergalactic medium temperature.

Photoionization models based on hydrogen ionization alone predict a
temperature of the intergalactic medium of $T_{IGM} \approx 1.2 \times
10^{4}$ K.  However, Ly$\alpha$ forest observations yield lines with
median broadening widths of $b=26-36$ km/s (Carswell et al. 1987, 1989;
Zhang et al. 1997;  Dav\'e et al. 1997) in the intermediate redshift
range of z=2-4. Because the thermal and differential Hubble flow
components in the total HI line broadening are of the same order of
magnitude (Zhang et al., 1998; Aguirre, 2002), one can infer a thermal
width range in the HI Ly$\alpha$ forest between $b_{th} = 13-18$ km/s.
The effects of the peculiar velocity can be ignored if the the focus is
at densities close to the cosmic mean. This value range in $b_{th}$ would
require the temperature in the intergalactic medium to be $T_{IGM} =
20,000 - 40,000$ K, a factor of $\sim 1.7-3.3$ above the temperature
inferred by hydrogen photoionization alone. As seen above, such an
increase in temperature can be reproduced by the \heii ionization
photoheating due to a hard ultraviolet spectrum.

\subsection{Perturbed Gas Energy Equation}

The post-processing of the simulation data dumps involves an update of
the local gas temperature by solving for the perturbed temperature in the
thermal energy equation. In the simplest approximation, we will assume
that the change in the net rate of the cell thermal energy density is due
to the balance between the heating by the ejected photoelectrons from
\heii ions and the \heiii recombination cooling, $\frac{1}{\gamma-1}
\frac{k \rho}{\mu m_{p}} \Delta \frac{\delta T}{\delta t} = G - \Lambda$,
where $\gamma=\frac{5}{3}$ is the gas adiabatic index, $\rho$ the local
gas density and $\mu$ the mean molecular weight. All quantities have
local proper values. In addition, in the notation followed here $G =
n_{HeII}G_{r}$ is the total photoheating rate, measured in
$ergs/s/cm^{3}$, which is the radiative rate $G_{r} \equiv \bar{G}$
mentioned above. The cooling rate, $\Lambda$, is a function of the local
gas temperature and density and is equal to $\Lambda \equiv
n_{e}n_{He_{III}} L(T)$. $L$ denotes the cooling function due to \heiii
recombinations and is given by $L(T) = 3.48 \times 10^{-26} T^{1/2}
T^{-0.2}_{3} (1+T^{0.7}_{6})^{-1}$ in units ergs cm$^{3}$/s (Cen 1992),
where $T_{n} = T/10^{n}$.

Recombination \heiii cooling is only one out of several processes that
cool the cosmic gas, the list of which is described in detail in Anninos
et al. (1997). The cooling rate coefficients, in parametric fits that
depend on the local gas temperature, due to excitation, ionization and
recombination of the primary chemical species along with bremsstrahlung
and Compton cooling are fully incorporated into the ENZO code.  We expect
that \heiii cooling would dominate the cooling processes inside the
\heiii bubbles when compared to the corresponding \heii recombination
cooling, because the \heii abundance is significantly reduced there to
fractions $\psi_{2} \simeq 10^{-6} - 10^{-3}$. In addition, the \heiii
recombination cooling rate, along with \hii recombination, dominate at
the low temperatures found in underdense cosmic regions $T \lesssim
10^{4}$ K over all other types. In a scheme where hydrogen is already
almost completely ionized at $\delta \lesssim 1$ ($\chi_{2} \approx 1$),
but helium is predominately only singly ionized ($\psi_{2} \approx 1$)
the thermal balance would be controlled by the heating and cooling of the
latter species' ionization.

However, our ability to accurately post-process the temperature field in
our scheme is limited by the fact that excitation and collisional
ionization cooling from processes involving \hi, \hei and \heii species
dominate the cooling curves at temperatures $T \gtrsim 10^{4}$ K, while
exhibiting very steep profiles in the temperature range $T \simeq 10^{4}
- 10^{5}$ K. An increase in the gas temperature by the photoelectrons
ejected in the \heii radiative ionization would cause an increase of the
cooling coefficients. Combined with the increase in the electron number
density this shift in the thermal balance could significantly increase
the cooling rate even though the fractional abundances in \hi, \hei and
\heii are small. The end result is that, by only including \heiii
recombination cooling in recomputing the gas temperature, we may
overestimate the value of the adjusted temperature. This upper limit in
the temperature estimate implies that the recombination time-scale in
Equation~(\ref{discr.y3}) is also an upper estimate and that the overall
propagation of the cumulative \heiii ionization front is faster than in
the case of a self-consistent calculation. However, such calculation
would require computing the ionization and thermal balance of all species
self-consistently and is beyond the scope of this paper. A numerical
scheme is under development that will allow us to do this in the near
future (Reynolds et al., {\em in prep}). Therefore, we conclude that the
epoch of \heii reionization may be placed at a later redshift than the
one we calculated here ($z_{reion} \sim 2.5$) even though we note that
such adjustments could be reversed or may not be necessary if the diffuse
recombination radiation is added in the calculation. Such radiation would
further ionize HI and HeI species suppressing their contributions to the
cooling curve. That may explain why, even under all of the assumptions
and approximations that we allowed and followed in this work, the end
result of our predicted redshift evolution of the \heii opacity
correlates well with observations of the \heii Ly$\alpha$ forest as we
shall show in \S\ref{obs}.

We proceed with further detailing our temperature update method.  If
$T^{(o)}$ and $T^{(1)}$ denote the cell temperatures before and after the
presence of \heii ejected photoelectrons, then we approximate the change
in the thermal energy equation as follows:

\begin{equation}
\frac{\delta T^{(1)}}{\delta t} \simeq \frac{\delta T^{(o)}}{\delta t} +
\frac{(\gamma-1) m_{p} \mu^{(1)}}{k \rho}(G^{(1)} - \Lambda^{(1)})
\label{eneqn2}
\end{equation}

In an explicit, time-discretized form, where the original temperature at
time $t^{n}$ is obtained by interpolation between the logarithm of the
local temperature in the two data dumps with redshifts that bound the
time evolution ($z_{1} \leq t^{n} \leq z_{2}$), Equation~(\ref{eneqn2})
becomes:

\begin{eqnarray}
T^{(1)}_{n+1} - T^{(1)}_{n} = T^{(o)}_{n+1} - T^{(o)}_{n} + 
(\delta t)_{n} (\gamma-1) \nonumber \\ 
\frac{1}{k} (\frac{m_{p}\mu}{\rho})^{n+\frac{1}{2}}
(n_{HeII}^{n+\frac{1}{2}} G^{n+\frac{1}{2}}_{r} 
-n_{e}^{n+\frac{1}{2}} n_{He_{III}}^{n+\frac{1}{2}} L^{n}) 
\label{eneqn_dc}
\end{eqnarray}

In Equation~(\ref{eneqn_dc}), time-centered quantities are computed as
$X^{n+\frac{1}{2}}=0.5(X^{n+1}+X^{n})$ and only for variables that we
know their value at the forward time $t^{n+1}$. Therefore, $L$ is
computed at time $t^{n}$ because it is a function of temperature which is
unknown at $t^{n+1}$. The update in temperature occurs after all
principal quantities of $E^{n+1}$, $n_{HeII}$, $n_{He_{III}}$ and $n_{e}$
are computed at $t^{n+1}$ by advancing the local solutions at the
implicit time-step of the radiation field discussed in \S\ref{chem}. The
quantity $\frac{m_{p}\mu}{\rho} = \sum n_{i}$ is equal to total number
density of the cosmic species plus electrons and is computed as follows:
$\frac{m_{p}\mu}{\rho} = [ \frac{13}{16} \frac{\rho}{m_{p}} + n_{e}
]^{-1} \simeq [0.9 \cdot 10^{-5} \Omega_{b} h^{2} (1+z)^{3} +
n_{e}]^{-1}$, where the expression for $n_{e}$ was provided in
\S\ref{he2r}.

In an ideal calculation, if a local grid cell is outside the \heiii
bubble at $t^{n}$ then $T^{(1)}_{n} = T^{(o)}_{n}$ and $n^{n}_{He_{III}}
\rightarrow 0$. If no direct ionizing radiation reaches that grid cell by
$t^{n+1}$ then $n^{n+1}_{He_{III}} \rightarrow 0$ and
Equation~(\ref{eneqn_dc}) would predict $T^{(1)}_{n+1} = T^{(o)}_{n+1}$.
In a cell where thermal equilibrium was reached during the original
calculation between $t^{n+1}$ and $t^{n}$ then during post-processing
$T^{(1)}_{n+1} = T^{(1)}_{n}$ is achieved only when the heating and
cooling terms balance out.

A point of concern is that there was no physical reason to suppress the
collisional ionization of \heii in the original simulation. Even if there
are no ionizing sources capable of radiatively ionizing \heii, a local
temperature of $T \geq 4.64 \cdot 10^{4}$ K, found in overdense regions,
may be enough to collisionally eject the 54.4 eV bound electron in the
\heii atom. Consequently, the temperature evolution $T^{(o)}(z)$ includes
such an effect and can skew the post-processed evolution $T^{(1)}(z)$.
This is evident if we assume that due to collisional ionization in the
pre-processed data $n_{He_{III}} \neq 0$ and at $t^{n}$, $X^{(1)}_{n} =
X^{(o)}_{n}$ Equation~(\ref{eneqn_dc}) would then predict $T^{(1)}_{n+1}
= T^{(o)}_{n+1} + (\delta t)_{n} W(n^{(o)}_{HeII},
n^{(o)}_{He_{III}},n^{(o)}_{e},T^{(o)})$ where $W \equiv (\gamma-1)
\frac{1}{k} (\frac{m_{p}\mu^{o}}{\rho})^{n+\frac{1}{2}} \times
(-1)~n_{e,o}^{n+\frac{1}{2}} n_{He_{III},o}^{n+\frac{1}{2}}
L^{n}(T^{(o)}) $

The last equation would unnecessarily recompute the temperature in
regions which lack radiative input but have significant collisional rates
in \heii $\leftrightarrow$ \heiii and therefore, $n_{He_{III},o} \neq 0$.  
To account for such discrepancy in the temperature evolution, each
updated local temperature is adjusted as $T^{(1)}_{n+1} = T^{(1)}_{n+1} +
(\delta t)_{n} W(n^{(o)}_{HeII},n^{(o)}_{He_{III}},n^{(o)}_{e},T^{(o)})$.
Although in doing so we improve upon the temperature evolution, the
collisional effect can never fully be readjusted because of
interdependency between all of the physical quantities.

\subsection{Results}

\begin{figure*}
\includegraphics[width=3in,height=3in]{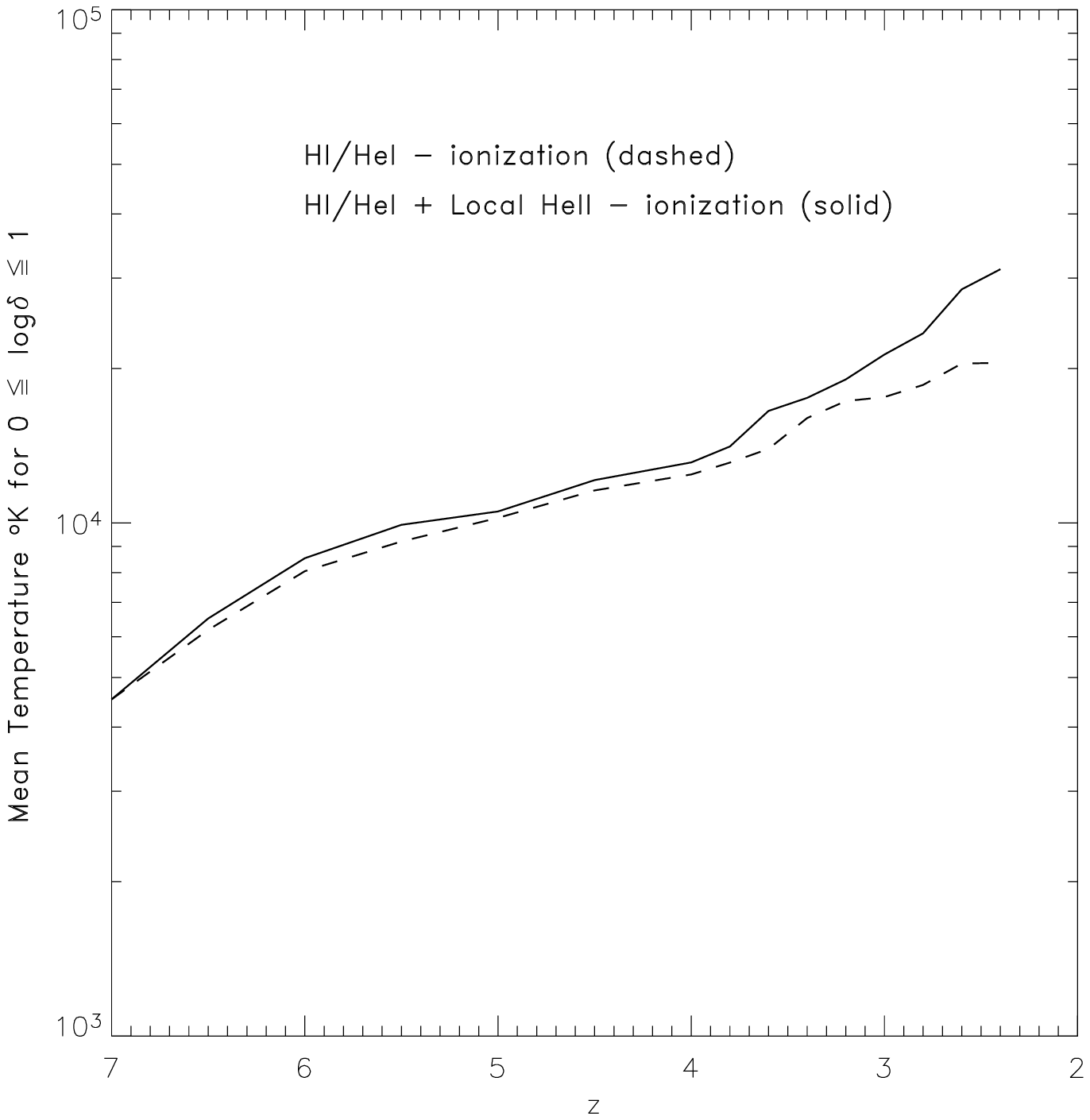}
\includegraphics[width=3in,height=3in]{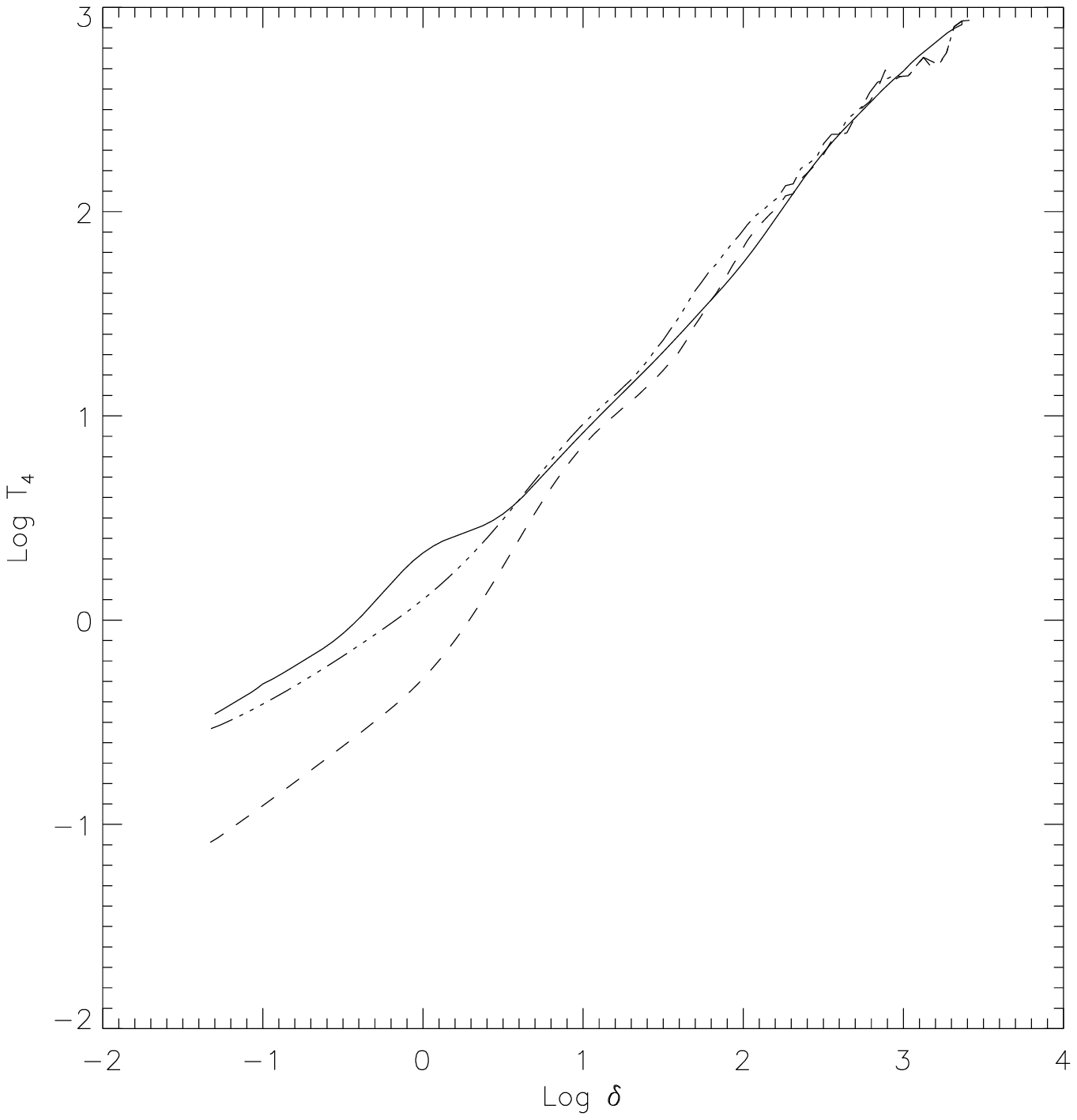}
\caption{Left Panel: Redshift of the temperature evolution. Shown as a solid (dashed) line
is the evolution in the processed (pure) case of the mean temperature in grid cells with
overdensities between $\delta = 1-10$. Right Panel: The median temperature distribution
in each logarithmic overdensity interval at z=2.5. It shows that
the processed (solid) and unprocessed (dashed) cases. The curves show that \heii reionization
primarily increases the temperature of the IGM in low to mildly overdense range.}
\label{tempevl}
\end{figure*}

In Figure~\ref{tempevl}, we show the results of our temperature
calculation.  In the left panel, we plot the evolution of the mean
temperature in overdensities $log(\delta) = 0 - 1$ (solid curve).
Overplotted is the evolution in the original calculation (dashed) to
showcase that at $z \lesssim 3.5$ the two profiles are deviate. This
demonstrates the effect of late \heii photoheating due to the rising
population of QSO's in the cosmic volume. The overdensity interval was
chosen in order to show that \heii reionization is a cosmic event that
primarily affects the diffuse and mildly overdense IGM where the the
temperature can get increased by about a factor of 2 at the end of the
calculation.  One should consider two effects that support this
conclusion. In regions of significant overdensity and therefore dark
matter potential, gas is heated by the gravitationally controlled free
fall compression to temperatures $T \gtrsim 10^{5}$ K. Therefore, the
effects of photoheating due to the photoelectrons ejected by radiative
ionizations of \heii only result in an insignificant fractional change.
In addition, the large electron density and gas density can be a source
of large optical depths and radiation trapping due to increased
recombination time scales for a modest increase in the temperature. As a
result the radiative energy density can significantly drop in such
regions which furthermore reduces the efficiency of \heii ionization.

\begin{figure*}
\includegraphics[width=3in,height=3in]{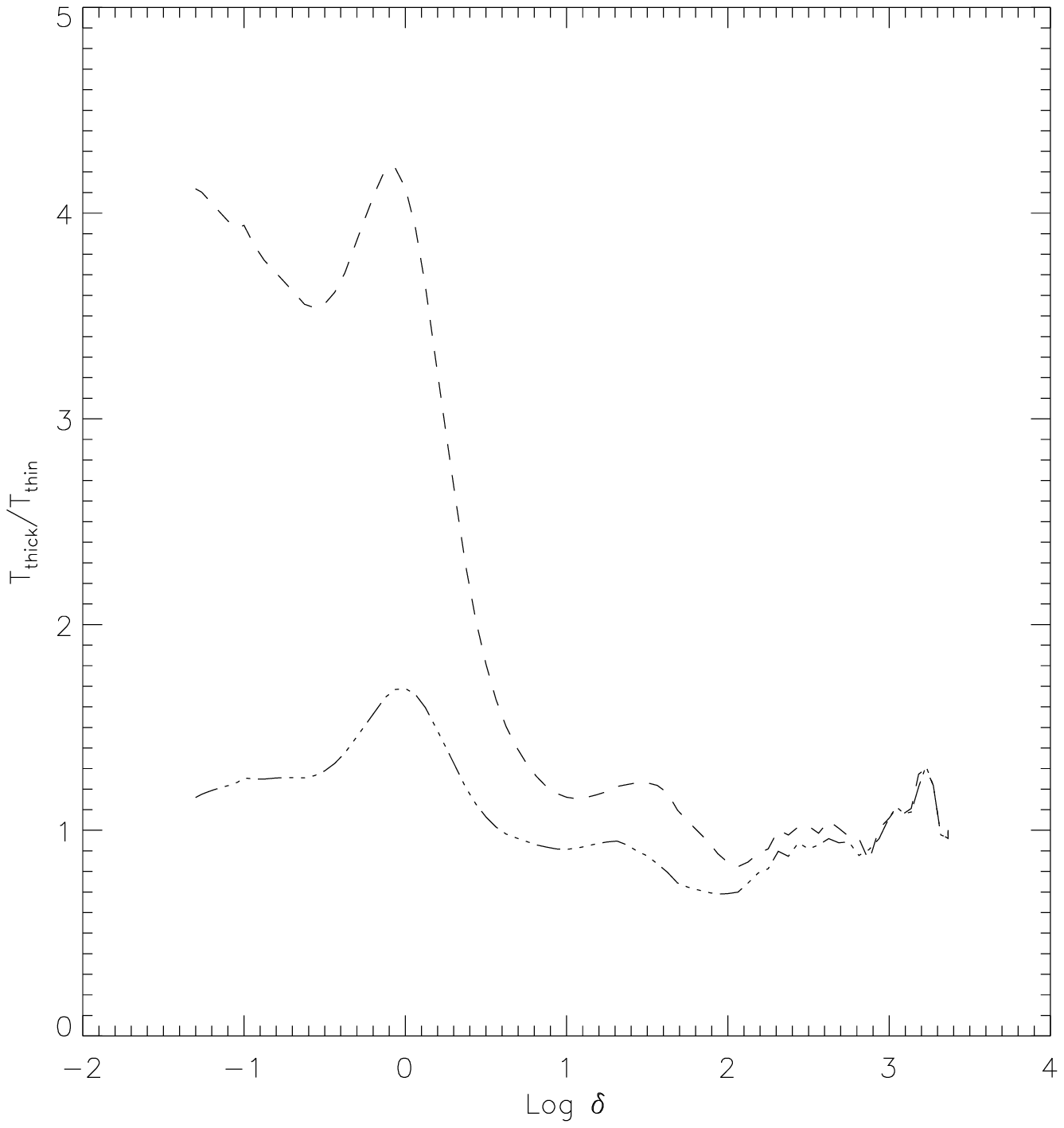}
\includegraphics[width=3in,height=3in]{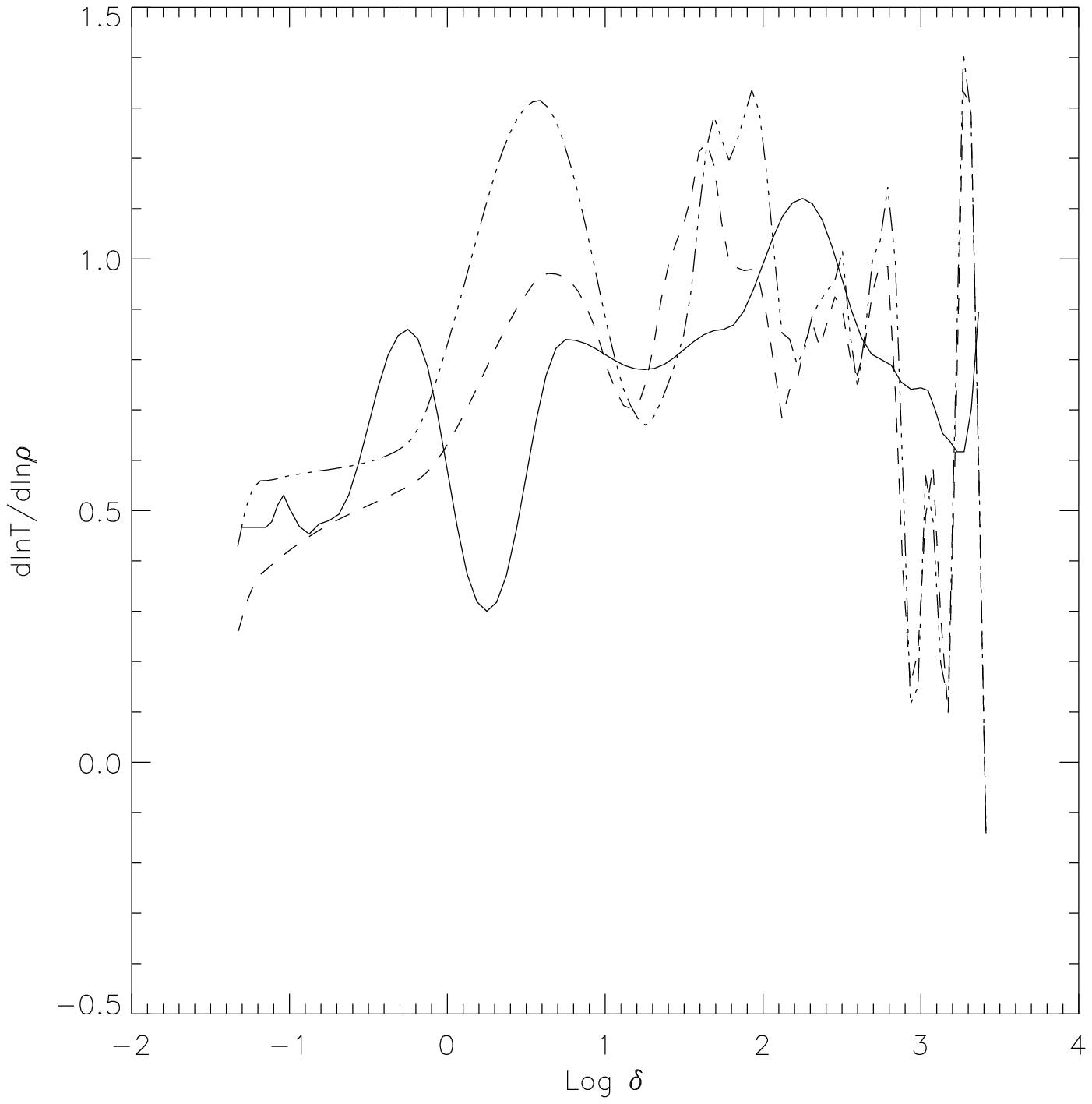}
\caption{Left Panel: From temperature-density plot in Figure~(\ref{tempevl}) we compute
the ratio of the postprocessed result over the input simulation (dashed) and the optically
thin simulation (dashed-dot). 
Right Panel: We plot the distribution of the logarithmic slope $\frac{dlnT}{dln\rho} = \gamma - 1 $
versus the gas overdensity density.}
\label{tempevl-2}
\end{figure*}

The optical depth effects are demonstrated in the right panel of
Figure~(\ref{tempevl}), where we show (z=2.5) the median temperature vs.
density (solid line) in the scatter plot between the two quantities
on the simulation grid. We obtain the
curve by binning the gas overdensity in logarithmic intervals and
computing the median gas temperature within each bin. For reference, the
relation in the original calculation is also shown (dashed lines). 
On the left panel of Figure~(\ref{tempevl-2}), we plot the ratio between the optically thick
optically thin calculations. The dashed lines shows the effect of
our postprocessing on the gas temperature. \heii photoionization contributes
primarily to the gas temperature at the mean and low gas densities. The effect is
diminished at higher overdensities where collisional ionization dominates.
In addition to comparing to the original calculation we also compare 
to a simulation where the chemical species abundunces are self-consistently computed
during the simulation due to the same homogeneous UV background and in the 
optically thin limit. The dot-dashed curves on the right panel of
Figure~(\ref{tempevl}) and the left panel of Figure~(\ref{tempevl-2})
trace the temperature density relation in that case and the ratio to our optically thick
calculation respectively. We find that our postprocessed temperature is higher by a factor
of $\simeq 1.7$ at the cosmic density level when compared to the optically thin 
calculation with a homogeneous UV background. This is consistent with the analytical
approximation between an optical thick and optical calculation derived in
Abel \& Haenhelt (1999). Finally, we plot on the right panel of Figure~(\ref{tempevl-2})
the relation between the slope of the equation of state versus the gas overdensity for
all three simulations. 

In conclusion, the increase in temperature is a manifestation of the
additional heating that results from photoelectrons ejected by \heii
ionizations in the redshift interval that corresponds to the rise in the
number density of sources emitting hard radiation. The physical
implication is that the late increase in temperature may well be the
reason why the galaxy luminosity function decreases at $z \lesssim 4$.  
A raise in the mean temperature due to the cosmic evolution of QSO
sources would increase the Jeans mass threshold by a factor of 2.2-5 if
we adapt an average increase in the temperature between 1.7-3 ($M_{jeans}
\propto T^{\frac{3}{2}}$). The latter would in turn suppress the further
formation of dwarf galaxies in the cosmic volume. However, we find that,
according to the right panel of Figure~(\ref{tempevl}), the fractional
increase in local gas temperature is on average not that large at higher
overdensities. If we exclude cosmic neighborhoods that are in close
proximity to the local UV sources, where the temperature increase is
significantly greater due to the high radiative energy density, then in
collapsed structures at $\delta \geq 100$ which are increasingly
self-shielding, photoheating is not as effective as shock heating due to
gravitational collapse. Nonetheless, a firm conclusion on that effect is
not possible in this work, due to the very coarse grid resolution that
does not adequately resolve the aforementioned structures.

\section{Signatures of \heii Reionization}\label{obs}

\subsection{Synthetic Flux Spectra}

The updated \heii density and gas temperature can be used to study the
effects of this inhomogeneous reionization scheme on the transmission of
the intergalactic medium transmission in the \heii Ly$\alpha$ restframe
wavelength. The objective is to compare with the \heii transmissivity
obtained from analyzing the currently available observed lines of sight.
We synthesize spectra of the \heii Ly$\alpha$ absorption at the rest
wavelength of 304~\AA~ along 300 random lines of sight (LOS). The number
of LOS was chosen in order to yield a less than 1\% fluctuation to the
mean transmitted flux by the end of the calculation.

The synthesis method is described in detail in Zhang et al. (1997). In
addition to \heii, we also compute the transmitted flux of the
corresponding \hi forest at the rest wavelength of Ly$\alpha$ 1216~\AA.
Each velocity pixel registers the local absorption in \heii and \hi and
therefore maps onto the same grid location for the two redshifted
wavelengths. The redshift interval of the spectrum output is set at
$\Delta z=0.2$. In addition to this interval being the redshift output
interval of the hydrodynamic simulation it is a long redshift path that
minimizes the sightline to sightline variance by forcing the transmission
path to cross and wrap through the volume boundaries along the same
directional vector about $\frac{\Delta z}{\Delta z_{cube}}$ times.
$\Delta z_{cube}=L~H(z)/c$ is the linear size of the cube in redshift
units where $H(z) = 100h ( (1+z)^3\Omega_{M} + \Omega_{\Lambda}
)^{\frac{1}{2}}$ is the Hubble constant at redshift z and L=100 Mpc
comoving. At $z=2.5$ and for h=0.71 the Hubble constant is $H(z=2.5) =
249.1$ Mpc/km/s which yields $\Delta z_{cube} = 0.083$ and $\frac{\Delta
z}{\Delta z_{cube}} = 2.4$, the number of wraps through the computational
volume.

The output from the hydrodynamic simulation is saved every $\delta
z_{dump} = 0.2$ and each optical depth integration is computed between
two data dumps at $z_{1}$ and $z_{2}$. The output spectrum is therefore
centered at $0.5 \times (z_{1} + z_{2})$, where $z_{2} = z_{1} - \delta
z_{dump}$. The input fields are species density (\hi and \heii), gas
temperature and the three peculiar velocity components, all assumed to be
frozen in the comoving frame of reference. However, we allow proper
evolution along the redshift path of the sightline for the densities
($\propto (1+z)^{3}$) and velocities ($\propto (1+z)^{-1}$). The spectra
are computed along lines of sight that sample the computational volume
continuously for $z \leq 6.1$. We do this by restarting the calculation
in the new redshift interval from the mesh location where the previous
calculation stopped and continue along the same directional vector. The
initial point of origin is randomly selected. After the completion of
each $\Delta z = 0.2$ segment we can paste all segments together to
obtain a continuous transmission line of sight. The latter can then be
resampled at intervals centered to suit our analysis.

At redshifts where absorption features can be recognized as blends of
individual lines, a deblending algorithm, described in Zhang et al.
(1997; 1998) based on fitting lorentzian profiles below a transmission
cutoff, can identify such lines and compute properties such as the column
density, broadening width (b-parameter) and equivalent width. We must
note, that as the opacity of the IGM increases with redshift and the
transmission spectrum is dominated by dark regions, the algorithm fails
in identifying the lines. Typically, good results are obtained just past
the reionization opacity tail which will be the focus in this analysis.

\begin{figure*}
\includegraphics[width=3in,height=2in]{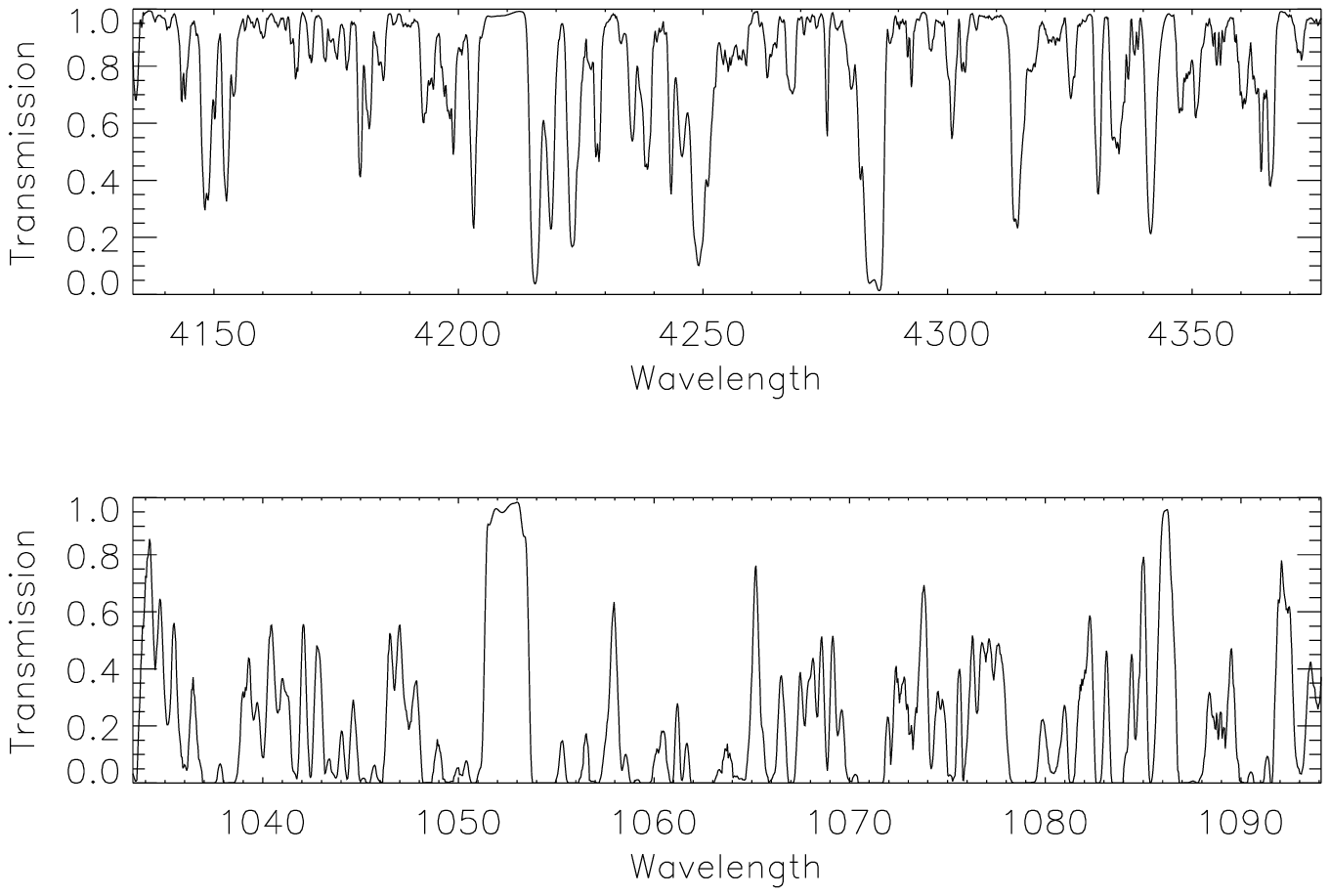}
\includegraphics[width=3in,height=2in]{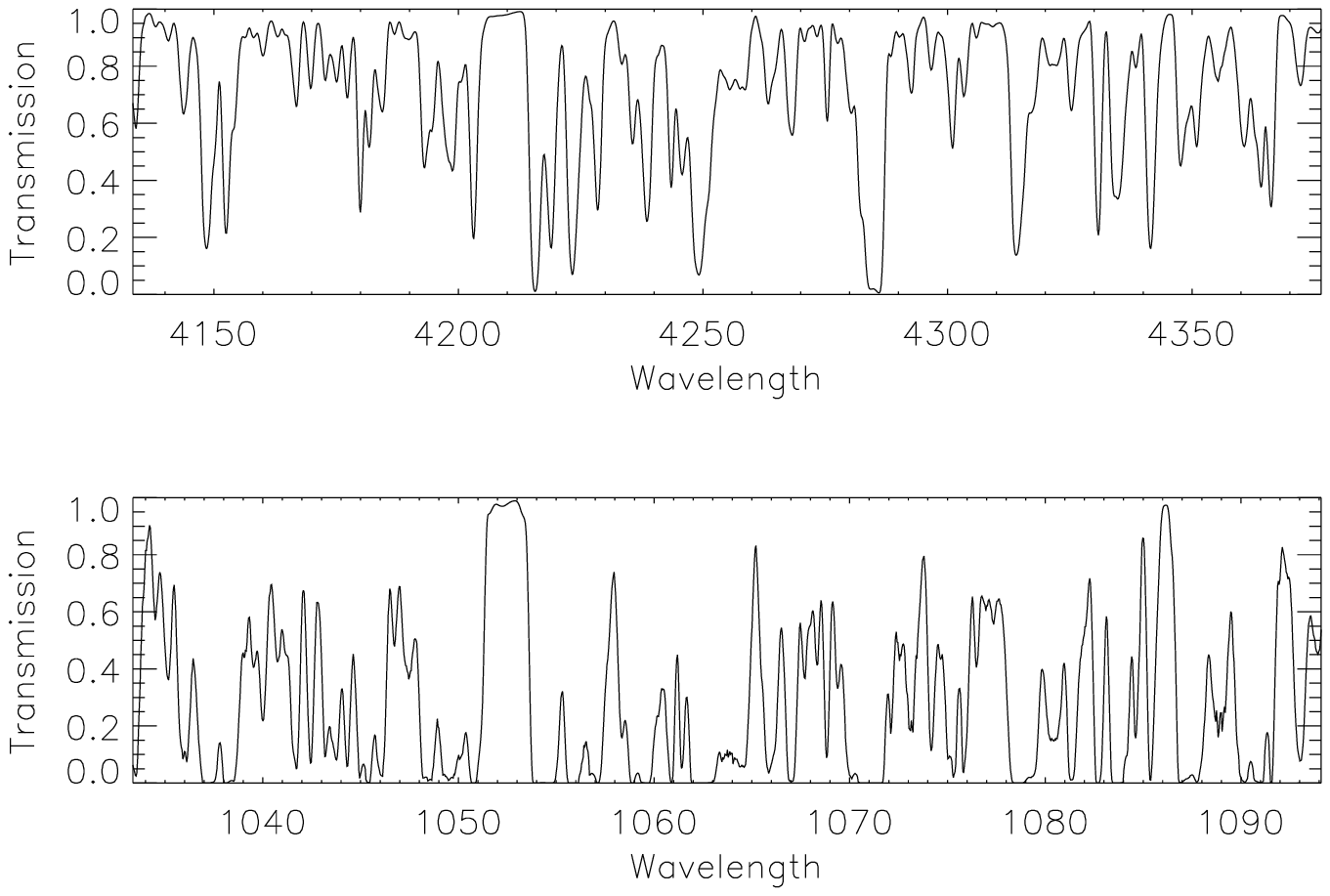}
\caption{
\heii and \hi Ly$\alpha$ spectra between $z=2.6-2.4$, at $\bar{z}=2.5$, 
along a randomly casted
sightline through the computational volume. The upper panels refer to the HI Ly$\alpha$
transmission. The lower panels refer to the HeII Ly$\alpha$ transmission. The horizontal axis
is converted to observed wavelength $\lambda = \lambda_{o} (1+z)$, where $\lambda_{o}$ is the
restframe wavelength of the resonant Ly$\alpha$ scatter. Left Panels: Uniform UVB ionizes and
heats the IGM in the optically thin approximation. HeII ionization and heating is included in
this calculation. Right Panels: The uniform UVB only ionizes \hi to \hii and \hei to \heii.
Photoheating therefore due to the uniform UVB only refers to the above processes. The
non-uniform point source radiative input from QSO type sources is used to photoionize \heii to
\heiii and to calculate the subsequent photoheating. Although the \hi spectra seem similar
between the two sets of calculations, ignoring the feedback on \hi abundance by the \heii
ionization results in underestimating the mean \hi Ly$\alpha$ transmission, compared to the
full simulation, by $\approx 8$\%.
\label{spclm}}
\end{figure*}

In Figure~(\ref{spclm}) we show an example sightline of \hi (top left
panel) and \heii (bottom left panel) Ly$\alpha$ transmission at z=2.5 and
compare with the results from an Enzo simulation that computes abundances
self-consistently in the optical thin limit (top and bottom right panels)
using a UVB from a mix of evolving quasar and galaxy populations (Haardt
\& Madau 2001).  Both transmission lines are casted from the same initial
point and along the same directional vector. We use the updated
temperature and the unperturbed neutral hydrogen abundance for
calculating the \hi transmission spectra. Although there are few visible
differences at first glance, the \hi spectrum in the postprocessed
temperature case has a continuum flux level below that of the optically
thin, self-consistent case. Since the mean transmitted flux is sensitive
to the number of pixels close to the continuum at low redshifts the line
of sight average transmitted flux in the post processed calculation is
below the value obtained from the self consistent calculation by about
$\approx 8$\%.  That was expected in our discussion in
Section~(\ref{h1r}).

In the absorption spectra along the randomly casted sightlines there is
an apparent lack of signatures marking the presence or not of the local
ionizing sources. The spectra in the examples shown in
Figure~(\ref{spclm}) look in general similar, although in some isolated
regions the local amplitude clearly differs from a uniform UV background.
This result is expected because at the time of reionization completion,
the mean absorption in a random sightline spanning 2.4 times the length of
the simulated cube is not sensitive to any local effects by the ionizing
sources but rather depends on the value of the ionizing radiation at the
level of the mean density and the density valleys.

\begin{figure*}
\includegraphics[width=3in,height=3in]{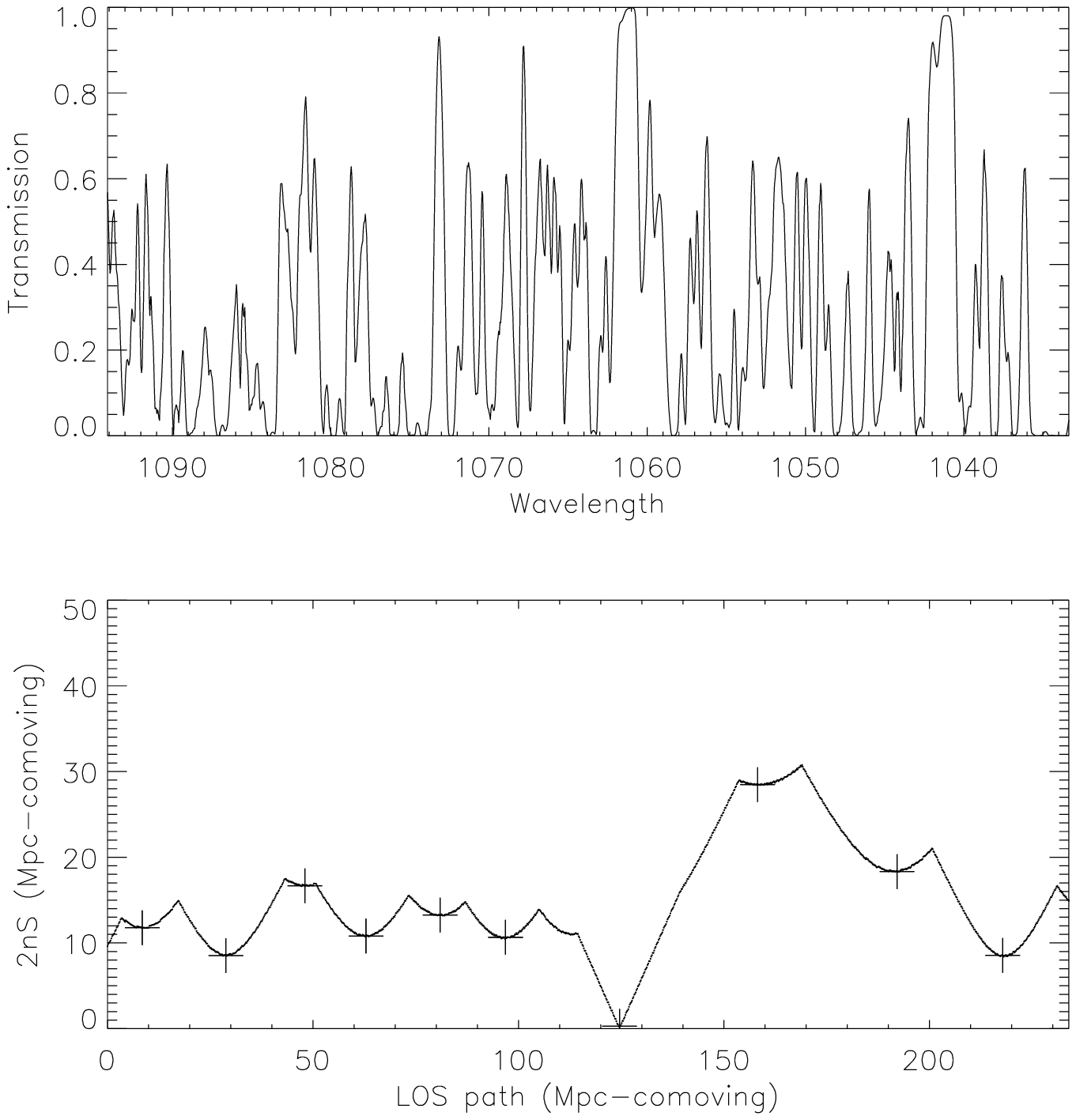}
\includegraphics[width=3in,height=3in]{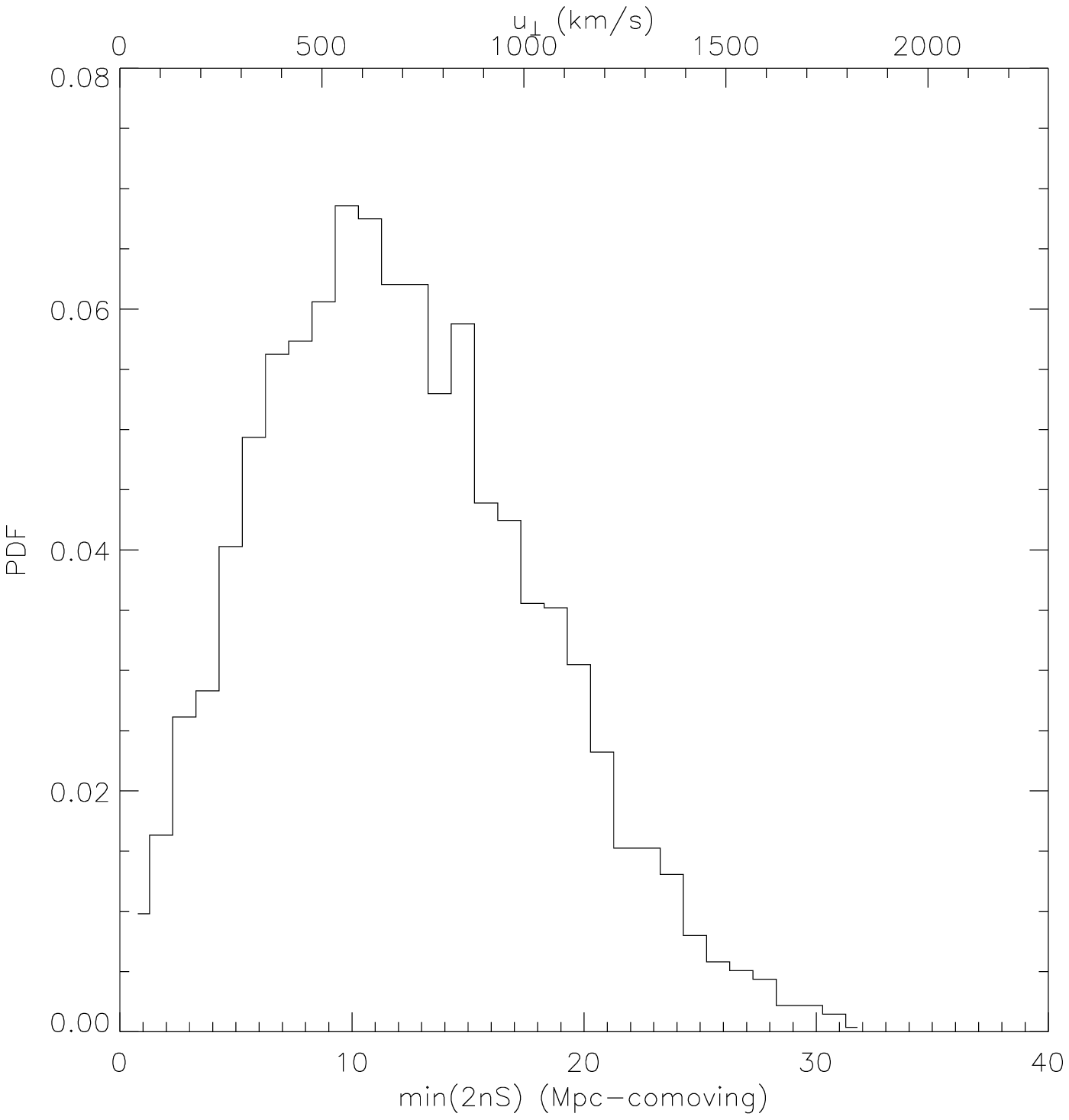}
\caption{Impact parameters of the random sightlines to the
point sources in the box at $\bar{z}=2.5$. At each point along 
the path of the sightline, we compute the distance to the closest
point source. As the trajectory crosses the computational volume,
the closest source changes depending on position. On the lower
left part of the figure, we plot this proximity distance for
a sightline that comes the closest to a point source, about
one grid resolution element. On the
top left panel, we show the \heii transmitted flux for that sightline.
Crosses mark the locations along the path that are closest to 
the point source. The alternating proximity sources are indicated
by the alternating parabolic profiles. On the right panel, we 
show the probability distribution of the minimum distance to
the sources, which we define as the impact parameter, from all random sightlines.
The asymmetry in the distribution is small, which is due to the near isotropic distribution 
of high density peaks at the scale of the simulated volume. 
\label{2nSfig1}}
\end{figure*}

We show on the left panels of Figure~(\ref{2nSfig1}) a sightline with a
path that takes the sightline about one mesh resolution element, $\approx
0.2$ Mpc comoving, away from a UV source. On the top left panel, we show
the Ly$\alpha$ \heii transmission versus observed wavelength. Below, we
plot the instantaneous comoving distance of each velocity pixel to the
closest UV source versus the comoving path length of the sightline.  The
last curve is a joint ensample of parabolic curves each having a minimum,
marked with crosses, that corresponds to the position of nearest distance
to the ionizing source.  Each time another UV source becomes closest to
the trajectory, the curve is marked by the beginning of an another
parabola. We note that at the location of closest proximity for that
sightline, at $\approx 120 $ Mpc comoving along the path, the
corresponding transmission is almost 1. Since the sources are placed at
the high density dark matter peaks, which they subsequently follow during
the dynamical evolution of the dark matter distribution, the absorption
at that location is due to the UV source proximity. However, at a part of
the spectrum about 1042~\AA, the high transmission there is due to a
highly ionized underdense region since the closest point source is more
that 20 Mpc comoving away. Therefore, from the information in the spectra
alone, we will not be able to distinguish between transmission in a region
close an ionizing source or transmission from an underdense region.

We will call impact parameter each nearest distance to an ionizing source
encountered by the sightline trajectory. On the right panel of
Figure~(\ref{2nSfig1}), we plot the sightline distribution of impact
parameters in bins of constant size 1 Mpc. The upper horizontal axis
converts comoving Mpc distances to proper separation velocities. The
distribution is negatively skewed with a peak frequency (mode) located at
$\approx 9$ Mpc ($\approx 550$ km/s). Due to negative skewness, the mean
is shifted to the right at a value of $\approx 14$ Mpc ($\approx 750$
km/s). The negative skewness is most likely due to the fact that the
distribution of the UV point sources on the grid is not isotropic but are
clustered according to the clustering properties of the host dark matter
halos. From the distribution in the right panel of Figure~(\ref{2nSfig1})
we can compute a typical range of impact parameters in the sightlines of
$\approx 2-22$ Mpc comoving contained within the values of the
distribution at 1/e the peak value. We are interested to determine
whether the transmission properties of our sightline sample is in any way
biased by the proximity to the ionizing sources or if we are mostly
sensitive to the values of the ionizing flux at the mean level.

Detailed analysis of the proximity effect around each source is beyond
the scope of this paper. However, we can impose an upper bound limit
based on the highest luminosity quasar in the computational volume at
z=2.5, $L^{max}_{4} = 2 \times 10^{44}$ ergs/s where $L_{4}$ designates
continuum ionizing luminosity above 4 Ryd. An estimate of a proximity
size distance can be obtained from the volume averaged energy density
above the \heii ionization threshold, $\bar{E}_{4}$. In our calculation
we have computed $\bar{E}_{4} = 7.7 \times 10^ {-17}$ ergs/cm$^{-3}$
which corresponds to a volume averaged \heii ionizing flux equal to
$\bar{F}_{4} = c \bar{E}_{4} = 2.3 \times 10^{-6}$ ergs/s/cm$^{-2}$. The
condition that the geometrical attenuation of the central emission equals
the total flux through the surface of a sphere centered at the source
yields an upper bound for this proximity distance of $d^{proper}_{HeII}
\leq (L^{max}_{4}/4 \pi \bar{F}_{4})^{\frac{1}{2}} \simeq 0.86$ Mpc
proper or $d_{HeII} \lesssim 3$ Mpc comoving. When compared to the
comoving length of the sightlines ($\approx 270$ Mpc comoving) we expect
that these regions do not have a large effect on the mean \heii
Ly$\alpha$ transmission value.

\begin{figure*}
\includegraphics[width=3in,height=3in]{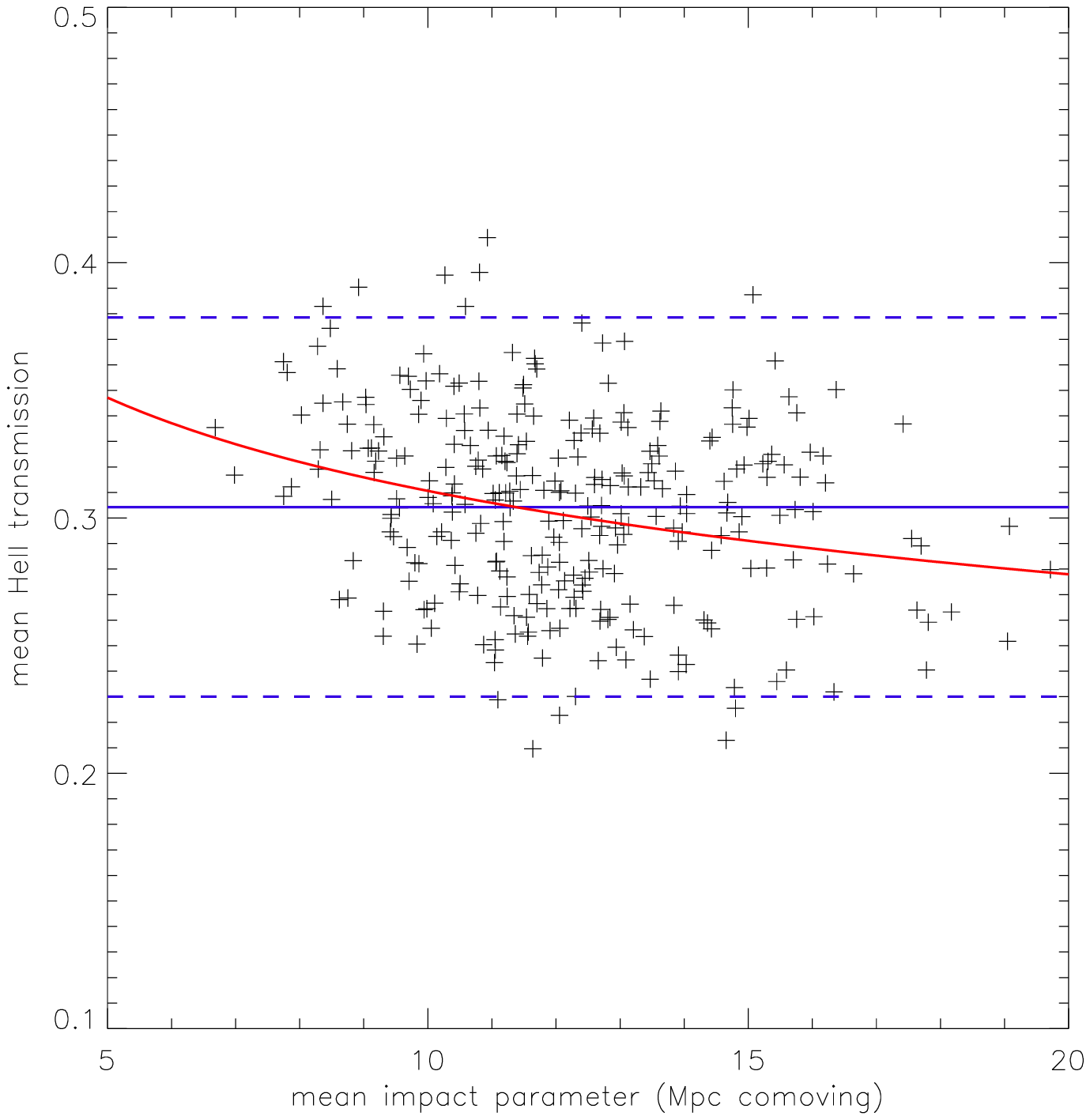}
\includegraphics[width=3in,height=3in]{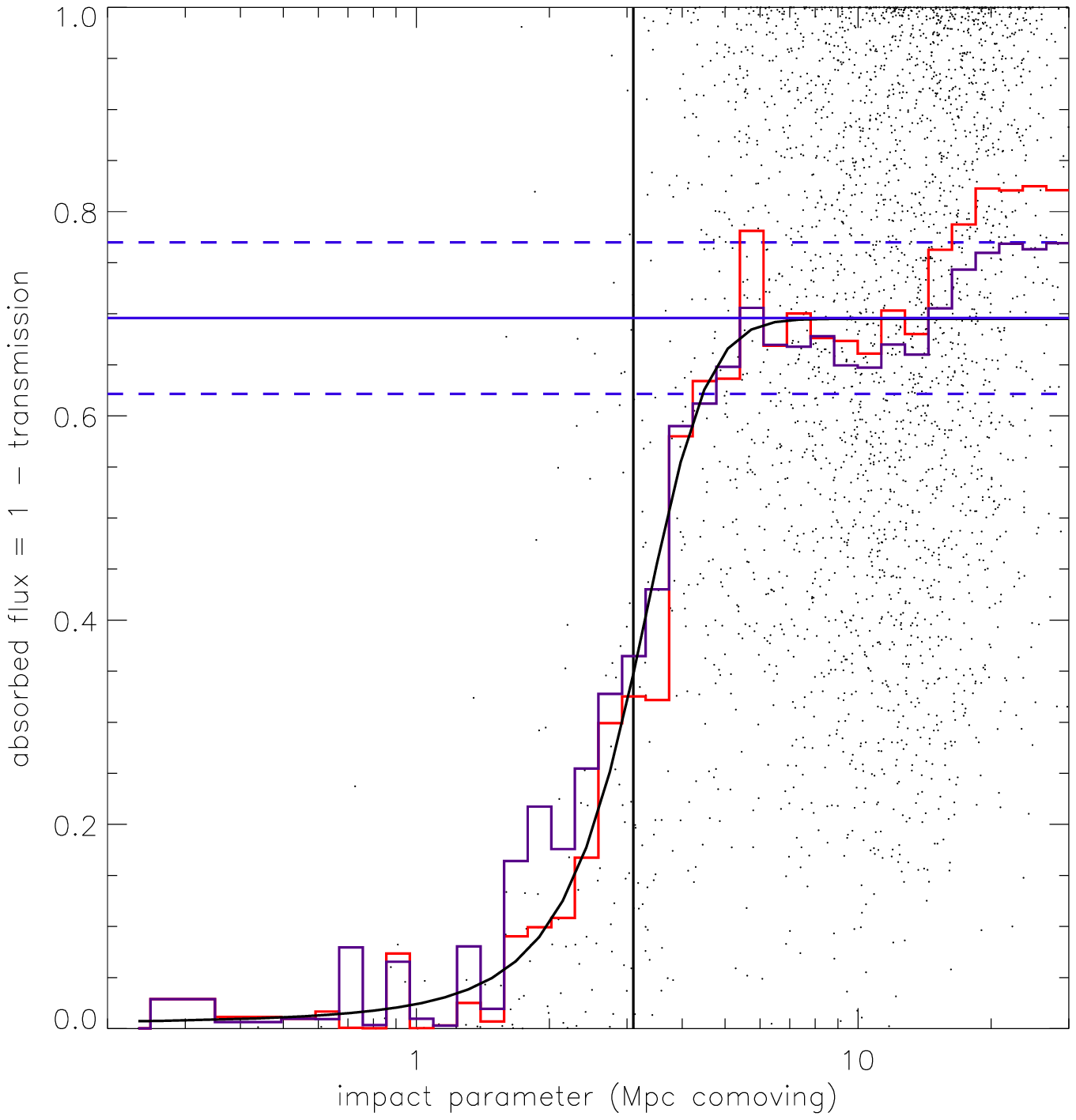}
\caption{
On the left panel, we show the scatter plot between the average impact parameter 
for each sightline in respect to the distribution of the 
ionizing point sources versus the sightline-specific mean flux. 
The red line is power law fit to the data with a slope of $s = -0.16 \pm 0.04$.
The blue solid (dashed) line shows the average ($2\sigma$) \heii Ly$\alpha$ transmission 
from all sightlines at $\bar{z}=2.5$.
The small negative correlation $r = -0.24$ suggests that the mean flux of a sightline is largely 
insensitive to the average proximity in respect to the distribution of ionizing sources along it's path.
On the right panel, we show the locally absorbed flux (y-axis) at the point of closest proximity to 
the ionizing source shown on the x-axis. Individual points correspond to the absorbed flux
at the pixel of closest distance. The red (blue) histogram is the median (mean) absorbed flux per 
constant logarithmic bins. Solid and dashed blue horizontal lines are the level of the mean absorption
in our sightline sample and the 2$\sigma$ standard deviation respectively. The solid line is
a Fermi-Dirac function that fits the median (red) histogram for impact parameters less that 10 Mpc 
comoving. The half point of the Fermi-Dirac function is at $d_{o} = 3.1$ Mpc comoving (shown as 
a vertical black solid line) and it has a skin width of 0.63 Mpc comoving. Impact parameters along the 
sightline of $\lesssim 4.5$ Mpc comoving, the point where the absorbed flux falls below the 
2$\sigma$ level of the mean absorption, show a steep sensitivity in the absorbed flux due to the 
quasar proximity. 
\label{2nSfig2}}
\end{figure*}

On the left panel of Figure~(\ref{2nSfig2}), we show a quantitative
estimate of that effect.  There, we scatter plot the average impact
parameter of the sightline versus the mean transmission in each sightline
(cross symbols).  The degree of their linear correlation, measured by the
Pearson correlation coefficient, is rather small, $r=-0.24$. The figure
does show that there is non-negligible negative correlation between mean
transmission and the average proximity to the distribution of ionizing
sources for each sightline.  A power law fit to the data in the form
$\bar{F}_{los} \propto \bar{d}_{ip}^{-s}$ where $\bar{F}_{los}$ is the
sightline specific mean flux and $\bar{d}_{ip}$ is the average impact
parameter yields $s = -0.16 \pm 0.04$ and is plotted over the individual
points. The curve lies within the two standard deviations levels of the
mean transmission. Therefore, we conclude that the proximity effects have
no statistical weight on the sightline transmissions. That is consistent
with our estimate of the upper limit on the proximity distance ($\simeq
3$ Mpc comoving) based on matching the attenuation of the point
luminosity to the volume averaged ionizing flux. Since the typical range
of the impact parameters is between 2-22 Mpc comoving our sightlines are
on average too far away from the point sources. This conclusion is also
supported by the right panel of Figure~(\ref{2nSfig2}) where we show the
scatter plot between the individual impact parameter values and the local
\heii Ly$\alpha$ absorbed flux at that distance from the point source
(points). The data show a tail of low absorption at small impact
parameters at distances $\lesssim 4.5$ Mpc comoving. The distribution of
absorption above $\gtrsim 4.5$ Mpc is consistent with the mean level of
absorption in the sightline sample, shown as a horizontal blue solid line
along with upper and lower blue dashed lines corresponding to the
$2\sigma$ level. To get a clear view of the trend we bin the horizontal
axis into constant logarithmic bins and compute the median (red
histogram) and mean (blue histogram) per bin. A Fermi-Dirac function in
the form $f(x) = \frac{1}{1+exp(\frac{(do-x)}{\mu}}$ is then fitted to
the median histogram with $do=3.1$ Mpc (point of half maximum) and
$\mu=0.63$ Mpc (skin width) for impact parameters up to 10 Mpc. The point
of half maximum is consistent with the previous estimate of the proximity
effect distance based on the luminosity and mean intensity of the UV
field. The figure clearly shows that the effect of source proximity on
the local absorbed flux are relevant at comoving distances of $\lesssim
7$ Mpc where the local absorption matches the mean \heii Ly$\alpha$
absorption in the IGM and becomes significant at impact parameters
$\lesssim 4.5$ Mpc comoving where the local absorption matches the lower
$2\sigma$ level of the mean absorption.

\subsection{Optical Depth Evolution}

The straightforward average of the flux, $\bar{F} = \frac{1}{Npx \cdot
nlos} \Sigma^{nlos}_{j}\Sigma^{Npx}_{i} F_{ij} $, from all pixels and
lines of sight per redshift interval, is a measure of the opacity of the
cosmic volume due to the absorption by the particular chemical species
(\heii and \hi in this case). In the notation followed, $Npx = 30,000$ is
the number of pixels per redshift interval, which yields a redshift pixel
resolution of $R_{z} = \delta z= 0.2/30,000 \simeq 6.6 \times 10^{-6}$,
and $nlos = 300$ is the number of random lines of sight. The redshift
resolution is fixed in our calculation, which results in a redshift
dependent spectral resolution of $R_{\lambda} = \frac{\lambda}{\Delta
\lambda} = \frac{1+z}{\delta z} = 1.5 \times 10^{5} (1+z)$. At $z = 3$
this yields a spectral resolution of 50 times higher than the designed
value of the spectrograph aboard FUSE.

\begin{figure*}
\includegraphics[width=3in,height=3in]{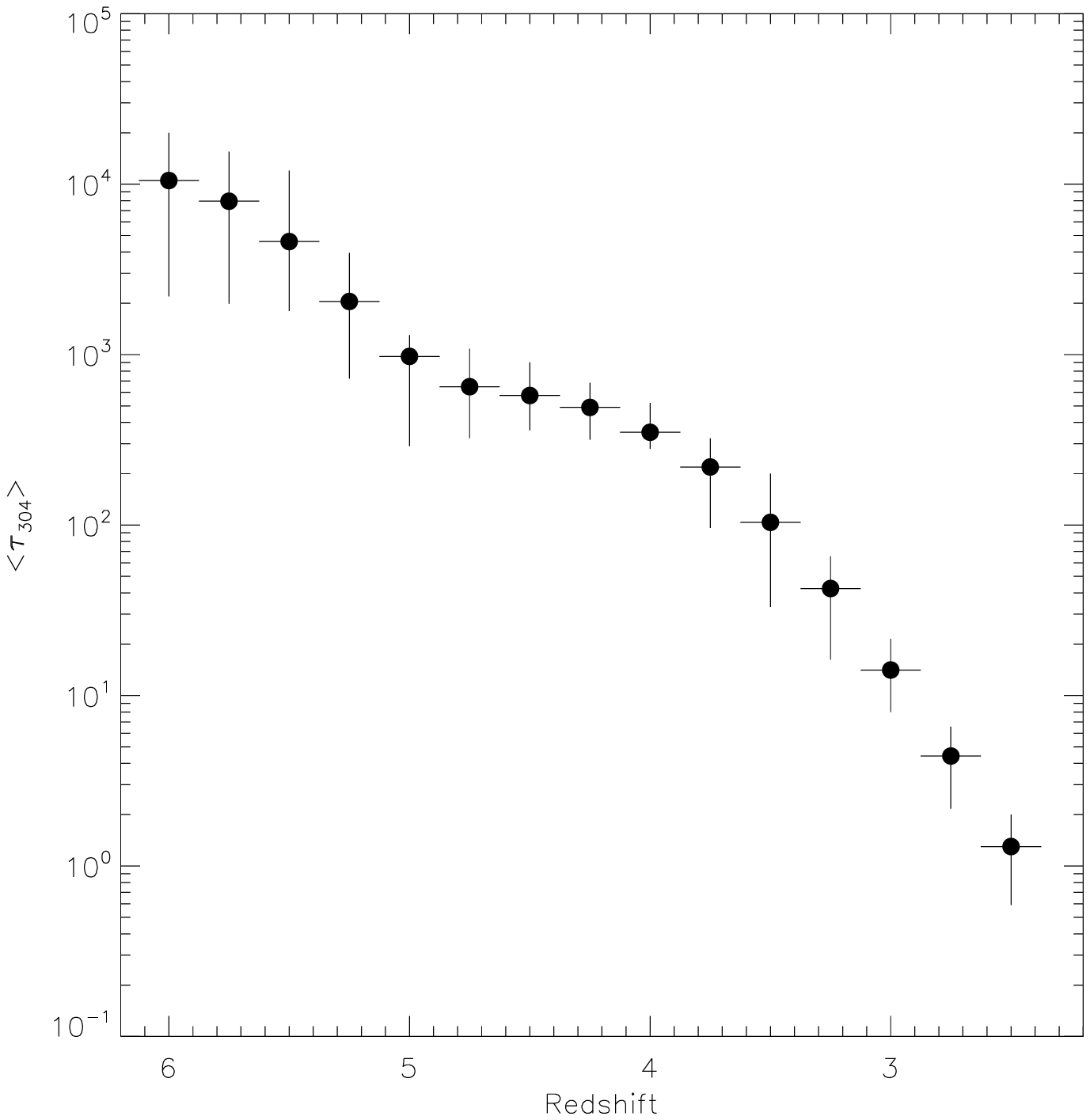}
\includegraphics[width=3in,height=3in]{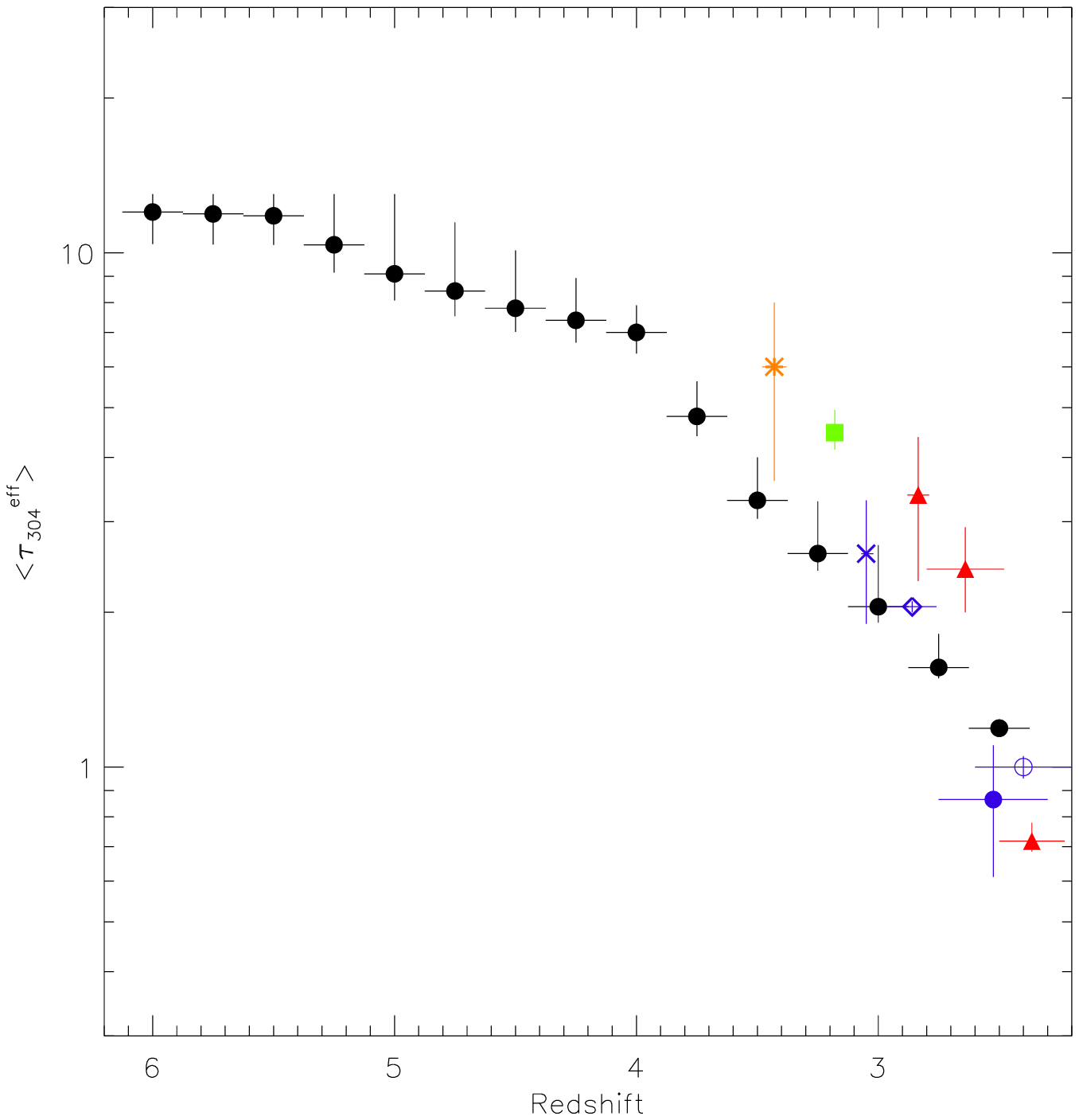}
\caption{
We compute the mean optical depth and effective optical depth at selected mean redshift points,
from $\bar{z}=6-2.5$ every 0.25 redshift units and in intervals of $\delta z = 0.25$ about the
mean redshift. Left Panel: Redshift evolution of the mean pixel Ly$\alpha$ \heii optical depth
averaged over all lines of sight. The error bars are 1$\sigma$ standard deviation to the sightline
average optical depth.  
Right Panel: Redshift evolution of the Ly$\alpha$ \heii effective
optical depth. The error bars in the simulation results are sightline-to-sightline errors 
estimated from the standard deviation to the average transmission from sightline-specific mean
flux. 
Overplotted are sightline measurements along HE2347-4342
(red triangles), HS1700+6416 (blue circles), HS1157+3143 (blue diamond), PKS1935-692 (blue cross), 
Q0302-003 (green square) and SDSSJ2346-0016 (orange star). Error bars to these points refer
to error estimates of the pixel flux along the sightline specific to the quasar. 
The HE2347-4342 data from Zheng et al. (2004b) have been resampled to a larger bin size to make it comparable 
to our data. The redshift profiles of the optical depth point statistics suggest a smooth
evolution of the \heii reionization epoch.
\label{taus}}
\end{figure*}

The evolution of the mean transmitted flux is typically represented by
the effective optical depth, defined as $\tau_{eff} = -ln(\bar{F})$. As
discussed in Paschos \& Norman (2005) (PN), the effective optical depth
is biased by high transmission gaps in the pixel flux distribution and
therefore will systematically yield lower values when compared to the
mean optical depth. The latter is defined as the raw average of the pixel
optical depth per redshift interval, $\tau_{mean} = \frac{1}{Npx \cdot
nlos} \Sigma^{nlos}_{j} \Sigma^{Npx}_{i} \tau_{ij}$, where $\tau_{ij}$ is
the pixel optical depth at redshift $z_{i}$ in the line of sight index
$LOS = j$. In Figure~(\ref{taus}), we show the redshift evolution of the
the optical depth of the 304~\AA~ line using both representations, mean
(left panel) and effective (right panel), which suggests a rather smooth
evolution leading to \heii reionization at $z \simeq 2.5$. Error bars
show the 1$\sigma$ standard deviation of the mean flux and optical depth
values between different lines of sight.  In the effective optical depth
case, standard deviation at more than 100$\%$ the mean flux value at $z
\geq 5$ yield negative lower bound values. A cutoff in the mean flux of
$F_{min} = 10^{-6}$ was therefore imposed in order to be able to compute
the natural logarithm.

The error bars also indicate a large degree of variance in the data at
redshifts $z \gtrsim 3.5-4$ consistent with PN which discussed properties
of \hi transmission during hydrogen reionization. There, the large degree
of variance during reionization was attributed to the presence of high
transmission gaps associated with underdense regions in the IGM which
ionize first. In this picture of inhomogeneous \heii reionization, the
high transmission segments at high redshifts are associated with \heiii
bubbles that the lines of sight intersect as they are cast through the
simulation box. However, this is not inconsistent with the conclusions
reached in PN because the \heiii bubbles primarily extend in underdense
to mean density cosmic regions.

Estimates of the observed \heii effective optical depth, shown in
Figure~(\ref{taus}), are comprised of the six known and analyzed
sightlines todate: HS1700+6416 (Fechner at al. 2006; Davidsen et al.
1996), HE2347-4342 (Zheng et al. 2004b; Kriss et al. 2001), HS1157+3143
(Reimers et al. 2005), PKS1935-692 (Anderson et al. 1999), Q0302-003
(Heap et al. 2000) and SDSSJ2346-0016 (Zheng et al. 2004a). Although very
few in number they suggest a lack of \heii Ly$\alpha$ forest transmission
at redshifts above $z \simeq 3$ and the presence of a trough there
consistent with the hypothesis of late \heii reionization. Upper bound
estimates suggest a steeply rising optical depth above $z \simeq 2.8$
while lower bound estimates infer a smoother reionization transition.
Lower redshift sightlines towards HS1700+6416 (Fechner et al. 2006) and
HE2347-4342 (Zheng et al. 2004b; Kriss et al. 2001) suggest an optical
depth of about 1 between $z \simeq 2.3$ and $z \simeq 2.7$. In this work,
we have calculated that between $z=2.6-2.4$, at $\bar{z} = 2.5$, a mean
transmitted flux value at $\bar{F} = 0.304 \pm 0.002$ which corresponds
to $\tau^{eff} = 1.190 ~\pm~ 0.007$. The error to the mean values is due
to sightline-to-sightline variance and is equal to $1\sigma/ \sqrt
N_{los}$. Observed effective optical depth estimates are
$\tau^{eff}_{obs} = 0.91 \pm 0.01$ for HE2347-4342~ averaged over $z
\simeq 2.3 - 2.7$ and a range between $\tau^{eff}_{obs} = 0.74 \pm 0.34$
and $\tau^{eff}_{obs} = 1.06 \pm 0.18$ at $\bar{z} \simeq 2.45$ for
HS1700+6416 (Fechner et al. 2006). If reionization is completed by $z
\simeq 2.5$, as suggested by the observed lines of sight, that would
correspond to an optical depth of $\tau_{eff} \simeq 1$. Our calculation
is about 1.5$\sigma$ above that value at that redshift. Furthermore, the
computed redshift evolution of the optical depth is smoother than
suggested by observations. We attribute such differences to the
limitations of our treatment and most notably ignoring diffuse emission
due to recombinations to the ground state of \heii. Such recombinations
increase the ionization of \heii and therefore lower the opacity. The
slope of the redshift evolution is affected by such omission because of
the diffuse emission's dependency on the gas temperature, $\propto
T^{-\frac{1}{2}}$. At earlier redshifts, the temperature is lower and
therefore recombinations to the ground state may contribute a significant
amount of \heii ionizing radiation. Nonetheless, due to the lack of a
self-consistent calculation we do not know how large that effect would be
on the slope of the optical depth redshift evolution.  We do anticipate
that the effect diminishes with redshift due to photoheating which raises
the IGM temperature.

From the right panel of Figure~(\ref{taus}) we can infer a general
agreement between the observed and computed values at redshifts that
evidently sample the tail of the \heii reionization epoch. That leads to
the conclusion that the final opacity distribution in \heii is largely
insensitive to the details of our numerical setup. Direct
photo-ionizations due to a rising number density of QSO sources appears
to be adequate to yield a value of the \heii abundance close to the
observed one at late redshifts.

\subsection{Line Statistics}

The largely ionized \heii by z=2.5 ~in our computational volume and the
update in the gas temperature due to photo-heating, allows for the
derivation of two standard statistical properties of a Ly$\alpha$ forest,
the column density and b-parameter distribution. Our grid resolution is
too coarse to draw any absolute conclusions about these distributions. In
order to resolve the \hi Ly$\alpha$ forest, a grid resolution of $\sim
40$ kpc is required (Bryan et al. 1999).  A larger grid resolution may
not be required for \heii lines if they primarily form in underdense
regions. However, comparison with a resolved \hi forest is desirable. Our
intent in this work is to look at the relative differences between the
two forests and draw conclusions that may stand the test of an improved
grid resolution in a future simulation.

\begin{figure*}
\includegraphics[width=3in,height=3in]{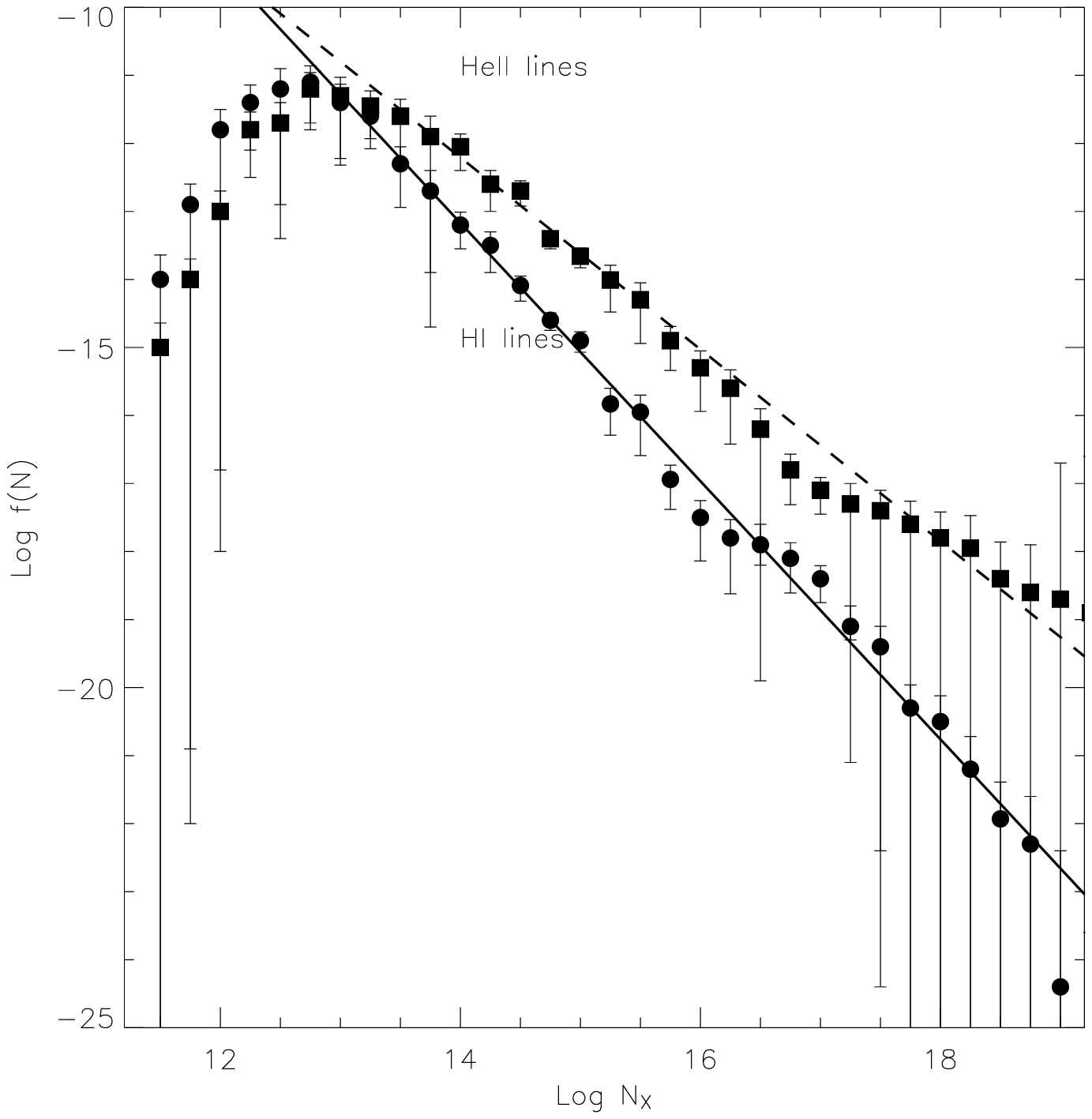}
\includegraphics[width=3in,height=3in]{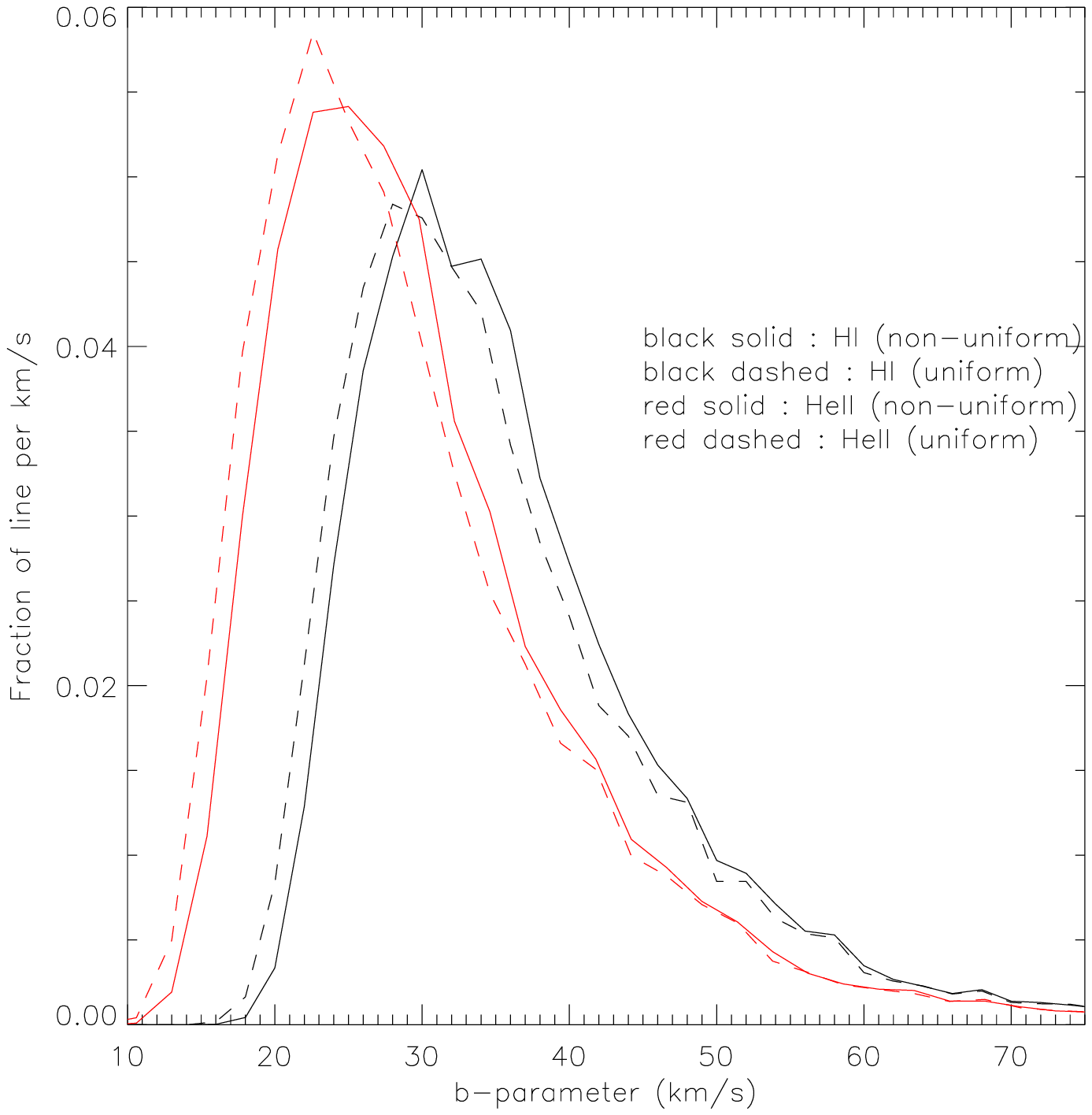}
\caption{Left Panel: Column Density Distributions, $f(N) = \frac{d^{2}N}{dzdN_{X}}$ of
\hi and \heii absorbers between $z = 2.6$ and $z=2.4$. Circle and square points
correspond to the \hi and \heii column density distribution respectively. Solid (dashed) straight lines (in log-log) 
show the power law fit to the \hi (\heii) distribution between column densities $N_{X}=10^{13.5}-10^{15.5}$ $cm^{-2}$.
The error bars show the 1$\sigma$ standard deviation to the number of lines per logarithmic column density bin equal to 0.25.
Right Panel: Broadening width distributions at $\bar{z}=2.5$ for \heii and \hi. The distributions are shown as
the fraction of absorption lines per km/s in bins of 2 km/s. Black and red colors correspond to \hi and \heii distributions
respectively. The dashed lines show a reference self-consistent calculation in the optical thin limit using an ionizing
uniform UVB. The peaks of non-uniform calculations are offset by $\simeq 1.25$ km/s to the right of the uniform results. 
\label{stat}}
\end{figure*}

Between z=2.6-2.4, at $\bar{z} = 2.5$, we identify the \heii and \hi
Ly$\alpha$ lines and compute their column density and b-parameter width.
The line identification was performed using the method described in Zhang
et al. (1998). We count lines per constant logarithmic column density
bins ($\Delta N_{X} = 0.25$)  and normalize per column density and
redshift interval, $dz = 0.2$, for the two species,
$\frac{d^{2}N}{dzdN_{X}}$ (X = \hi, \heii). The result is shown in the
left panel of Figure~(\ref{stat}). The \hi distribution flattens at $Log
N_{HI} \sim 12.5$ which is due to the coarse grid resolution. A similar
flattening occurs at $Log N_{HeII} \sim 13.5$. In addition, the \heii
column density distribution shows a decrease in the slope and large error
bars as the column density increases. The error bars refer to the
1$\sigma$ standard deviation of the identified line counts per
logarithmic bin. By $Log N_{HeII} \gtrsim 17.5$ the dramatic increase in
the error and hardening of the slope is interpreted as the effect of
regions of \heii that are not fully ionized by z=2.5. Such regions occupy
about 10\% of the computational volume and are a source of large optical
depths and therefore column densities in lines of sight that pass through0
them. The two profiles suggest that for column densities in the range
13.5 - 15.5 we identify 10-100 times more \heii than \hi lines. This
result is consistent with a high resolution cosmological simulation that
resolves the two forests (Zhang et al. 1997).

Power law fits to the column density distributions in
Figure~(\ref{stat}), $f_{X}(N)dN_{X} = \beta N^{-\alpha_{X}}_{X} dN_{X}$,
yield $\alpha_{HI} = 1.89 \pm 0.14$ and $\alpha_{HeII} = 1.41 \pm 0.12$
in the logarithmic column density range of 13.5 - 15.5. The error
estimates are derived from the least squares fit and do not take into
consideration any propagated error in the individual column density bins
which is sensitive to the bin size. The Ly$\alpha$ forest is well studied
in the literature, however results that are pertinent to the slope of the
column density distribution typically refer to either the range of $log
N_{HI}(cm^{-2}) = 12.5-14.5$ or to an average slope between the minimum
to maximum value in the column density sample.  From distributions
obtained in simulations (Zhang et al. 1997; Jena et al. 2005) and
observations (Petitjean et al. 1993; Rauch et al. 1997; Kirkman et al.
1997) we have estimated a range in the slope of the distribution between 
$N_{HI} = 10^{13.5}-10^{15.5}$ cm$^{-2}$ of $\alpha_{HI} = 1.7 - 1.85$ for redshifts between z=2-3. 
The slope for the \hi column density distribution 
obtained in this work is slightly above. That is primarily due 
to the poor grid resolution that underestimates the 
number of \hi absorbers with high column densities. To a lesser extent,
we also expect a steepening in the distribution because we overestimate
the \hi opacity for lower column density absorbers in our numerical setup.
However, we note that within the error estimate of the power law fit, 
the value of the slope we computed here overlaps with the range of published values.

The estimated slope of the \heii Ly$\alpha$ column density distribution,
in the optical thin limit, using uniform photoionization models, ranges
between $\alpha_{HeII} \approx 1.5 - 1.6$ in the redshift interval z=2-3 
and for $logN_{HeII} = 13.5-15.5$ (Zhang et al. 1997). 
In this work, we have calculated the slope of the distribution at $\bar{z} = 2.5$ 
to be $\alpha_{HeII} = 1.41$ which is below the previous estimates.
As we shall show below the \heii absorbers correspond to physically extended IGM structures 
and therefore the \heii column density distribution is less sensitive to 
our coarse grid resolution than hydrogen. The smaller slope is due to the 
opacity effects of the inhomogeneous ionizing radiation. 
Overdense regions will have higher opacity to \heii ionizing radiation
compared to a uniform photoionization model. That in turn results in
lines migrating to higher column density bins which softens the slope of the distribution
compared to an optically thin calculation.

We now proceed to investigate the effects of non-uniform photoheating in the distribution
of the line broadening widths which is the observable that closely traces
the thermal state of the IGM. The b-parameter distribution, computed here
as the fraction of lines per km/s is shown for the two species on the
right panel of Figure~(\ref{stat}) (solid lines).  The distributions were
computed from lines with column densities in the range $Log N_{X} = 13.5
- 15.5$ for both species.  The selected range aims to avoid the
undersampling of weak lines due to the coarse grid resolution.
For reference, we also plot the b-parameter distribution computed from a
uniform photoionization optically thin simulation due to a Haardt \&
Madau (2001) UVB model from a mix of quasar and galaxy populations
(dashed lines). The median line broadening widths derived in the
non-uniform UV case are $b^{HI}_{med} = 34.23$ kms$^{-1}$ and
$b^{HeII}_{med} = 28.16$ kms$^{-1}$ for the \hi and \heii lines
respectively. For comparison, the median values in the uniform UVB case
are $b^{HI}_{med}(u)= 32.98 $ kms$^{-1}$ and $b^{HeII}_{med}(u) = 26.89$
kms$^{-1}$. When we compare the non-uniform and uniform results we note
that the peaks of the distributions are offset by $\simeq 1.25$ km/s in
both species. This is expected according to the right panel of
Figure~(\ref{tempevl}) where the median temperature per logarithmic
overdensity bin is plotted. Differences in the overdensity dependent
median temperature distribution arise at overdensities $\Delta \lesssim
3$. If we adopt a relation between \hi column density and overdensity in
the form of $\Delta \geq 10 (\frac{N_{HI}}{10^{15}})^{2/3}
(\frac{1+z}{4})^{-3}$ (Schaye et al. 2003) then we can compute that at
z=2.5 a \hi column density of $10^{13.5}$ cm$^{-2}$ is due to
overdensities of $\Delta \gtrsim 1.5$. At such overdensities we predict
the biggest difference in the temperature between the non-uniform versus
the uniform cases. The increase in temperature by a factor $\simeq 1.6$
is consistent with a shift in the peak of the b-parameter distribution.
Note however, that the distributions at broadening widths larger than 50
km/s are indistinguishable which may be due to collisional ionization
dominating the line opacity.

Since the forests are unresolved on the simulation grid, we make no claim
on whether the computed b-parameter values have any relevance to the real
universe.  We can however compare the median values of the two species.  
Relative to hydrogen the median values infer that on average the \heii
lines have about $\approx 82$\% the broadening width of the \hi lines.
For pure thermal broadening, the heavier by a factor of 4 helium atoms
would yield a width only half the corresponding size of the hydrogen
line, $b^{HeII}_{th} = \frac{1}{2} b^{HI}_{th}$. Our calculation suggests
that the broadening due to the Hubble differential flow and peculiar
velocities dominate the line formation in the \heii forest. That in turn
would indicate that the \heii lines primarily form in underdense and
cosmic mean density regions. Our result is consistent with the
conclusions in Zheng et al. (2004) where they calculated that along the
HE2347-4342 transmission line $b_{HeII} \simeq 0.95 b_{HI}$ which also
supports a Hubble dominated absorption line broadening for \heii.

\subsection{$\eta$-Parameter Evolution}

\begin{figure*}
\includegraphics[width=6in,height=4in]{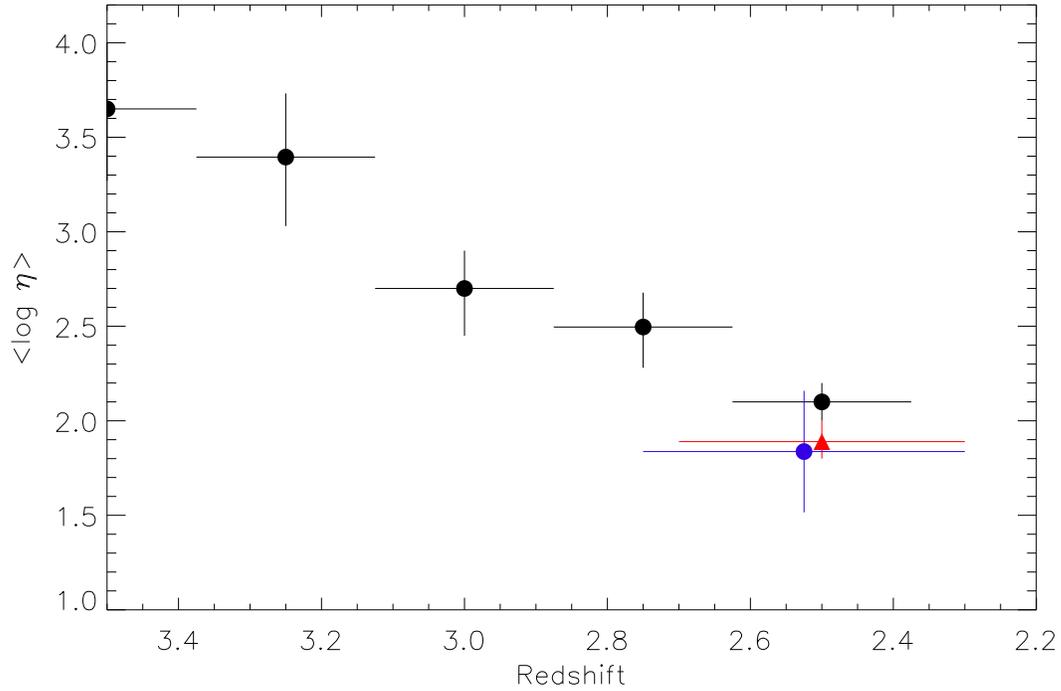}
\caption{Redshift evolution of the $\eta$ parameter in this simulation.
Results are shown as $<log \eta>$ per redshift interval and are obtained
as the line column density ratio of \heii over \hi 
at low z ($z < 3$) and estimated
from the ratio of optical depths at higher redshifts. 
The filled triangle shows the average estimate between $z=2.3-2.7$
towards HE2343-4342 from Kriss et al. (2001). We also show the data from
Fechner et al. (2006) (blue towards HS1700+6416 collapsed into a single 
redshift interval from z=2.3-2.75. Our calculation yields 
systematically larger values, a
result due to the larger estimate of the effective optical depth 
by $\bar{z} = 2.5$.  
\label{eta}}
\end{figure*}

We conclude this section, by computing the redshift evolution of the
$\eta$-parameter through the R-factor, which are defined in
\S\ref{whylate}.  In Figure~(\ref{eta}), we plot the redshift evolution
of the $\eta$-parameter in the range $3.5 \leq z \leq 2.5$ in redshift
intervals of $\delta z = 0.2$, averaged over all lines of sight. Direct
computation of the quantity not only requires knowledge of the column
density for each individual line but also knowledge of the local
association between the \hi and \heii absorption features. However,
because the transmission of the \heii Ly$\alpha$ line exhibits a trough
at redshifts $z \gtrsim 3$, obtaining the column density value for
individual \heii lines is not feasible there due to the inability to
deblend the absorption features in the spectrum. In addition, whatever
correlation between the \hi and \heii lines exists at $z \lesssim 3$ it
would be incomplete because not all \hi lines can be associated with
\heii absorption features and vice versa (Kriss et al. 2001).  Our
biggest problem though in this calculation lies in the separate treatment
of hydrogen and helium ionization. The latter coupled to a low resolution
simulation can be a source of significant bias against the true
correlation between the \heii and \hi absorbers.

Therefore, we will approximate here the line of sight and redshift
interval average of the $\eta$-parameter as $<log(\eta)> \simeq <log(4
\times \frac{\tau_{HeII}}{\tau_{HI}})> = <log(4 \times R)>$ where R
stands for the R-factor. Through this approximation, the identification
of individual lines and the derivation of column density through their
gaussian profiles are not required. Instead, in each redshift interval we
compute the average of the ratio between the local \heii and \hi pixel
optical depths along the transmission line.  The values for each line of
sight are then averaged to obtain a mean value for the $\eta$-parameter.
We plot the redshift evolution of such calculation in Figure~(\ref{eta}).
Because the errors due to pixel averaging are large, they are ignored.
Instead, the error bars in the figure represent the 1$\sigma$ standard
deviation due to line of sight averaging. For reference, we overplot the
computed value of the $\eta$-parameter in the spectrum of HE2347-4342
from Kriss et al. (2001) and also data from Fechner et al. (2006) (blue
circle) towards HS1700+6416. We obtained the latter by computing the mean
$log\eta$ from the published data in the range z=2.3-2.75. At z=2.5 our
computed value $log\eta = 2.1 \pm 0.1$ is above the observed values which
is mainly due to our larger estimate of the \heii Ly$\alpha$ optical
depth at that redshift. As with the optical depth calculation, the
limitations of our treatment result in an overestimate. However, we came
close enough to the observed values at redshifts later than the \heii
transmission trough, to reinforce our conclusion that our limited
calculation is able to reproduce the much of the observed signatures of
\heii reionization.

\section{Summary \& Conclusions}

We have simulated the late inhomogeneous reionization of \heii by quasars
and the attendant photoheating of the IGM including opacity effects.  We
post-process baryonic density fields from a standard optically thin IGM
simulation with a homogeneous galaxy-dominated UV background which
reionizes \hi and \hei at z=6.5. Quasars are introduced as point sources
throughout the 100 Mpc simulation volume located at density peaks
consistent with the Pei luminosity function. We assume an intrinsic
quasar spectrum $J(\nu) \propto \nu^{-1.8}$ and a luminosity proportional
to the halo mass. We evolve the spatial distribution of the \heii
ionizing radiation field at h$\nu$ = 4, 8, and 16 Ryd using a
time-implicit variable tensor Eddington factor radiative transfer
scheme which we describe. Simultaneously, we also solve for the local
ionization of \heii to \heiii and the associated photoheating of the gas.
This distinguishes our calculation from the work of Sokasian, Abel and
Hernquist (2002) who simulated inhomogeneous \heii reionization but did
not study the thermal evolution of the gas.

We find that the cosmic evolution of the QSO population causes the
individual \heiii regions to overlap and subsequently ionize 90 \% of the
volume at \heiii ionization fractions $\psi_{3} =
\frac{n_{He_{III}}}{n_{He}} \gtrsim 10^{-5}$ by $ z \simeq 2.5$. In
addition, to the \heii and \heiii number density calculation, we also
update the local gas temperature due to the thermalization of the
photoelectrons ejected by the HeII ionization process. As expected, 
the temperature is higher compared to the unprocessed simulation with 
the difference increasing rapidly at redshifts $z \lesssim 3$.
Relative to a self-consistent optically thin simulation where 
\heii is also photoionized by a homogeneous UV background, we find that
optical depth effects result in an increase in the temperature 
of the intergalactic medium at the cosmic mean density level by
a factor of $\approx 1.7$ at $z=2.5$.
The results of our temperature calculation are consistent
with analytic and numerical predictions of the HeII heating effect (Abel
\& Haehnelt 1999; Schaye et al. 2000; McDonald et al. 2000). Finally, we
trace the redshift evolution of the \heii Ly$\alpha$ transmission using
randomly casted synthetic spectra through the simulated volume. The analysis of the mean
transmission allows for the derivation of the effective optical depth of
the 304~\AA~ line.

We have calculated that at $\bar{z} = 2.5 \pm 0.1$, the average of pixel
flux among all lines of sight yields a value for the mean transmission of
$\bar{F} = 0.304 \pm 0.002$ which corresponds to $\tau^{eff}_{304} =
1.190 ~\pm 0.007$. The length of each sightline ($\Delta z=0.2$) is longer 
than the size of the simulation in order to minimize the transmission 
variance across the redshift path. We compute the error only due to the
sightline to sightline variance and find that at z=2.5 
the 1$\sigma$ standard deviation is $11\%$ the mean \lya flux.
When we compare to estimates from \heii
forest spectra observed with the FUSE (Far Ultraviolet Space
Explorer) satellite and HST at the same redshift interval we find that the
our value of the effective depth is comparable but slightly above the observed values.
We attribute the disagreement to our approximate treatment of the inhomogeneous
ionizing radiation that ignores the diffuse component and only focuses on the
point sources' input.

In addition to the mean transmission estimate, we also compute the column
density and b-parameter distribution of the identified \heii Ly$\alpha$
lines and compare them to the corresponding statistical properties of the
HI Ly$\alpha$ lines. We find that an optically thick calculation results in 
extra heating that shifts the b-parameter distributions to higher broadening widths.
In addition and within the limitations of our coarse numerical
resolution, which does not resolve the \hi forest, our calculation shows that
the median \heii b-parameter at $\bar{z} = 2.5$ is $82\%$
the HI broadening width. The latter suggests that the Hubble differential
flow may be the dominant line broadening mechanism which in turn
indicates that \heii lines primarily form in underdense to cosmic mean
density regions.

ACKNOWLEDGEMENTS

This work was supported by NSF grants AST-9803137, AST-0307690 and
AST-0507717. The simulations were carried out on the TITAN IA64 Linux
cluster at the National Center for Supercomputing Applications and on the
IBM Blue Horizon system at San Diego Supercomputing Center under LRAC
allocation MCA098020.

\appendix

\section{Species Concentration Equation in an Expanding Universe}

For a single species medium and in proper coordinates the rate equation
that governs the abundance of ionization state i is given by the
following equation:
                                                                                                         
\begin{equation}\label{eqrates1}
\dot{n}_{i} = - 3 \frac{\dot{a}}{a} n_{i} +
\Sigma_{j} \beta_{ij} n_{j}
\end{equation}
                                                                                                         
In Equation~(\ref{eqrates1}), $n_{j}$ denotes the number density of
ionization states of the species in ionization or recombination coupling
to state i. The ratio $\frac{\dot{a}}{a}$ is a function of redshift and
is equal to the redshift value of the Hubble constant.
                                                                                                         
Dividing by the total number density of the species, $n_{T}$, we get:
                                                                                                         
\begin{eqnarray}
\dot{n}_{i}/n_{T} =  - 3 \frac{\dot{a}}{a} n_{i}/n_{T}+
\Sigma_{j} \beta_{ij} n_{j}/n_{T} \Rightarrow \nonumber \\
\dot{n}_{i}/n_{T} =  - 3 \frac{\dot{a}}{a} y_{i} + \Sigma_{j} \beta_{ij} y_{j}
\label{eqrates2}
\end{eqnarray}
                                                                                                         
The LHS of Equation~(\ref{eqrates2}), can be rewritten as follows:
                                                                                                         
\begin{eqnarray}
\frac{1}{n_{T}} \frac{\partial}{\partial t} n_{i} = \frac{\partial}{\partial t}
(\frac{n_{i}}{n_{T}}) - n_{i} \cdot \frac{\partial}{\partial t} (\frac{1}{n_{T}})  = \nonumber \\
\frac{\partial}{\partial t} y_{i} + \frac{1}{n^{2}_{T}} \dot{n}_{T} n_{i} =
\dot{y}_{i} - 3 \frac{\dot{a}}{a} y_{i}
\label{eqrates3}
\end{eqnarray}
                                                                                                         
Plugging the result back into Equation~(\ref{eqrates2}), we get $
\dot{y}_{i} = \Sigma_{j} \beta_{ij} y_{j}$, which shows that when solving
the rate equation using ionization fractions, instead of number
densities, the equation is independent of the Hubble expansion term.
However, coefficients $\beta_{ij}$ that are pertinent to recombinations
to state i depend on the electron density which is computed on the proper
frame of reference.

\section{Effect of Late Reheating on Hydrogen Ionization Balance}
                                                                                                         
In this calculation, we estimate the neutral hydrogen fraction if an
additional source of ionization and heating is introduced due to the full
photoionization of \heii by a local UV radiation followed and the ejection
of an additional electron per helium atom, assumed to be already singly
ionized. The unmodified neutral fraction is labeled as Case~(I).
Case~(II) represents the updated hydrogen abundance.

\begin{figure*}
\includegraphics[width=6.5in,height=3in]{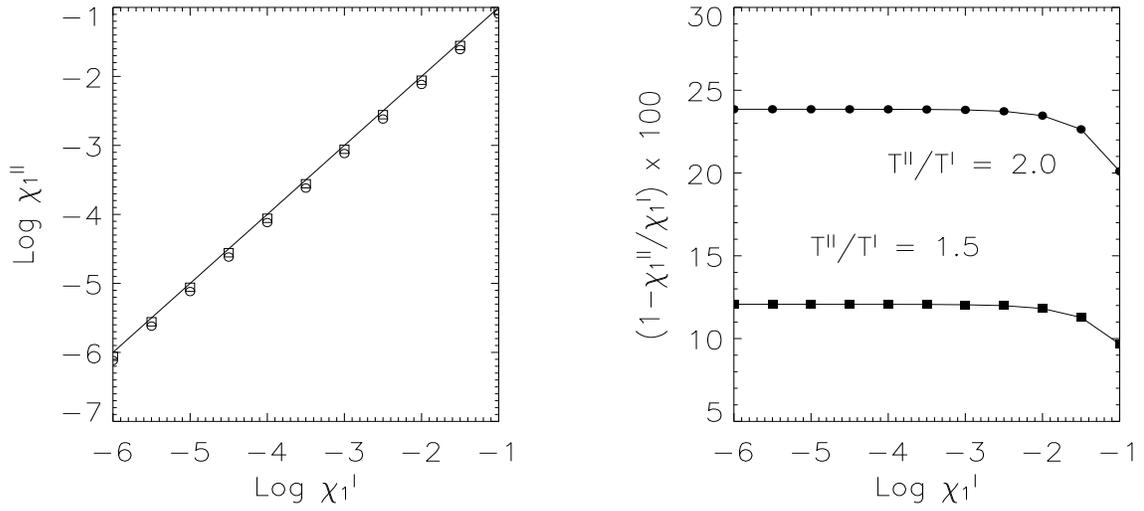}
\caption{Left Panel: $Log \chi^{I}_{1}$ (input) vs. $Log \chi^{II}_{1}$ (output)
for $T^{II}/T^{I} = 1.5$ (squares) and $T^{II}/T^{I} = 2.0$ (circles).
Right Panel: Percentage change in the hydrogen neutral fraction vs.
$Log \chi^{I}$.}
\label{chi12}
\end{figure*}

In Figure~(\ref{chi12}), Equation~(\ref{hieqn3}) was solved for two
values of temperature increase, $T^{II}/T^{I} = 1.5 - 2$. On the left
panel we show the neutral hydrogen fractions in Case~(II) for a range of
Case~(I) inputs. The open squares (circles) represent the smallest
(largest) temperature increase. The solid line stands for no difference
between the two cases (slope of 1). The calculation shows that the
largest temperature increase results in a greater departure from the
straight line, although the degree of such departure becomes smaller at
larger neutral fractions.
                                                                                                         
This is shown on the right panel of Figure~(\ref{chi12}) where we plot
the percentage change in the hydrogen neutral fraction. We see a fixed
degree of change between $\chi^{I}_{1} = 10^{-6} - 10^{-2.5}$ and a rapid
decrease at $\chi^{I}_{1} \gtrsim 10^{-2.5}$. For the range of
temperature increase used in the calculation we can then determine that a
total \heii photoionization would lead to an adjustment of the hydrogen
neutral fraction by 12-24\%. This corresponds to an overestimate of
$\chi_{1}$, in the absence of \heii photoheating, by $\approx 14 - 32$\%.
                                                                                                         
\begin{figure*}
\includegraphics[width=3in,height=3in]{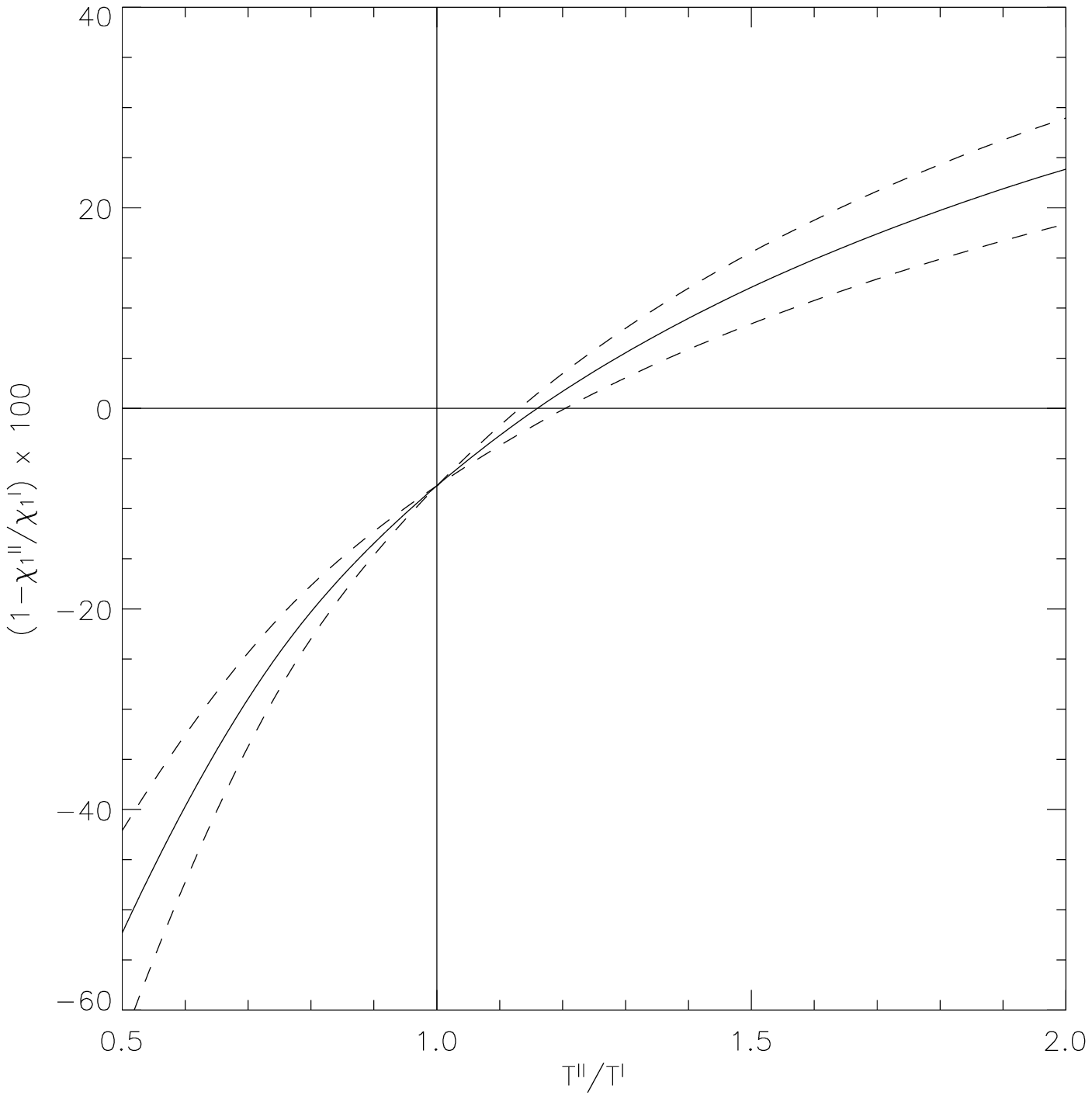}
\includegraphics[width=3in,height=3in]{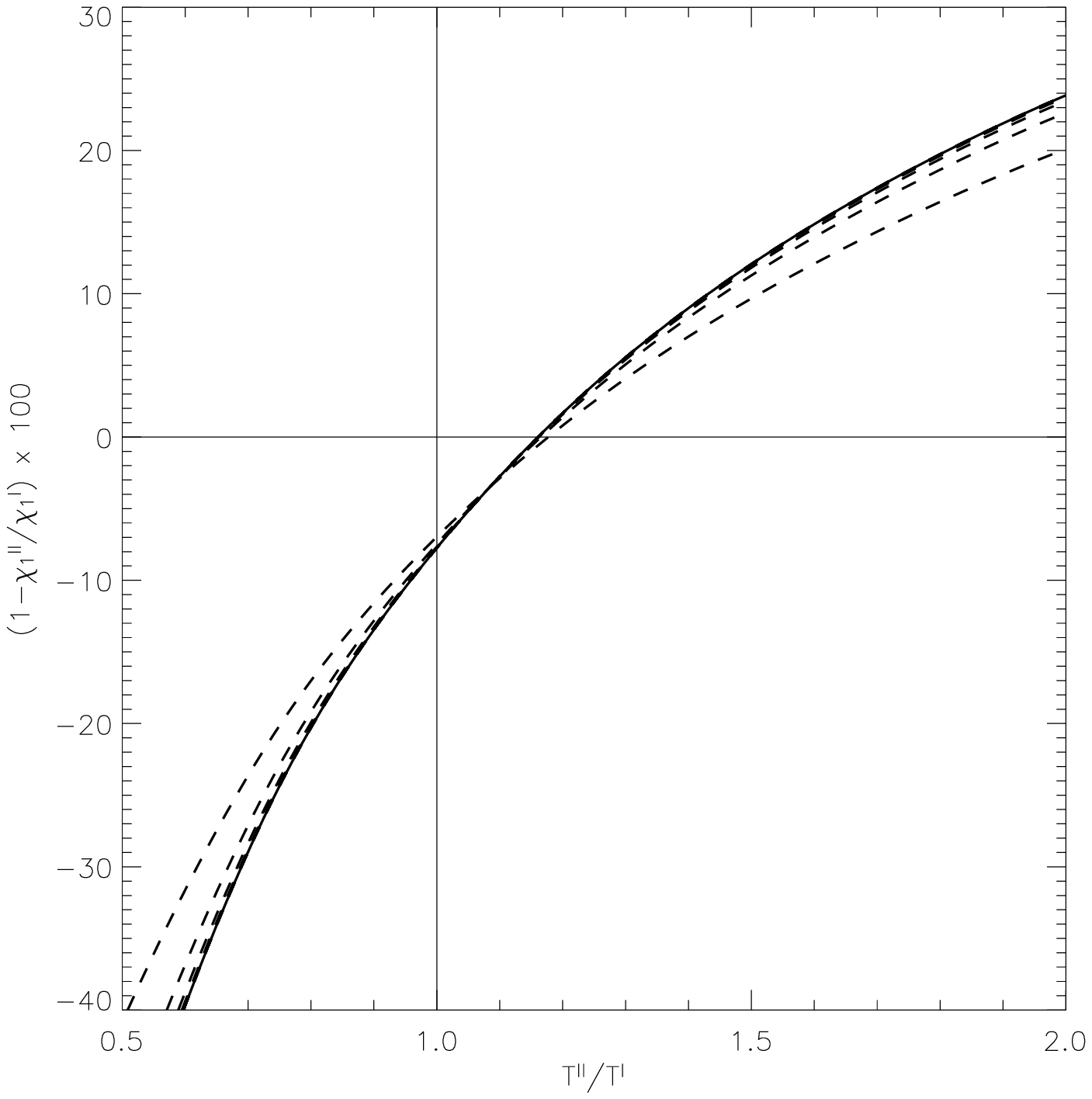}
\caption{Left Panel: Percentage change vs. $T^{II}/T^{I}$ for three
values of the exponent in $\alpha^{(1)}_{R} \propto T^{-\beta}$.
Solid line: $\beta = 0.5$. Upper dashed line at $T^{II}/T^{I} \geq 1$: $\beta = 0.6$.
Lower dashed line at $T^{II}/T^{I} \geq 1$: $\beta =0.4$.
Right Panel: Percentage change vs. $T^{II}/T^{I}$ for $10^{-6} \leq \chi^{I}_{1} \leq 10^{-1}$.
Curves on the upper quadrant from upper to lower (dashed lines) represent
an increase in the hydrogen neutral fraction.}
\label{devtemp}
\end{figure*}

The degree at which any temperature change affects the hydrogen
ionization balance is sensitive to the functional form of the
recombination coefficient.  In general, the coefficient steepens at
temperatures above $T \simeq 10^{4}$ K where the slope becomes a function
of temperature itself. The results shown in Figure~(\ref{chi12}) are
derived with a constant slope of $\beta = 0.5$ which is roughly valid at
$T \simeq 10^{4}$ K.  To explore the effects of the slope on the
adjustment of the hydrogen neutral fraction, due to \heii photoheating,
we show on the left panel of Figure~(\ref{devtemp}), the results of the
calculation for an input neutral fraction of $\chi^{I}_{1} = 10^{-5}$,
$\beta = (0.4, 0.5, 0.6)$ and for a range of temperature ratios
(0.5-2.0).

The upper dashed line at $T^{II}/T^{I} \geq 1$ represents the largest
$\beta$ value (0.6) and indicates that a steepening of the temperature
exponent leads to an increase in the neutral hydrogen fraction
adjustment. This trend is reversed at $T^{II}/T^{I} < 1$ (cooling).  The
latter temperature adjustment range is not related to any \heii
photoheating effects but it was a numerical investigation of the
solution. However, one can make the argument, depending on whether the
UVB drops after a certain redshift, that additional cooling terms, such
as collisional ionization cooling ($He_{II} + e^{-} \rightarrow He_{III}
+ 2e^{-}$) and recombination radiation cooling ($He_{III} + e^{-}
\rightarrow He_{II} + \gamma$), might lower the temperature below the
unperturbed value. However, such possibility is small at least up to $z
\simeq 2$.

We also note from the left panel of Figure~(\ref{devtemp}) that not all
values of a temperature increase lead to a positive neutral hydrogen
fraction adjustment ($1 - \frac{\chi^{II}_{1}}{\chi^{I}_{1}} \geq 0$). In
the $\beta = 0.5$ case (solid curve) a small increase of the temperature
by less than 15
adjusted to a larger value and therefore $1 -
\frac{\chi^{II}_{1}}{\chi^{I}_{1}} \leq 0$. This a result of the
increased electron number density due to the additional photoelectron
ejected from the \heii atom. In this narrow range of temperature increase
the greater electron number density dominates the recombination rate,
which consequently increases to shift the ionization balance towards more
neutral hydrogen.

This effect depends on the initial neutral fraction, as shown on the
right panel of Figure~(\ref{devtemp}). There, we plot the family of
solutions of Equation~(\ref{hieqn3}) ($\beta = 0.5$) for the range of
neutral fractions ($10^{-6} \leq \chi^{I}_{1} \leq 10^{-1}$)  and
temperature adjustment factors ($ 0.5 \leq T^{II}/T^{I} \leq 2.0$). The
solid line represents the solution obtained for $\chi^{I}_{1} = 10^{-5}$
from the left panel. The plot shows that as the neutral fraction
increases the adjustment decreases for $T^{II}/T^{I} \geq 1$. As
discussed in the previous paragraph, the range of $T^{II}/T^{I}$ where
the electron density dominates the recombination rate shifts to the right
towards lower ionization fractions.

\section{Radiative and Collisional Rates}

In this section we document the radiative and collisional rates we used
in our simulation. For the full range of chemical reaction rates
incorporated into the cosmological hydro code Enzo please see Anninos et
al. (1997) and Abel et al. (1997).
                                                                                                         
\subsection{Collisional Ionization and Radiative Recombination of Singly
Ionized Helium}
                                                                                                         
Fits are accurate within 1\% in the temperature range $1-10^{8}$ K.
                                                                                                         
$He^{+} + e^{-} \rightarrow He^{++} + 2e^{-}$ Abel et al. (1997)

In the following fit temperature T is in units of eV: $T =
\frac{T}{11600~K}$

\begin{eqnarray}
\Gamma^{ion}_{2}/n_{e} \equiv k_{5} =
exp[-68.71050990+43.93347633 ln(T) \nonumber \\
- 18.4806699 ln(T)^{2} + 4.70162649 ln(T)^{3} -
0.76924663 ln(T)^{4} +  \nonumber \\
8.11042 \times 10^{-2} ln(T)^{5} -
5.32402063 \times 10^{-3} ln(T)^{6} + \nonumber \\
1.97570531 \times 10^{-4} ln(T)^{7} - 3.16558106 \times 10^{-8} ln(T)^{8}]~cm^{3}s^{-1} \nonumber
\end{eqnarray}
                                                                                                         
$He^{++} + e^{-} \rightarrow He^{+} + \gamma$ Cen (1992)
                                                                                                         
$\alpha^{2}_{R} \equiv k_{6} = 3.36 \times 10^{-10} T^{-1/2} (\frac{T}{1000~K})^{-0.2}
(1+(\frac{T}{10^{6}~K})^{0.7})^{-1}$ $cm^{3}s^{-1}$
                                                                                                         
Temperature is expressed in K.
                                                                                                         
\subsection{Photoionization cross section of Hydrogen and Helium II}
                                                                                                         
\begin{eqnarray}
H + \gamma \rightarrow H^{+} + e^{-} \nonumber \\
He^{+} + \gamma \rightarrow He^{++} + e^{-} \nonumber
\end{eqnarray}
                                                                                                         
The cross section for both rates are expressed through the same
functional form because singly ionized helium is a hydrogen like atom
(from Osterbrock 1974).
                                                                                                         
$\sigma = \frac{A}{Z^{2}} (\frac{\nu}{\nu_{Z}})^{4}
\frac{ exp[4-4(arctan \epsilon)/\epsilon] }{1-exp(-2\pi/\epsilon)}$
                                                                                                         
$A = 6.30 \times 10^{-18}$ $cm^{2}$, $\epsilon = (\frac{\nu}{\nu_{Z}}-1)^{1/2}$ and
$\nu_{Z} = 13.6 \times Z^{2}$ ($Z = 1,2$).
                                                                                                         
For consistency with the notation used in the main text, we need to clarify that we
we label $\nu_{1} \equiv \nu_{(Z=1)}$, $\nu_{2} \equiv \nu^{HeI}_{(Z=2)}$ and
$\nu_{3} \equiv \nu^{HeII}_{(Z=2)}$.
                                                                                                         
For numerical convenience we use an approximation to the full \heii cross
section in the form $\sigma_{HeII} = \frac{A}{4}(=\sigma^{o}_{HeII})
(\frac{h\nu}{h\nu_{3}})^{-3}$, where $h\nu_{3} \equiv 54.4$ eV. Such an
expression allows an analytic calculation and fast reconstruction, via a
polynomial fit, of the energy density profile between four frequency
points. The power law approximation of the cross section is estimated to
be accurate to within $\approx 3$\% (Haehnelt \& Abel 1999).

\newpage

\clearpage
\newpage

\clearpage
\newpage

\clearpage
\newpage

\clearpage
\newpage

\clearpage
\newpage

\clearpage
\newpage

\clearpage
\newpage

\clearpage
\newpage

\clearpage
\newpage

\clearpage
\newpage

\clearpage
\newpage

\clearpage
\newpage

\clearpage
\newpage

\end{document}